# Anomalous Edge Contrast in Magnetic Force Imaging of 2D Topological Insulators

Timothy W. Carlson[1,2], Motahhare Mirhosseini[1,2], Allison Swyt,[1,2] Gabriel Marcus[1,2,3], Daniel Lazega[1,3], Scott Mcguire[1,3], Lindsey J. Gray[1,2], Anthony Walsh[1,2], Swathi Kadaba[1,2], Maria Koleśnik-Gray[4], C.C. Dun,[1,5] Vojislav Krstić[1,2,4], D.L. Carroll[1,2,*]

[1] The Nano and Quantum Technologies Laboratory, Wake Forest University, 95W 32nd St. Winston Salem NC 27105, USA
[2] Department of Physics, Olin Physical Laboratory, Wake Forest University, Winston Salem NC 27109, USA
[3] Quoherent Inc. 2860 Lyndhurst Avenue, Winston-Salem, NC, 27103, USA
[4] Department of Physics, Friedrich-Alexander-Universität Erlangen-Nürnberg (FAU), Staudtstr. 7, 91058, Erlangen, DE
[5] Lawrence Berkeley National Laboratory, Berkeley, CA, 94720, USA

* Corresponding author: carroldl@wfu.edu

## Abstract

Magnetic force imaging (MFM) of 2D topological insulators, such as crystalline plates of $Bi_2Te_3$ and $Bi_2Se_3$, reveals anomalously large contrast around the plate edges even under ambient imaging conditions. And while there are several known contrast mechanisms in MFM, these cannot easily explain the magnitude of contrast observe for such systems. In this work we use multi-parameter magnetic force phase imaging together with a simple analytical model to suggest that the observed magnetic edge-features can be associated with two sources: a *pseudo*-Faraday effect in the conductive edge channels of the topological system and magnetic scattering in those channels caused by the proximal, localized field of the tip. These results suggest novel MFM contrast mechanisms unique to topological systems and point to the utility of multi-parameter MFM techniques in studies of topological-material-based circuit elements.

Topological insulators (TIs) are promising for applications in the coming "quantum revolution" [1,2,3] and 2D TIs [4,5,6] are particularly important in this regard. Here we focus on the set of 2D chalcogenide compounds with strong spin-orbit coupling. These $\mathbb{Z}_2$ topological class compounds are gapped ($E_{gap} \sim$ 0.1-0.3 eV), characterized by time-reversal symmetry (TRS) [7], and well-described by the BHZ model. [8] Crystals of representative, layered $M_2X_3$ structures are straight forward to synthesize. [9] Five covalently bonded atomic planes in this structure form a quintuple layer (QL) of 1 nm thickness. For undoped $Bi_2Te_3$ or $Bi_2Se_3$, 1 - 6 QLs are 2D topological insulators. Larger layer numbers typically drive hybridization between layers and the formation of a 3D topological system. In 2D, "topological edge states" occur as a continuum or band of states in the bulk gap but localized to the edge of the crystal as seen in angle resolved photoelectron spectroscopy (ARPES). [10] Magneto-transport in 2D mercury telluride (HgTe) quantum wells, as well as 2D $Bi_2Se_3$ flakes, has shown spin-momentum locking associated with their 1D perimeter. [10]

Local, real-space probes of topological states in 2D TIs are intriguing, but face challenges. Attempts to resolve topological edge states have included cryogenic scanning tunnelling microscopy and spectroscopy (STM/STS), which shows bandgap features in the local density of states (LDOS) at the edge of $Bi_2Se_3$ crystals. [11] Similarly, Sn and Cd-doped $Bi_2Te_3$ 3D crystals have shown electron density oscillations near the cleaved edges of the crystal [12] which are proposed to be associated with topological states at those edges. While such studies do demonstrate electronic features within the bulk bandgap localized to the crystal edge, direct interpretation of STS-derived LDOS on edges can be confounded by local hybridization with support substrates. [13] Near-field scanning optical microscopy (NSOM), has been used to isolate local spin and charge currents [14] and suggests that plasmon oscillations in the $Sb_2Te_3$ might be supported by highly conductive topological states. However, ambiguity arises due to interactions between conduction band electrons and crystal boundaries. [15] But, most experimental work on 2D TIs [16] are performed in cryogenic, non-perturbative environments, and lack spatial resolution, making tests of expected robustness difficult. [17] Importantly, the temperature behavior of such topological systems is directly related to the entropic manifestation of quantum information within such systems, [18] meaning local experimental probes of 2D TIs in a *noisy regime* are applications relevant.

Edge states of 2D topological systems are typically associated with spin-momentum locking and persistent current flow in the presence of applied fields. So, such crystals may have some magnetic response when examined with magnetic probes such as magnetic force microscopy (MFM). Such magnetic response should be robust to temperature perturbations due to topological protections of the states. [5] However, we might also expect for the fields associated with commercial MFMs to be too small to induce any such response from 2D TIs.

The 2D TI is a Quantum Spin-Hall system with conductive edge channels in which currents flow in both directions equally, as Kramers pair states, when no field is applied. Applied magnetic fields scatter current through spin accumulation ($\mathbf{B}_{||}$) and time-reversal symmetry (TRS) breaking ($\mathbf{B}_{\perp}$). Coupling to the helical edge states opens an energy gap $\Delta_z = g_s \mu_B B$, ($g_s$ is the electron g-factor, $\mu_B$ is the Bohr magneton). [19] Counter-propagating electron helical edge states mix and when the Zeeman energy exceeds the intrinsic spin-orbit coupling gap, it induces elastic backscattering. [20] So, both $\mathbf{B}_{||}$ and $\mathbf{B}_{\perp}$ can drive a persistent current which depends on the size of the field applied. [21] For $\mathbf{B}_{\perp}$, $\Delta_Z \sim E_{Th}$ (Thouless energy). Since $E_{Th} \sim \hbar D/L^2$ where D is the diffusion constant, L is the length of the system, for the case of few layer $Bi_2Te_3$ we can expect $E_{Th} \sim 10$ μeV – 100 μeV. The Zeeman energy of a (0.01 T) or (100 G) MFM tip is ~ $5.79 \times 10^{-5}$ eV. This combines with any spin-accumulation effects of $\mathbf{B}_{||}$ to create a persistent charge current. [22] Indeed, calculations [23] show persistent currents of ~ 300 nA/channel can be expected for homogeneous fields of only 10% of the flux quantum $\Phi/\Phi_0 < 0.1$ [see Figure 6 of reference 24]. But scattering from this mechanism will not be strong at the field strengths of an MFM tip, leading to extremely small phase-imaging contrast (estimated in SI). [24]

The edge current density in one direction can be estimated (SI) as ~ 10's of μA. [25] If a localized field is introduced, a portion of the edge state electrons are scattered introducing an asymmetry for current in one direction. Localized field sources such as magnetic nanoparticles, proximal to the edge of a 2D TI yields such non-TRS scattering. [26, 27] Such mechanisms also lead to significant magnetoresistance. [26] In MFM imaging, the MFM tip is also proximal with the edge of the 2D TI crystal being imaged. We suggest the small field of the tip will induce magnetic scattering locally in the 1D channels of the perimeter of the $Bi_2Te_3$ plate.

Localized magnetic tip scattering at the 2D plate edge means that persistent currents may occur even when applied fields are smaller than the Zeeman field discussed above. But a commercial MFM's sensitivity to such effects should be weak since the currents would be small. Thus, as a rough estimate, it seems plausible that ambient MFM might be able to resolve the fields from edge states, but that such effects are likely to be dominated by other contrast mechanisms such as dielectric response (see estimates in SI).

Curiously, this is NOT what is observed. Edge contrast on 2D TIs in ambient MFM imaging is surprisingly large and can be linked directly to a magnetic response of the system. In this work we demonstrate that this enhanced contrast effect is unique to topological materials. We show that the enhancement derives almost entirely from a "Faraday-like" mechanism that originates with the tip dither.

The details of the solution synthesis route for our $Bi_2Te_3$ plates are given in the (SI). The synthesis produces plates within the range of 1 to 2 μm in diameter and 3 to 30 nm thick which is believed to span the range from 2D to 3D behavior. Sample characterization provided in SI provides a basis of comparison to materials of previous reports. [28] Specific attention is given to creating stoichiometric layers as it is well known that some synthesis routes can result in Te capping between these layers. [29]

Scanning Probe Microscopy (SPM) was carried out in ambient laboratory conditions (25°C RH 50%) using a Digital Instruments (Bruker) D3100 for atomic force microscopy and phase calibration, and a JEOL 5200 SPM for point-by-point magnetic force microscopy. Point-by-point interleaving lift mode is where the cantilever is lifted at each point of the scan. This is chosen in these studies as opposed to the more common line-by-line approaches as it is less susceptible to drift when imaging with an open environment cantilever at the slow scan speeds made necessary by Faraday effects. Freshly cleaved HOPG support substrates were used for analytical MFM. The substrates were characterized for magnetic response. Other substrates such as cleaved mica and thermally grown $Si/SiO_2$ were used for comparison.

**Figure 1a** shows the imaging geometry. Data from a ~800 nm diameter ~ 10 nm thick $Bi_2Te_3$ plate on HOPG, are shown in **Figure 1b and 1c** as a 3D rendering at low pixel resolution and in **Figure 1d and 1e** as a 2D height map at higher pixel resolution.

**Figure 1b** and **1c** are low resolution height and phase images, the individual pixels can be seen and a direct correspondence can be made at each point. **Figures 1d** and **1e** are a higher resolution topographic-phase image pair as is normally seen in publications. In **Figure 1e**, the phase shift on the HOPG from free cantilever oscillation to imaging height: $\Delta\phi \sim -1^o$, consistent with literature for HOPG. [30] At the sample's edge a shift of $\Delta\phi \sim +3^o$ occurs <u>from the HOPG substrate</u> ($+2^o$ from free oscillation). Once the tip is clearly over the plate itself and away from the edge, $\Delta\phi \sim -0.5^o$ from free oscillation. Reproducibility of phase determination is $\sim 0.1^o$. The plane of the 2D crystal is attractive to the tip whereas the edges are repulsive. [31] Careful examination reveals a slight asymmetry in the edge contrast with the edge on the right-hand side stronger than the contrast on the left-hand side. The scan direction goes from left to right in this image. Changing the scan direction to right to left changes flips the asymmetry.

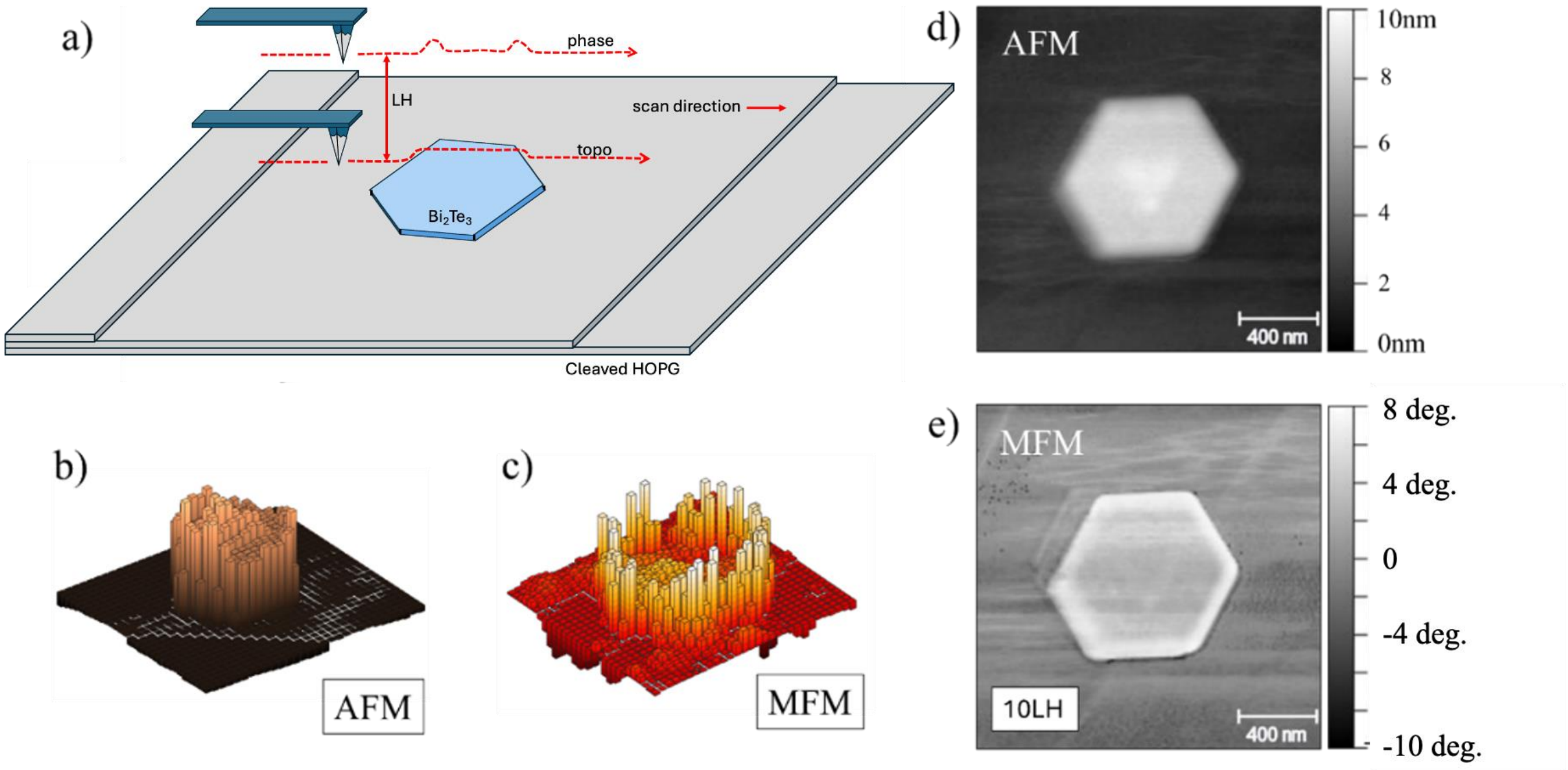

**Figure 1**: Point-by-point-lift magnetic force microscopy **a)** The measurement scheme, **b)** A 1.5 x 1.5 μm$^2$ topographical image (AFM) is shown with a (30 x 30) pixel resolution. Each column in **b)** has a corresponding column in **c)** which is the phase measured each location at a 10nm LH above the plate. **d)** and **e)** are the images of **b)** and **c)** now in 256 x 256 pixels and viewed in a 2D projection. The support substrate is freshly cleaved highly oriented pyrolytic graphite (HOPG).

In **Figure 1**, the contrast at the $Bi_2Te_3$ plate edge, where we expect topologically protected states, is larger than what our rough estimates above propose and this is what we wish to understand. There are published models for contrast in MFM that estimate electrostatic and magnetic forces. [32] Applying these to the image gives edge currents that are unrealistic. (see SI) Also, most of these models predict $\Delta\phi \sim 1/z^2$ which is not reproduced by our data (**Figure 2**).

The contrast enhancement for $Bi_2Te_3$ is seen for both the Bruker instrument using line-by-line lift mode as well as the JEOL using point-by-point imaging modes. It is seen for $Bi_2Se_3$ plates of similar dimension. However, images of topologically trivial dielectric plates of similar dimensions, such as $MoS_2$ or metal plates such as Au, do not show this enhanced contrast at the plate edges. These comparisons are made in **Figures S23, S24, S25** in the (SI). Note that dielectrics such as $MoS_2$ show some edge contrast as expected from established models [32] but the magnitude observed in $Bi_2Te_3$, $Bi_2Se_3$, and $Sb_2Te_3$ is significantly larger than predicted by standard models.

To address this, we propose that the interaction between MFM probe and a 2D topological insulator can be thought of in terms of these components:

1. a capacitive attraction between tip and substrate which differs from that between tip and sample,

2. a "persistent current" established by magnetic scattering when the tip's localized dipole field is brought near the edge,
3. an induced (Faraday) current in the conducting edge channels due to the tip dither,
4. an induced (Faraday) current associated with the scan, (scan speed dependent)
5. damping of the tip oscillation from:
    a. dielectric dissipation in the $Bi_2Te_3$,
    b. magnetoresistance due to local-field scattering in edge channel,
6. interactions between conduction band electrons and edge states. (not considered in this study)

Multi-parameter imaging where Δϕ is observed as the imaging conditions of height, z, and tip oscillation amplitude, A, are varied, allows mechanisms of phase shift to be inferred. From reference [33]:

$$\Delta\phi = -\frac{Q}{k}\left[\frac{dF(x,y,z)}{dz}\bigg|_{magnetic} + \frac{dF(x,y,z)}{dz}\bigg|_{electronic}\right] \quad (1)$$

$k$ is the cantilever's spring constant. $Q$ is the quality factor of the oscillator. For MFM imaging systems with induced currents and fields, torsional and damping effects can alter $k$ and $Q$ near the sample.

**Figure 2** maps the variation in phase contrast as oscillation amplitude (A) and lift height (z) are varied across the plate **(Figure 2a** and **Figure 2b** respectively**)**. The oscillation amplitude (A) in **Figure 2a** is changed over the range of only ~ 2nm (roughly 10% of the imaging amplitude). The phase shift is measured for each amplitude of oscillation with the tip held constant at z = 100 nm. This is done for each pixel of the image in **Figure 2**. With the cantilever positioned over the sample edge Δϕ ~ - 1° to - 2° is observed for A ~ 18.2 nm and Δϕ ~ 4° for A ~ 29.3 nm. This is the phase shift from the free oscillation far from the sample to the tip placed 100 nm above the sample.

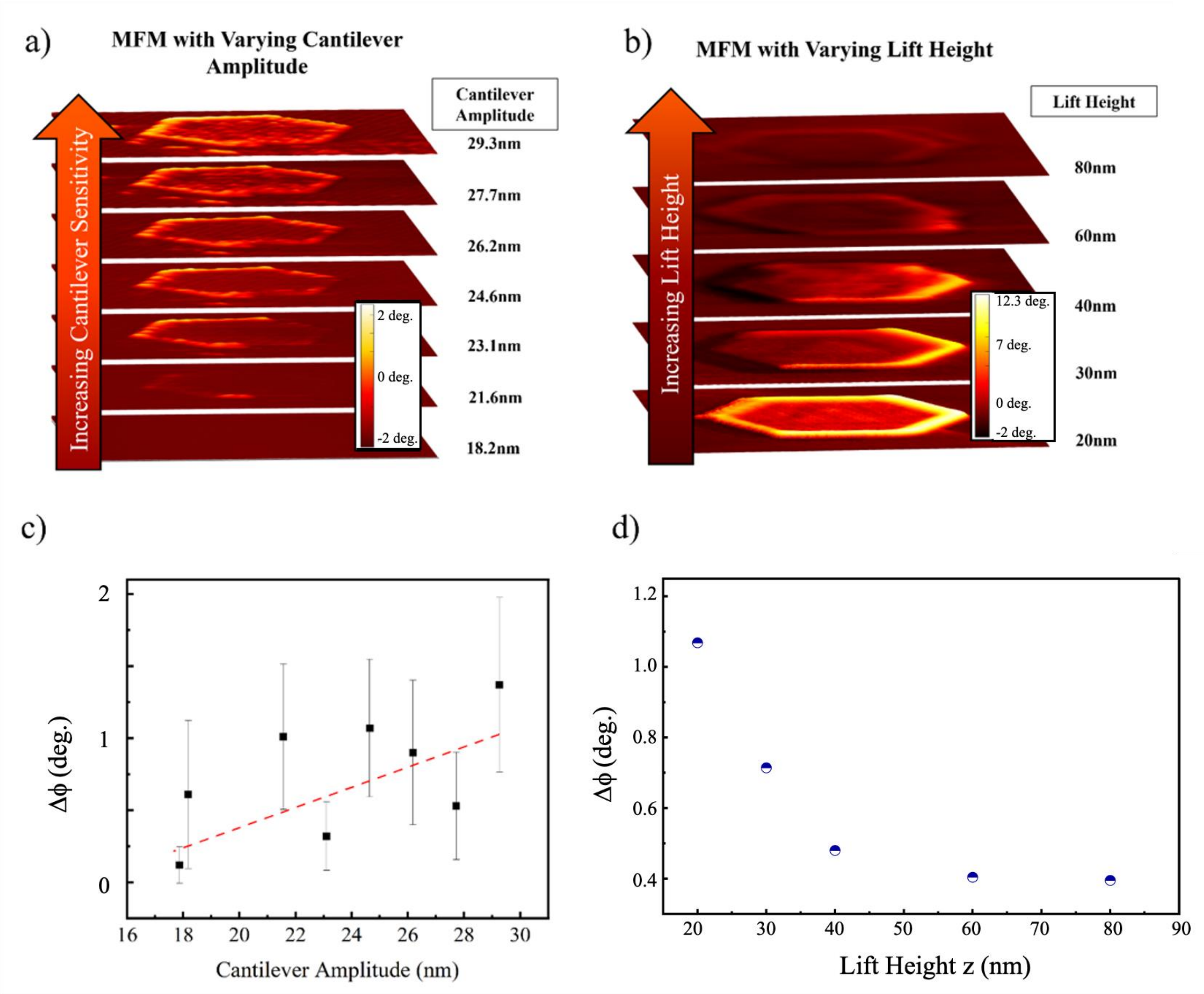

**Figure 2**: $\Delta\phi$ as imaging parameters are varied. **a)** $\Delta\phi$ at z = 100nm (LH) vs. cantilever oscillation amplitude A. Plate is ~ 20 nm thick, and 1 µm dia. **b)** $\Delta\phi$ as a function of z, with drive amplitude A ~ 20.6 nm. Plate is ~ 6 nm thick, and 1 µm dia. The scan direction has been reversed to show the change in the asymmetry of the edge contrast. **c)** Averaged $\Delta\phi$ from the plate edge vs. A, z = 100nm **d)** Averaged $\Delta\phi$ from plate edge vs. z. A ~ 20.6 nm.

The larger the amplitude of oscillation, the more dF/dz begins to look like a finite difference evaluated at the mean value of the swing amplitude, $\Delta F/\Delta z|_{mean}$. The phase shift seen by the MFM at larger amplitudes will be larger and small changes in A should give an approximately linear response in phase shift. [34] But, from **Figure 2c,** $\Delta\phi/\Delta A \neq$ constant. To understand this, we suggest that the dithered magnetic tip induces currents in the conductive edges of the plates, through scattering and Faraday mechanisms. So, when A is reduced, the tip will move more slowly through its equilibrium position and dB/dt decreases. Thus, the current will also decrease. The combination of field change and sensitivity change means a non-linear response results.

In **Figure 2, a)** and **b)**, Δϕ around the perimeter is not constant. Even though there are numerous scattering events in the process of imaging the edge, we expect the current to occur in a continuous loop. Thus, asymmetries in loop images should reflect only asymmetries of the tip. However, this asymmetry occurs only in the scan direction and in proportion to the tip velocity of the scan. Thus, it appears as though the lateral motion of the tip ($v_{||}$) across the sample creates its own dΦ/dt through the edge loop. This generates opposite currents in the loop as the tip approaches and retreats from the loop; together with the Faraday inducing dither. The effect is quantified as a function of scan speed in **Figure S4**. For slow scan speeds, averaging the phase shift around the circumference allows the asymmetry to cancel out. **Figure 2 c)** and **d)** show this averaged phase, obtained from the images in **Figure 2 a)** and **b)**, as a function of the imaging amplitude and the LH, (z), respectively.

The two image sets in **Figure 2** are taken on plates with different thicknesses with **Figure 2 a)** ~ 20 nm thick, and **Figure 2 b)** ~ 6 nm thick. Both are roughly 1 μm in diameter. Normally, we would expect the 20 nm thick plate (20 QL of the $Bi_2Te_3$) to express 3D TI behavior and currents would be induced in the metallic current states across its surface [35]. But in this case, we clearly see a bright repulsive edge and an attractive "bulk." Contrast, is the difference between background and features on the sample, will appear to change as dielectric forces changes and magnetic forces change from sample to substrate. The images are scaled the same, but note that the background can shift with z, as well as the edge.

Δϕ generally, doesn't depend strongly on thickness. There are two explanations consistent with this observation. The first is that it is a 2D TI. A nearly non-dissipative current *exists on the edge but not across the surface*. There will be only small, strongly dissipative currents due to conduction band electrons at room temperature across the surface. This occurs if the synthesis route results in weaker than expected inter-layer interactions, making the stack of 20 QLs behave like 20, 2D TIs. [35] Then the difference in thickness gives a different edge current density on the thick plate and the thin plate. But the same overall current will be induced for the same imaging geometries and conditions since the dB/dt term is the same.

Second, the 20 QL system is a 3D TI. The current density induced in the surface states are 20X smaller than the current density seen along the edge. Topological surface states would be nearly non-dissipative, but they spread out across the conductive surface creating a weak field of a sheet current as opposed to the edge ring's strong toroidal field.

The attractive capacitive force dominates on the surface in both cases. The difference will be the Δϕ vs. z and Δϕ vs. A functions. First, we examine the edge response. **Figure 2 c)** presents Δϕ vs. A along the edge of the sample. **Figure 2 d)** shows Δϕ vs. z also along the edge. Δϕ is taken relative to free-oscillation.

**Figure 3** below shows $\Delta\phi_{surface}$ vs. z the contrast between the center of the plate and the background HOPG.

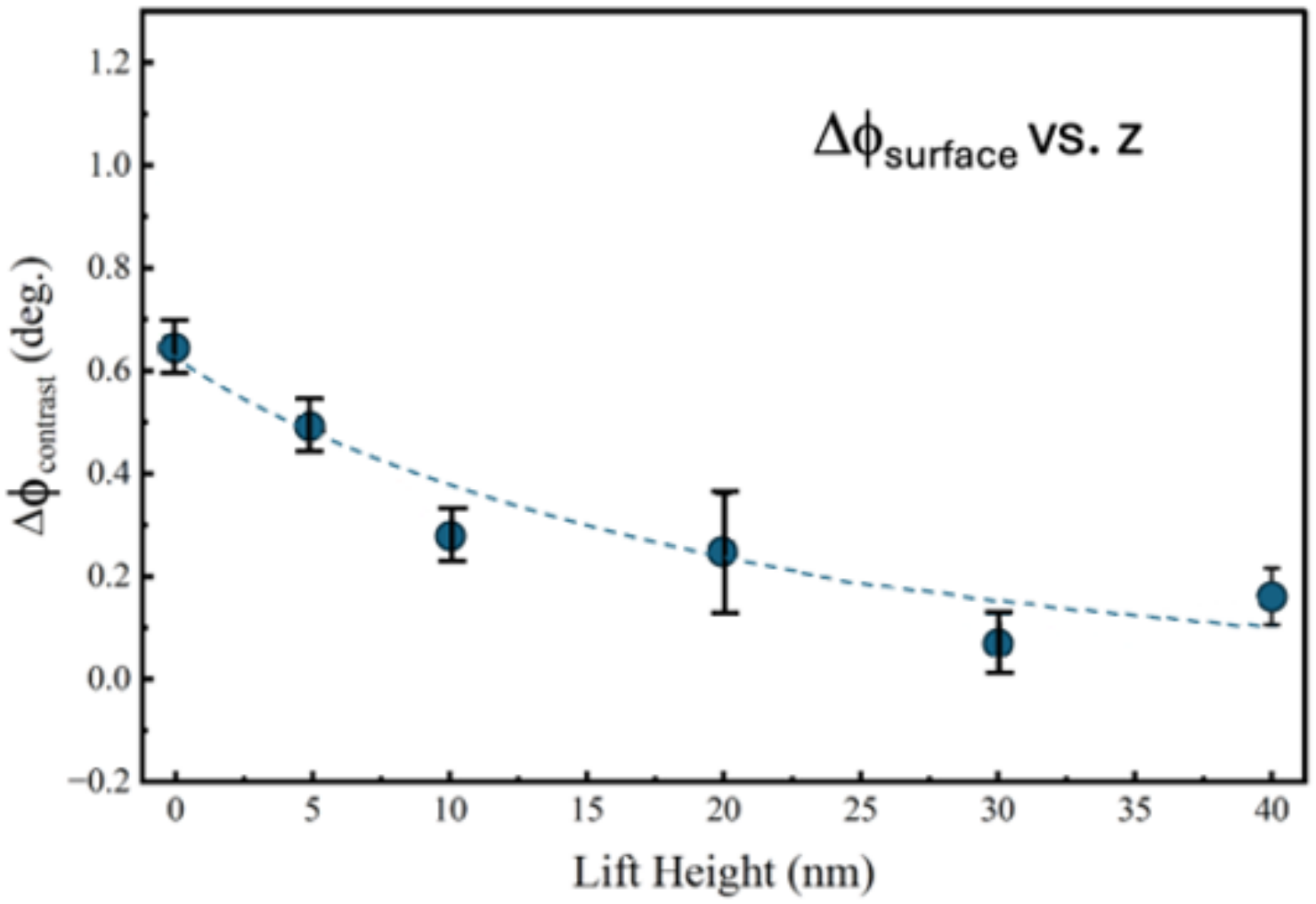

**Figure 3:** $\Delta\phi_{surface}$ shown in contrast to the background HOPG. Since HOPG is typically between -2.0 degrees and -1.0 degrees, the difference between the phase shift over the plate surface and the HOPG is positive. So as $z \to 0$ $\Delta\phi - \Delta\phi_{HOPG} \to 0.65$ degrees, meaning $\Delta\phi \sim -1.35$ degrees.

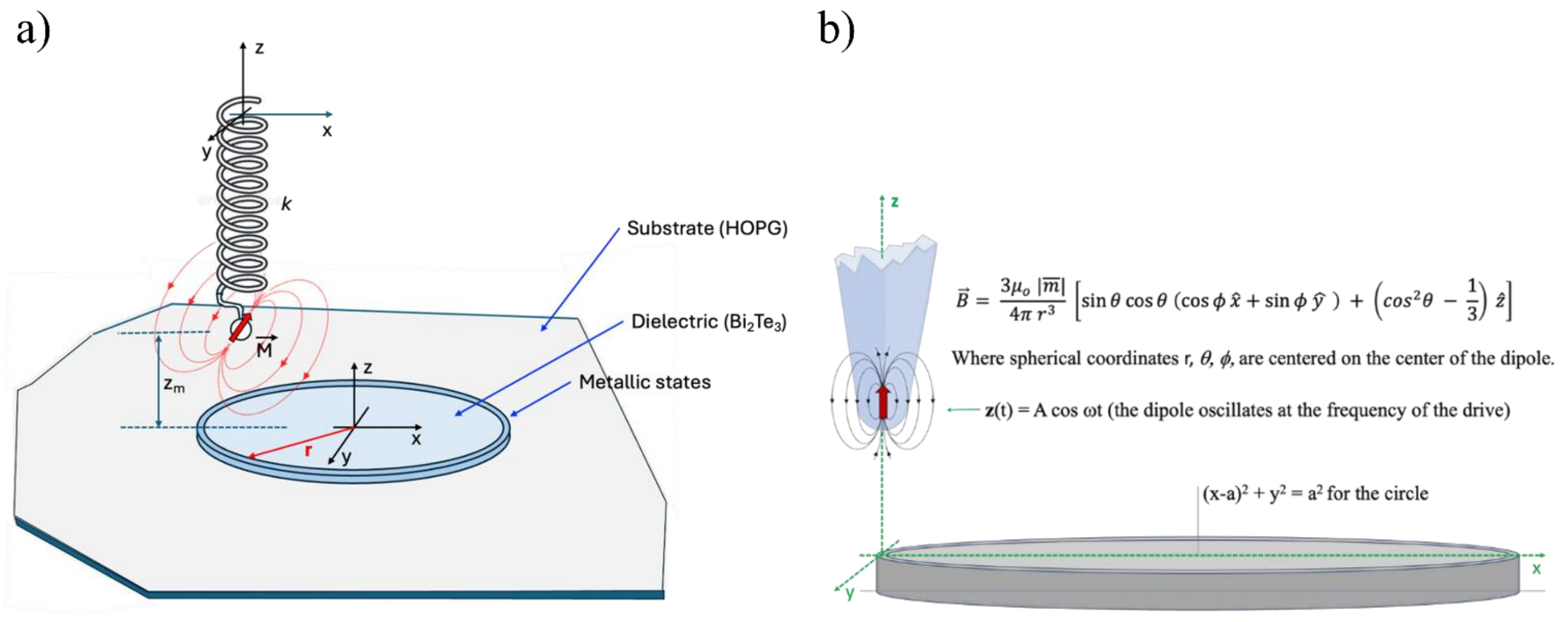

**Figure 4:** The geometry used for determining the magnetic and electrostatic field components of the force gradient. **a)** The capacitor formed by the semi-spherical metal-coated tip and the metallic support substrate. There are two dielectrics in this system, the air ($\varepsilon_0$) and $Bi_2Te_3$ ($\varepsilon_r$). The current ring is treated as an isolated wire. **b)** The tip is a small magnetic dipole, and the current ring is large with a spatially extended magnetic field. We can treat the dither of the tip as a simple harmonic time-dependent motion in z.

The expressions for the $\Delta\phi$ of magnetic effects and the capacitive effects are derived in detail in the SI and used here for comparison to data. Referring to **Figure 4,** we model the system as a dielectric disk with a conductive edge-loop. The electrical forces in MFM are: $F_{electro} \sim F_{VdW} + F_{cap.}$ We ignore short-range Van der Waals interactions.

Starting with the capacitive and magnetic interactions on the HOPG and on the surface of the plate. The tip, HOPG, and $Bi_2Te_3$ form a partially filled dielectric capacitor. [36]

For z < t, the phase contrast between the HOPG and dielectric plate due to capacitive effects is approximated by:

$$\Delta\phi_{elec} \sim -\frac{Q_{HOPG}}{k}\frac{\pi\,\varepsilon_o R\,(U_c + U_a)^2}{z^2} - \frac{Q_{BiTe}}{k}\frac{\pi\,\varepsilon_{BiTe} R\,(U_c + U_a)^2}{t^2} \tag{2}$$

where $R$ is the radius of the tip. $U_c$ is the contact potential, $U_a$ is any additional applied potential. t is the thickness of the $Bi_2Te_3$ plate, and $z$ is the lift height. $Q_{HOPG(BiTe)}$ is the Q factor near the substrate. For z < t, this can be approximated as:

$$\Delta\phi_{elec} \sim -\frac{Q_{HOPG}}{k}\frac{\pi\,\varepsilon_o R\,(U_c + U_a)^2}{z^2} - \frac{\pi^2 A^2\,R}{\omega\tan(\delta)} \tag{3}$$

$$\sim -\,\alpha\frac{(U_c + U_a)^2}{z^2} - \beta\,A^2 \tag{4}$$

$\alpha = Q_{HOPG}\pi\,\varepsilon_o R/k$, $\beta = \frac{\pi^2 R}{\omega\tan(\delta)}$. For z < t, this agrees with **Figure 3** as derived in SI.

If the plate surface has topological, non-dispersive states, $\Delta\phi$ will have an additional magnetic component: [37]

$$\Delta\phi_{\mathrm{mag\,BiTe}} \sim -\frac{1}{2}\frac{z_c^4}{z_m^4} \tag{5}$$

$$z_c^4 \equiv \frac{3}{64\pi}\frac{\omega}{k}\frac{{m_{eff}}^2}{\rho_{BiTe}}$$

$z_m$ is the distance to the magnetic dipole of the tip ~ z, the position of the tip apex. $\rho_{BiTe}$ ~ resistivity, $m_{eff}$ ~ effective dipole moment. The negative sign comes from damping and from our estimates in SI, this term is of the same order as the capacitive $\alpha$-term. This is not obvious in our data, **Figure 3** and **SI,** suggesting that the thicker layered crystals may behave as a 2D system.

For magnetic interactions on the edge of our samples, we use the approximation of [38]. $z_m \geq$ tip diameter. Plate diameter ~ 1 µm > $z_m$ . The edge current gives a phase shift (SI):

$$\Delta\phi = \frac{Q}{k}\frac{dF_z}{dz} \sim -\frac{I\,m_{eff}}{2\pi}\frac{Q}{k}\left[\frac{1}{z^3}\right]\;;\; x = 0 \tag{6}$$

For the tip positioned over the edge of the plate $\Delta\phi$ vs. $1/z^3$ should be linear. But the data of **Figure S16** has two distinctly different slopes for different imaging height regimes. I ~ 79 nA for 20 nm < LH < 40 nm and I ~ 1.04 μA for 40 nm < LH < 80 nm. Apparently current shifts two orders of magnitude with a small change in LH. The larger current seems to occur with the tip furthest from the substrate. Neither seem physical.

However, I could be "induced" through a Faraday component and a direct magnetic scattering component. The "topological edge channels" in a TI are highly conductive. At room temperature there will be a small carrier concentration in the conduction band of $Bi_2Te_3$. The Faraday induction component into the edge channels is approximated as:

$$I_{Faraday} = \frac{3m_{eff}\gamma z_{\mathrm{m}} r}{8\pi\rho(z_{\mathrm{m}}^2 + r^2)^{5/2}}\dot{z}_{\mathrm{m}}$$

γ is a geometrical factor related to the plate shape. There is also a damping factor ($\delta$) such that: $Q \rightarrow Q_{edge}$.

$$Q_{edge} \sim \frac{4\pi^2\rho m\omega r^6}{{m_{eff}}^2(4z^2 - r^2)}$$

$\rho$ = resistivity, r = radius of plate (**Figure 4b**). [34] Yielding:

$$\Delta\phi_{Faraday} \sim -\frac{Q}{k}\frac{3A\omega m_{eff}^2\, r^{-5}}{16\pi^2\rho(\mathrm{x}^2+1)^{5/2}}\left[\left(\frac{\mathrm{x}}{[4+\mathrm{x}^2]^{\frac{3}{2}}}\right)+\left(\frac{4\mathrm{x}}{[4+\mathrm{x}^2]^2}\right)-\left(\frac{16x^3r^{-1}}{(4+\mathrm{x}^2)^3}\right)-\left(\frac{3x^3}{[4+\mathrm{x}^2]^{\frac{5}{2}}}\right)\right] \quad (8)$$

written in terms of the far-field Q and $x = r/z_{\mathrm{m}}$ .

Adding **equation 6,** associated with any $I_{persistent}$ and **equation 8** Faraday related, together we get the overall contrast phase shift of magnetic components. In the limit A or $\omega \rightarrow 0$ only the $I_{persistent}$ term is left. In **Figure 2c**, this trend was not very linear. As we go across a larger span of z heights it converges to something more linear. In **Figure 4** $\Delta\phi$ vs. A, for three different lift heights, extrapolates to the y-intercepts which should be the $\Delta\phi_{mag}$ of $I_{persistent}$. If we assume that the highest LH would have the least amount of scattering (**Figure 4b**), then it should represent the smallest possible $\Delta\phi$ and Q (free). At 200nm LH, the "persistent current" is calculated to be $I_{persistent} \sim$ 802 nA. This is similar to theoretical predictions for $\Phi/\Phi_o \sim 0.2$ ($\Phi_o$ is a quantum of flux) in HgTe. [39]

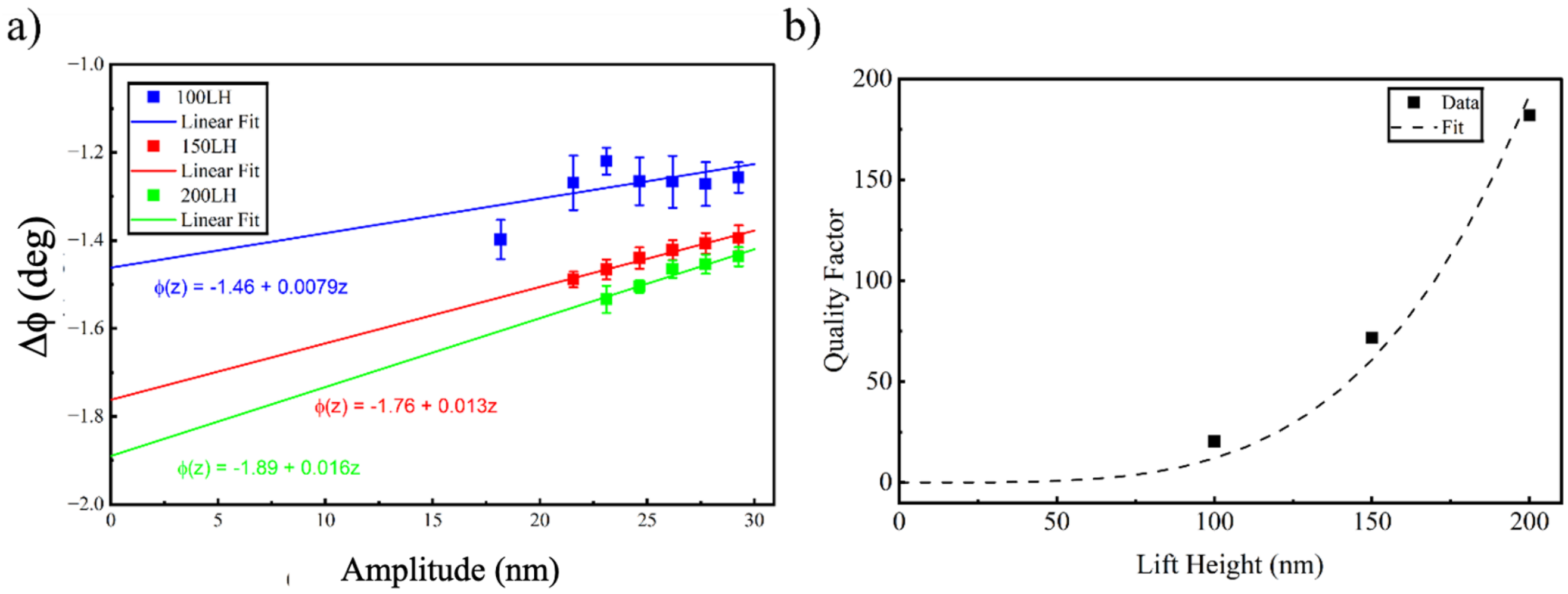

**Figure 4: (a)** Δϕ vs. A on the edge at $z_m$ = 100nm, 150nm, and 200nm. **(b)** Plot of Q vs. $z_m$ (calc. from **(a)**) with best fit line (dashed).

Note: the scattering strength of the proposed mechanism for "persistent" currents should depend on the lift height. So, the persistent current from $\lim_{A \to 0} \Delta\phi_{mag}$ will also depend on height **Figure 4(a)**.

In summary, to explore edge contrast in. The anomalous contrast enhancement in ambient MFM phase images, around the sample edges of 2D chalcogenide TI's, can be identified with Faraday induction of currents together with scattering mechanisms associated with the localized tip field. Using a simple formalism we find that the signature of these effects can be estimated to within an order of magnitude to be consistent with our expectations of topological states found in literature for such systems.

**Acknowledgements**
Financial support for this study came from the Wake Forest University Nano and Quantum Technologies Laboratory (NQTL). The authors thank Dundappa Mumbaraddi at the Wake Forest University Chemistry department for performing the XPS and PXRD measurements. The authors also thank Chris Winkler at NC State for assistance with HRTEM and EELs work at the Analytical Instrumentation Facility (AIF) of North Carolina State University, which is supported by the State of North Carolina and the National Science Foundation (Award No. ECCS-2025064).

**Author Contributions**
D.L.C. C.C. Dun experimental design, data analysis, modeling, and manuscript preparation,
T.C. EELs data reduction,
T.C., M.M., D.L., S.M. MFM data collection,
A.S. MFM calibration,
M.M., G.M., S.M. materials synthesis,
G.M. XPS, EDAX
S.K., M.K.-G., V.K. MFM, EELs data analysis, Raman data collection
L.G., A.W. initial MFM data used to determine parameters for this study.

**Competing interests**
The authors declare no competing interests.

**Additional Information**
The online version contains extensive supplementary material available at (URL to be inserted upon publication).

**Data availability**
All data are available from the corresponding authors upon reasonable request.

**References**

---

[28] Dun, C., Hewitt, C. A., Jiang, Q., Guo, Y., Xu, J., Li, Y., Li, Q., Wang, H., & Carroll, D. L. Bi2Te3 Plates with Single Nanopore: The Formation of Surface Defects and Self-Repair Growth. Chem. Mater. 30, 1965–1970, (2018), DOI: https://doi.org/10.1021/acs.chemmater.7b04985, Dun, C., Hewitt, C. A., Li, Q., Xu, J., Schall, D. C., Lee, H., Jiang, Q., & Carroll, D. L. 2D Chalcogenide Nanoplate Assemblies for Thermoelectric Applications. Advanced Materials 29, 1700070, (2017), DOI: https://doi.org/10.1002/adma.201700070, Gray, L. J., Swathi, K., Mumbaraddi, D., Carlson, T. W., Marcus, G., & Carroll, D. L. Solvothermal synthesis of crystalline 2D bismuth telluride with an isoelectronic dopant. APL Mater. 12, (2024), DOI: https://doi.org/10.1063/5.0221583, Marcus, G. E., Carlson, T. W., Swathi, K. & Carroll, D. L. Solution-based iron doping of solvothermally grown 2D hexagonal bismuth telluride. APL Mater. 12, (2024), DOI: https://doi.org/10.1063/5.0220681

[29] Sharifi, T., Yazdi, S., Costin, G., Apte, A., Coulter, G., Chandrashekar Tiwary, & Ajayan, P. M. Impurity-Controlled Crystal Growth in Low-Dimensional Bismuth Telluride. Chem. Mater. 30 (17): 6108–6115, (2018) DOI: https://doi.org/10.1021/acs.chemmater.8b02548

[30] Spemann, D., Han, K.-H., Esquinazi, P., Höhne, R., Butz, T., Ferromagnetic microstructures in highly oriented pyrolytic graphite created by high energy proton irradiation, Nuclear Instruments and Methods in Physics Research Section B: Beam Interactions with Materials and Atoms, Volumes 219–220, Pages 886-890, ISSN 0168-583X, (2004), DOI: https://doi.org/10.1016/j.nimb.2004.01.182, Martínez-Martín, D., Jaafar, M., Pérez, R., Gómez-Herrero, J., & Asenjo, A. Upper Bound for the Magnetic Force Gradient in Graphite Phys. Rev. Lett. 105, 257203, (2010), DOI: https://doi.org/10.1103/PhysRevLett.105.257203

[31] This presumes the sign of the force is preserved through the derivative. We anticipate the gradient to always be such that flux decreases away from the surface.

[32] Bleszynski-Jayich, A. C., Shanks, W. E., B. Peaudecerf, E. Ginossar, F. von Oppen, Glazman, L., & Harris, J. G. E. Persistent Currents in Normal Metal Rings. Science 326, (2009), DOI: https://doi.org/10.1126/science.1178139, Feng, Y., Mirzadeh Vaghefi, P., Vranjkovic, S., Penedo, M., Kappenberger, P., Schwenk, J., Zhao, X., Mandru, A.-O. & Hug, H. J. Magnetic force microscopy contrast formation and field sensitivity. Journal of Magnetism and Magnetic Materials 551, (2022), DOI: https://doi.org/10.1016/j.jmmm.2022.169073, Lohau, J., Kirsch, S., Carl, A., Dumpich, G. & Wassermann, E. F. Quantitative determination of effective dipole and monopole moments of magnetic force microscopy tips. Journal of Applied Physics 86, 3410–3417, (1999), DOI: https://doi.org/10.1063/1.371222, Okamoto, K., Imura, T., Abo, S., Wakaya, F., Murakami, K., & Nagao, M. Resistivity measurement for non-magnetic materials using high-order resonance mode of MFM-cantilever oscillation. Jpn. J. Appl. Phys. 64, (2025), DOI: 10.35848/1347-4065/adbfa0, Schaffer, T. E., Radmacher, M. and Proksch, R. Magnetic force gradient mapping. J. Appl. Phys. 94, 6525–6532, (2003), DOI: https://doi.org/10.1063/1.1623926, Neves, C. S., Quaresma, P., Baptista, P. V., Carvalho, P. A., Araújo, J. P., Pereira, E., & Eaton, P. New insights into the use of magnetic force microscopy to discriminate between magnetic and nonmagnetic nanoparticles. Nanotechnology 21, (2010), DOI: 10.1088/0957-4484/21/30/305706, Boyer, T. H. The Force on a Magnetic Dipole. American Journal of Physics 56, (1988), Hartmann, U. Magnetic Force Microscopy. Annu. Rev. Mater. Sci. 53–87, (1999).

[33] Sarid, D. Scanning Force Microscopy: With Applications to Electric, Magnetic, and Atomic Forces. Oxford University Press, New York, (1994), DOI: https://doi.org/10.1093/oso/9780195092042.001.0001

[34] Wakaya, F., Oosawa, K., Kajiwara, M., Abo, S. & Takai, M. Detection of nonmagnetic metal thin film using magnetic force microscopy. Appl. Phys. Lett 113, (2018), DOI: https://doi.org/10.1063/1.5079763

[35] Kealhofer, D. A., Galletti, L., Schumann, T., Suslov, A., Stemmer, S. Topological Insulator State and Collapse of the Quantum Hall Effect in a Three-Dimensional Dirac Semimetal Heterojunction. Phys. Rev. X 10, (2020), DOI: https://doi.org/10.1103/PhysRevX.10.011050, Chong, S. K., Liu, L., Watanabe, K., Taniguchi, T., Sparks, T. D., Liu, F., & Deshpande, V. V. Emergent helical edge states in a hybridized three-dimensional topological insulator. Nature Communications 13, (2022), DOI: https://doi.org/10.1038/s41467-022-33643-9

[36] Hartmann, U. Magnetic Force Microscopy. Annu. Rev. Mater. Sci. 53–87, (1999), DOI: https://doi.org/10.1146/annurev.matsci.29.1.53

[37] Wakaya, F. Detection of nonmagnetic metal thin film using magnetic force microscopy, Appl. Phys. Lett. 113, 261601, (2018), DOI: https://doi.org/10.1063/1.5079763

[38] Boyer, T. H. The Force on a Magnetic Dipole. American Journal of Physics 56, (1988), DOI: https://doi.org/10.1119/1.15501

[39] Michetti, P. & Recher, P. Bound states and persistent currents in topological insulator rings. Phys. Rev. B 83, 125420, (2011), DOI: https://doi.org/10.1103/PhysRevB.83.125420

# Supporting Information

# Anomalous Edge Contrast in Magnetic Force Imaging of 2D Topological Insulators

Timothy W. Carlson[1,2], Motahhare Mirhosseini[1,2], Allison Swyt,[1,2] Gabriel Marcus[1,2,3], Daniel Lazega[1,3], Scott Mcguire[1,3], Lindsey J. Gray[1,2], Anthony Walsh[1,2], Swathi Kadaba[1,2], Maria Koleśnik-Gray[4], C.C. Dun,[1,5] Vojislav Krstić[1,2,4], D.L. Carroll[1,2,*]

[1] The Nano and Quantum Technologies Laboratory, Wake Forest University, 95W 32nd St. Winston Salem NC 27105, USA
[2] Department of Physics, Olin Physical Laboratory, Wake Forest University, Winston Salem NC 27109, USA
[3] Quoherent Inc. 2860 Lyndhurst Avenue, Winston-Salem, NC, 27103, USA
[4] Department of Physics, Friedrich-Alexander-Universität Erlangen-Nürnberg (FAU), Staudtstr. 7, 91058, Erlangen, DE
[5] Lawrence Berkeley National Laboratory, Berkeley, CA, 94720, USA

* Corresponding author: carroldl@wfu.edu

In this work, we present evidence of topological states existing on the edge of 2D $Bi_2Te_3$ in ambient laboratory conditions as revealed by magnetic force microscopy. To do this:

1. Solution grown 2D $Bi_2Te_3$ micro-crystals with specific properties pertaining to stoichiometry, order, defect and impurity concentrations, oxidation, and interlayer interactions within the layered crystal are needed. Thus, sample crystals must be "defined" by their characterization in accordance with standards in the field. (you can't just call them a topological insulator and expect repeatable science)
2. The field of MFM, as with all scanning probes, is wrought with misinterpretations. It is ineffective to interpret an AFM/MFM/ EFM, etc. phase image without a model in mind that is unique to what is being imaged. Thus, a model for understanding and estimating the factors that influence MFM phase imaging of a <u>2D topological insulator</u> must be developed and fit to data. [1]
3. Finally, there must be some statistical and error analysis for such noisy environments. Since we have chosen to understand these systems in an environment that is hot, with air interacting with the probe cantilever and tips, across many different sample crystals, it is essential to understand how variations in such noise will complicate our outcomes and conclusions.

The supporting documentation, characterizations, and derivations are given here. It is more extensive than usual for *supporting information*, but this is necessary to fully clarify our findings.

### 1.0 Estimates of Currents and Fields

In the BHZ description of $Bi_2Te_3$, when there is no applied bias or magnetic field, charge currents in the edge channels flow in both directions equally, as Kramers pair states. A simple semi-classical estimate of this current density in one direction, based on Drude carrier drift and assuming no contacts or doping, is approximated as: $J_{edge} \sim \mathrm{n}e\bar{v}_d$ where $\bar{v}_d$ is the average drift velocity of the electrons in the channel, n is the number density of electrons and $e$ the elementary charge. Of course, this is a little misleading since the topological edge currents should be protected and so dissipationless. So, the mechanism leading to $\bar{v}_d$ is a bit ambiguous at this point.

The total edge current in the i$^{th}$ edge channel will be $I^{i}_{edge} \sim n_i e\bar{v}_d$, assuming a fully a 1D channel. In the edge channels only the gap states are participating, and we ignore conduction band and valence band electrons. However, we note that this is at room temperature and so there are thermally promoted conduction band electrons that will interact with the topological gap states by filling them. For each state of the two gap dispersion curves, there is a single electron (in one direction) on the circumferential loop and so n ~ 1/ $\pi d$ where $d$ is the approximate diameter of a plate. There are N of these channels: $N = \sum_{i=1}^{N} n_i$ from the number of unit cells around the perimeter. Taking $d \approx 1\ \mu m$ (typical for colloidally grown plates) and an edge spacing of ~ 1nm, N ~ 3 x $10^3$. Because $\bar{v}_d$ is of the same order as $v_F \approx 10^5 m\ s^{-1}$, over the narrow energy range of the bulk gap. Thus, $I_{edge} \approx N\ ev_F/(\pi d)$ which works out to be in the 10's of μA range, for each direction of current flow. [2] This crude estimate assumes nearly all the edge band dispersion curve states are filled at room temperature and that the average drift velocity is about equal to the Fermi velocity and the Dirac point sits in the middle of the gap. These assumptions are not strictly true, but they are close. And this number isn't very different from that found for a 1μm$^2$ cross-section Cu wire (~ 4 μA). Naïvely, if a portion of the edge state electrons are scattered through some non-TRS respecting mechanism like a magnetic impurity, [3] it will introduce an asymmetry for current in one direction and lead to a persistent current. For significant scattering, a commercial MFM can have the sensitivity to measure the resulting magnetic field. [4] We would expect this to require a large part of the electrons in the edge state gap bands.

What kind of field is needed to do this? The 2D topological insulator is a Quantum Spin-Hall system. If any magnetic field is applied to such a system the $\mathbf{B}_{||}$ forms a spin pump, yielding spin accumulation. The $\mathbf{B}_{\perp}$ component means the flux threads the loop of the plate perimeter, giving a Zeeman field that can act to break time-reversal symmetry (TRS) and couple to the helical edge states. This field behaves as an exchange field that opens an energy gap $\Delta_z = g_s \mu_B B$, where $g_s$ is the electron g-factor and $\mu_B$ is the Bohr magneton in the otherwise gapless edge spectrum. [5] The counter-propagating electron helical edge states get mixed and when the Zeeman energy exceeds the disorder potential (the intrinsic spin-orbit coupling gap), it induces elastic backscattering that quenches the ballistic persistent currents, annihilating the edge states. [6] Depending on the direction and magnitude of the Zeeman field, the system transitions into a Quantum Anomalous Hall state (Chern insulators with chiral states) or a trivial insulator.

Thus $\mathbf{B}_{||}$ and/or $\mathbf{B}_{\perp}$ both can drive a persistent current (orbital magnetization), but what you get depends on applied field strength. There are three regimes relative to the Zeeman energy $\Delta_Z$ to consider: [7]

1. ($\Delta_Z < E_{Th}$): The Zeeman effect acts as a minor perturbation. Persistent currents show a perfect ($\Phi_0$ = h/e) periodicity, driven purely by the orbital Aharonov-Bohm phase.
2. ($\Delta_Z \sim E_{Th}$): The Zeeman term distorts the Dirac cones, altering the energy-momentum dispersion and modulating the amplitude of persistent charge and spin currents.
3. ($\Delta_Z > E_{Th}$): The Zeeman gap exceeds the Thouless energy ($E_{Th}$), backscattering between edges becomes dominant. This destroys the topological protection and suppresses the amplitude of the persistent current.

Since $E_{Th} \sim \hbar D/L^2$ where D is the diffusion constant, L is the length of the system, for the case of few layer $Bi_2Te_3$ we can expect $E_{Th}$ ~ 10 μeV – 100 μeV. The Zeeman energy of a (0.01 T) or (100 G) magnetic force microscopy (MFM) tip is ~ 5.79 x $10^{-5}$ eV. This might indicate an *Intermediate Zeeman Energy* regime (# 2). Importantly, an MFM tip applies both a $\mathbf{B}_{\perp}$ and a $\mathbf{B}_{||}$ at an isolated point on the loop, resulting in distortion of the Dirac cone and altering the gap dispersion locally as well as a spin pump.

This means that the application of small magnetic fields, such as those associated with MFM tips, near the perimeter of a $Bi_2Te_3$ plate, should induce some magnetic scattering that doesn't respect TRS. $\mathbf{B}_{||}$ and $\mathbf{B}_{\perp}$ will result in different symmetry breaking and spin accumulation effects locally, and they add to create the scattering in the 1D channels. It is reasonable to assume however, that this scattering will not be strong at the field strengths of an MFM tip. Indeed, Paolo Michetti and Patrik Recher [8] show persistent currents of ~ 300 nA/channel can be expected for homogeneous fields of only a few percent of the flux quantum $\Phi/\Phi_0 < 0.1$ [see Figure 6 of reference 8].

## 2.0 Synthesis and characterization of crystalline $Bi_2Te_3$ plates

This work utilizes solvothermal synthesized $Bi_2Te_3$ plates as described in our previous publications [1, 2]. Briefly, bismuth nitrate pentahydrate ($BiH_{10}N_3O_{14}$) and sodium tellurite ($Na_2TeO_3$) precursors with sodium hydroxide (NaOH) as a reducer are added to a flask under reflux and vigorously stirred in an ethylene glycol solution for one hour. Then, polyvinylpyrrolidone (PVP) is added, and the entire solution is stirred overnight in an argon atmosphere at 180 °C. Once the reaction is complete, the product is centrifuged multiple times to eliminate unreacted components and crystals of very large or very small sizes. The final solution thus contains $Bi_2Te_3$ crystals with similar lateral dimensions.

The $Bi_2Te_3$ plates are examined using transmission electron microscopy z-contrast (CTEM) and selected area diffraction (SAD), high resolution scanning transmission electron microscopy (HR-STEM) with electron energy loss spectroscopy (EELS), scanning electron microscope (SEM) based energy dispersive x-ray (EDAX), X-ray photoelectron spectroscopy (XPS), Raman spectroscopy, and x-ray diffraction (XRD). The characterization was carried out within days of synthesis. Plates were exposed to atmosphere in their dry state for only brief periods while sample loading or ambient imaging. Otherwise, they were stored in a suspension of alcohol and PVP. This was done to reduce the effects of oxidation and aging which may play a role in the formation of edge phases of Te-O.

The resulting growth products are a well faceted layer-by-layer structure of quintuple unit cells that are a stacking of the 2D sheets as van der Waals heterostructures. The atomic registry and L-b-L epitaxy is evident by the precise alignment of edges and general lack of moiré patterns in electron microscopy of the materials. **Figure S2**, confirms a high degree of crystallinity, stoichiometric samples with multiple epitaxial quintuple layers. There is low sample-to-sample variability suggesting that defect densities, epitaxy, intercalates, impurities all remain constant. Highly conductive states are inferred at the edges of the crystal plates, through valence electron energy loss spectroscopy (**Figure S3**), as seen in previous studies of solution-grown chalcogenide crystals [9]. Specifically, edge state EELS features are associated with localized plasmons as discussed in the literature [10]

## 2.1 Microscopy

To determine thickness of the plates as synthesized, AFM was used. For these studies two different AFM microscopes were used: a JEOL JSPM 5200 and a Digital Instruments/Bruker 3100. Both were setup for AFM and MFM measurements. Generally, freshly cleaved HOPG substrates were used except where specified. Because of the different forces encountered in imaging of these materials, optimization of scanning parameters in terms of gain and force were set for each set of experiments. Results between the two microscopes were correlated in phase response to lift height, scan speed, and oscillation amplitude. Tips used where from MDT-NT and were generally Si with gold reflective surfaces. Co/Cr coatings were used for the MFM tips. Instruments were calibrated using standard calibration grids available from the manufacturer.

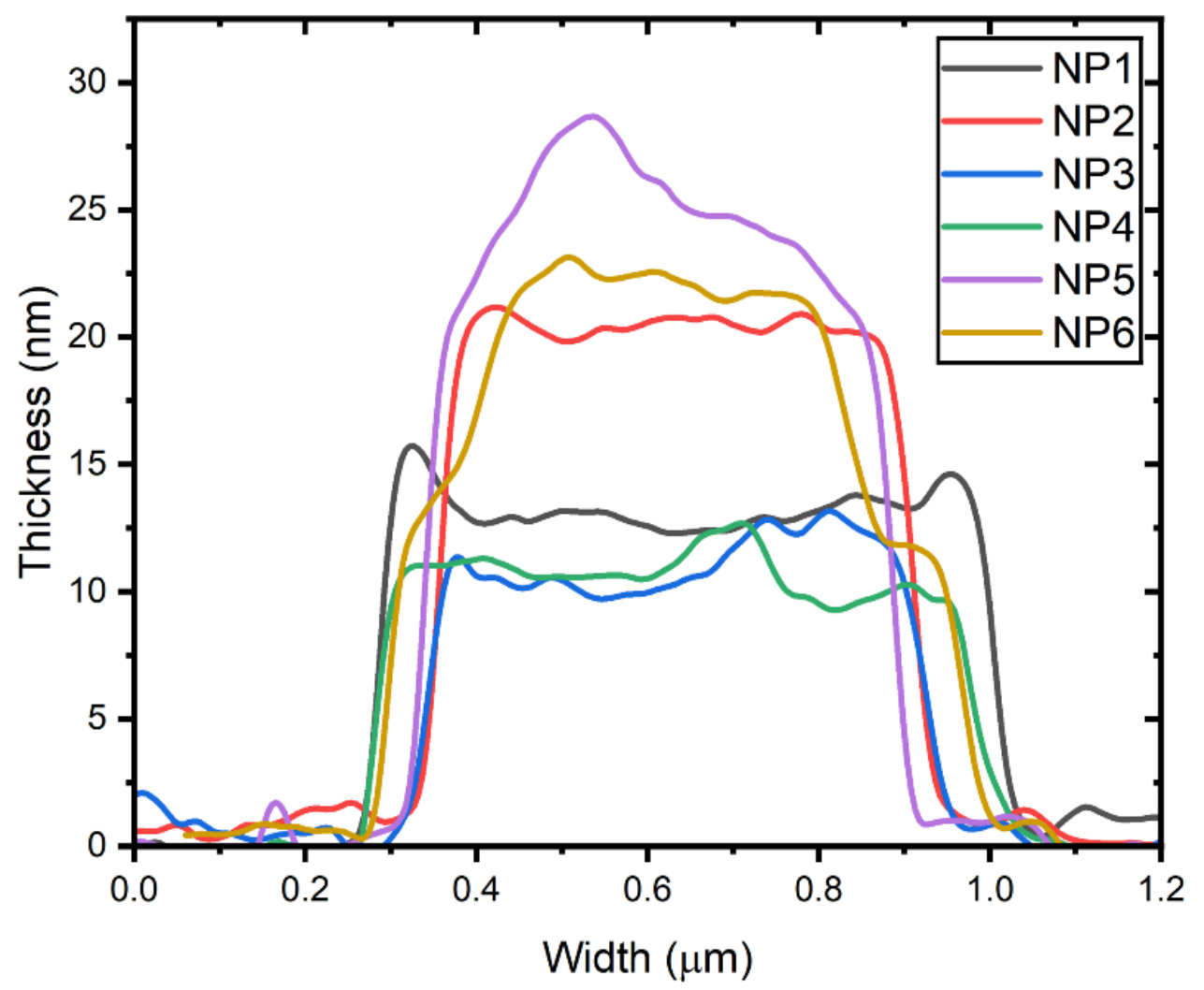

**Figure S1:** AFM line scans of topographical images of $Bi_2Te_3$ plates (NP).

### 2.2 CTEM and STEM Imaging

To ensure the structure and crystallinity of the $Bi_2Te_3$, the resulting nanomaterial structure was characterized thoroughly. A Thermo Fisher Titan 80-300 STEM using a 200 keV beam irradiation was used. First, the average plate diameter was 0.68 ± 0.08 µm as obtained from 6 plates (**Figure S1 (a)**). At high resolution, high-angle annular dark-field (HAADF) images show defect-free crystalline plates (**Figure S2 (a-c)**) with a crystal plane spacing of 0.35 nm, which corresponds to the (110) plane. The selective area diffraction (SAED) further confirms the typical $Bi_2Te_3$ set of crystal planes, with the brightest ring diameter in the diffraction pattern measuring 3.34 $nm^{-1}$, indicating an interplanar distance of 0.59 nm (**Figure S2 (d)** ). [11] Next, energy dispersive analytical X-ray analysis (EDAX) was conducted, and we see that the overall stoichiometry of $Bi_2Te_3$ was consistent with the atomic ratios measured (**Figure S2 (g)**).

### 2.3 XPS, EDAX, and Raman

X-ray photoelectron spectroscopy (XPS) [RBD/PHI 570, Al K-alpha line Ex = 0.25 eV base pressure $10^{-11}$ Torr] was carried out to examine bonding characteristics and determine the phase of the $Bi_2Te_3$. This is shown in **Figure S2 (h)** with two tellurium peaks (3d5/2, 3d3/2) at 578 eV and 589 eV and the bismuth peaks (4f7/2, 4f5/2, 4d5/2, 4d3/2) at 165 eV, 171 eV, 448 eV, and 472 eV consecutively, as expected for $Bi_2Te_3$. Finally, micro-Raman measurements [Horiba Lab Spec using a 532 nm line at 210 µW] show no significant excess of Te, Bi or other features associated with defects (**Figure 2 (i)**). Therefore, suggesting exceptionally clean and crystalline $Bi_2Te_3$. [12]

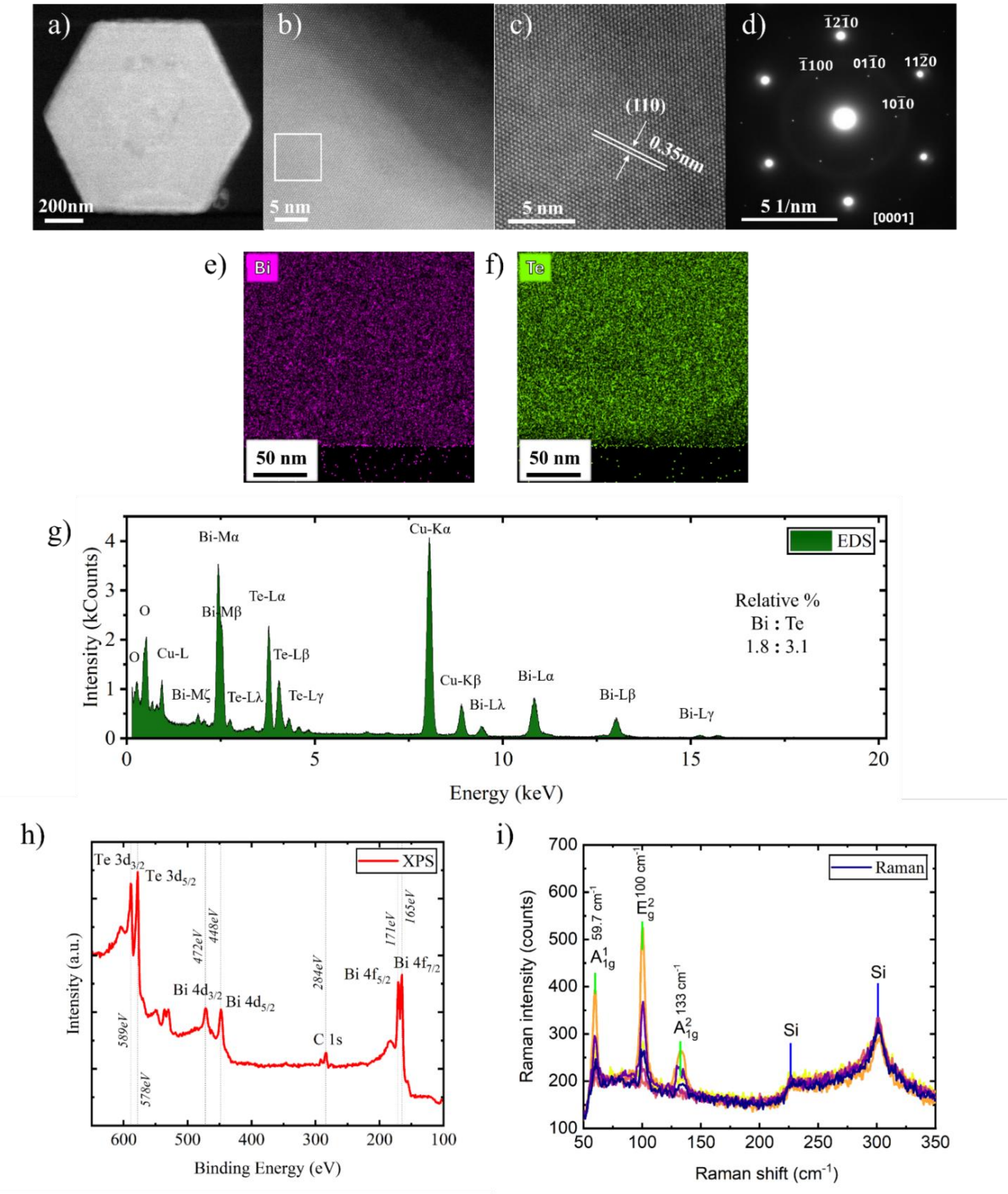

**Figure S2:** (a-d) HAADF and SAED images show a high degree of crystallinity in our synthesis, with relatively little lattice strain. (e-g) EDAX shows the materials to be stoichiometric, generally. (h) The phase of the materials is further confirmed by XPS to be $Bi_2Te_3$. (i) Raman spectroscopy also confirms the formation of defect-free $Bi_2Te_3$.

## 2.4 EELS of Edge States

EELS is increasingly being used to probe the spatially resolved plasmonic modes of low-dimensional materials for photonic applications [13]. In studies comparing the EELS peak energies of $Bi_2Te_3$ plates with the dielectric loss function derived from atomic-structure calculations a one-to-one correspondence with spin–orbit coupling was revealed. This correspondence indicates a directional dependence on electron scattering that could originate only from the plate's topological character. In extension to this study, the authors also observed that the EELS peak vanished when the plate started becoming amorphous [14]. Keeping in mind that the precursors used in our chemical synthesis were different compared to those of the referenced studies, we performed a similar EELS analysis to define the spatial resolution of the topological states of the $Bi_2Te_3$ plate.

A Thermo Fisher Titan 80-300 probe aberration-corrected STEM with a Gatan Tridiem high-resolution GIF, operating at 80 kV accelerating voltage was used for EELs analysis. The beam was aligned with a 19.3 mrad convergence semi-angle and 71 mrad collection semi-angle. The energy resolution of the beam was 25 meV, and the beam energy loss up to 45 eV was acquired. The sample was prepared by drop casting the $Bi_2Te_3$ plates from the ethanol solution onto Lacey Formvar Carbon reinforced TEM grids from Ted Pella. After sufficient air-drying, the plates were directly inserted into the HR-TEM, avoiding annealing under $Ar/H_2$ because of the fragility of the TEM grids. Plate thickness was extracted following the methods by *Williams and Carter* [15]. EELS spectra were sampled from a 14 x 14 pixel area on the EELS maps using the Digital Micrograph Software by Gatan. A fitted logarithmic tail method was used for the zero-loss peak subtraction using a fitting range from 0.25 eV to 3 eV. The edge spectrum was acquired 5 nm from the edge of the plate, while the center spectrum was acquired in the center of the plate, equidistant from each edge.

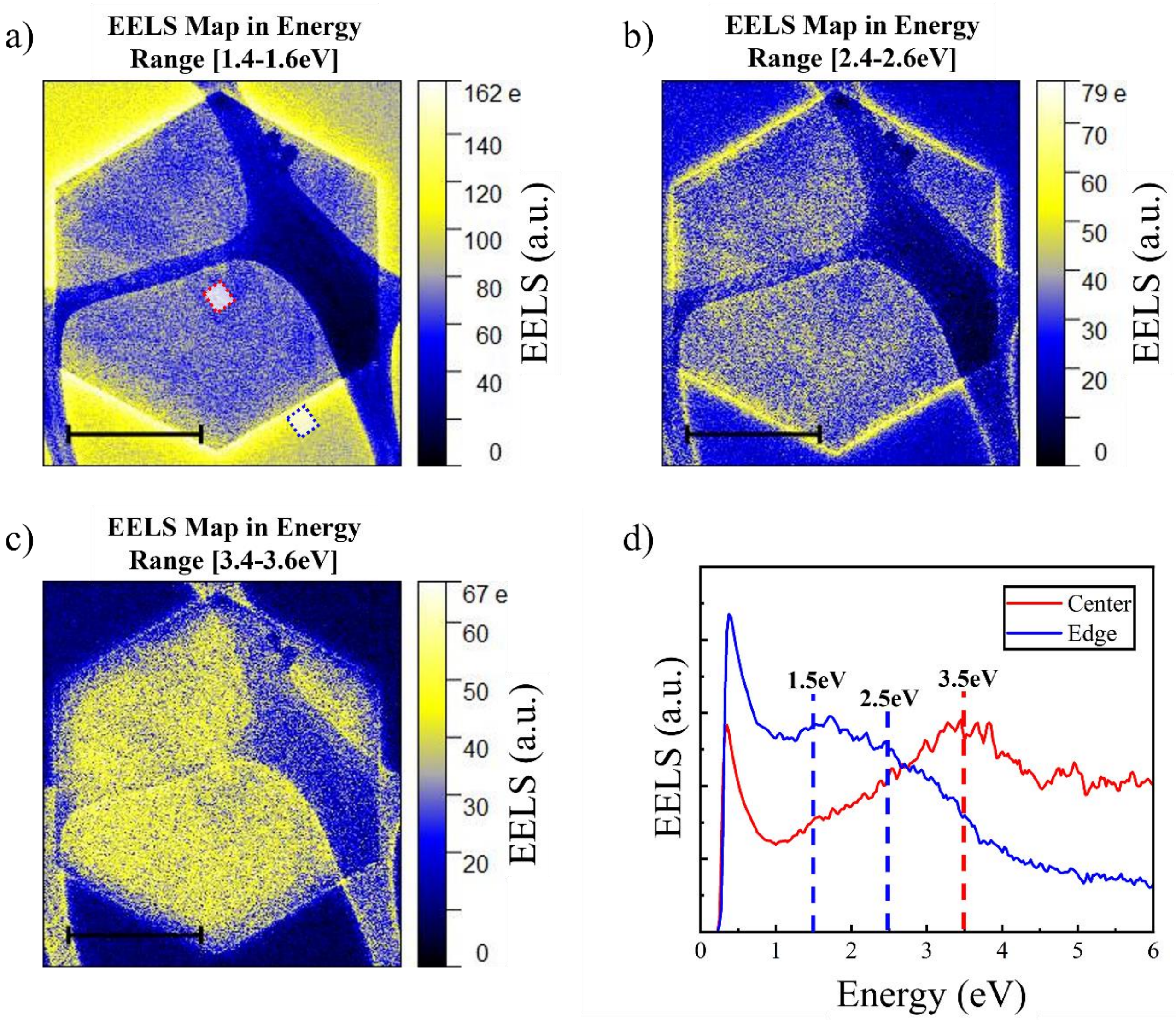

**Figure S3**: EELS on $Bi_2Te_3$ plates. (a-c) Energy-filtered EELS maps of a $Bi_2Te_3$ plate with a thickness of ~24 nm. EELS maps were created with pixel contrast set to an energy range of 0.25 eV and centered on the energies 1.5 eV, 2.5 eV, and 3.5 eV. Scale bars in the inset are 200 nm. g) Zero-loss-peak-subtracted EELS spectrum obtained from the center and the edge of the plate.

**Figure S3** shows the EELS maps of a 24 nm thick $Bi_2Te_3$ plate. As the electron beam is moved from the edge to the center of the plate, we see three distinct energy loss bands centered at 1.5, 2.5 and 3.5 eV electron energy loss. In energy-filtered imaging, a noticeable uniform bright edge contrast was observed at 1.5 eV (**Figure S3 (a)**) and 2.5 eV (**Figure S3 (b)**). As energy loss increases to 3.5 eV (**Figure S3 (c)**), the contribution from the center starts increasing, showing a clear resolution of the edge and surface contribution of the plasmon. We also confirm the spatial contrast by comparing the three spectral intensities at the center and the edge of the plate **Figure S3 (d)**). Our observation matches previous reports, of such plasmonic modes at the edge and across the surfaces of $Bi_2Te_3$. [16] However, these energies are greater than the band gap of the material, so the naturally occurring hyperbolic electromagnetic response of the topological insulator might also allow for the support of Dyakonov plasmons. [17] Thus, the use of these plasmon features as a signature of a 2D or 3D topological material is ambiguous for thick plates.

## 3.0 MFM

Magnetic force microscopy (MFM) is an intermediate contact surface probe, that uses an atomic force microscope (AFM) cantilever with a magnetized metallic coating (usually Co) on the tip. In a typical application, the probe is used to make two different measurements at each pixel of the image. The first is a topographic measurement using the magnetized tip as an AFM to determine the height of the surface relative to some setpoint and thereby creating a profile the sample's surface. This is done using the so called "tapping mode" with the feedback system engaged. Then, at each point, the microscope also lifts the magnetic tip above the surface, referred to as the lift height (LH), and dithers the tip for multiple cycles. The response of the tip oscillation relative to its driving signal allows any forces acting on that tip at this position to be determined. Combining the topographic information produces an AFM image of the surface. Combining the oscillation responses of the cantilever, pixel by pixel, for a constant height above each topographic point, a phase-, frequency-, or amplitude- difference image is generated. This is directly proportional to the flux gradient of forces experienced by the tip at this height. Phase is the most common variable to plot (known confusingly as a *phase image*). To interpret MFM phase imaging, all tip-substrate interactions must be considered. However, the topological insulator systems present a combination of forces and dissipative effects typically not encountered in MFM.

The MFM determines phase $\Delta\phi$ using lock-in techniques (homodyne detection). This directly relates to vertical, lateral, and torsional force gradients experienced by the oscillating cantilever (**equation 1**). The restoring force in AFM/MFM are due to cantilever bending modes. Typically, the system is setup to detect oscillation responses along some z axis of the cantilever, onto the detector. So, we are primarily concerned with the restoring forces associated with bends along $l$ the length of the cantilever and ignore torsional bending which twists the cantilever in other directions. Moreover, for relatively small forces and small driving amplitudes we can also focus only on the primary bending mode and ignore higher order vibrational modes. [18]

*An important note of clarification:* the LH is measured from the bottom of the cantilever's tip swing, so the feedback loop must be active across the scan, and the cantilever must adjust its equilibrium position of oscillation to accommodate the increased swing amplitude. We can estimate the effect of the slight changes in cantilever position from **Figure 2(b)**, where we examine the phase as LH is changed. $\Delta\phi$ shows a predictable decrease as the tip moves away from the surface. Simultaneously, we can conclude that a 2 nm shift in the equilibrium position is not large enough to significantly change $\Delta\phi$, although for larger shifts in oscillation amplitude, it might. We expect that this interdependence would lead to a $\Delta\phi$ reduction since the tip would move away from the surface (on average) as the amplitude of oscillation increases. Therefore, from our data, we conclude that for small changes in the amplitude, the effects of feedback and adjustments to cantilever position are second order at most.

For our MFM and topographic imaging, HOPG was used as a support substrate due to its atomically smooth surface and homogeneous carrier density [19]. The plates dispersed in ethanol were pipetted onto a freshly cleaved substrate surface. After air drying for 15 minutes, the substrates were imaged in the JEOL 5200, a conventional scanning probe microscope. We obtained a range of thicknesses from 6-30 nm (**Figures S1, S7, S8, S12, and S13**). Plate thickness was measured using the non-contact imaging mode. From our X-ray diffraction analyses along with theoretical predictions, a single $Bi_2Te_3$ quintuple layer (Te1-Bi-Te2-Bi-Te1) is 0.76nm thick [20] suggesting that the thinnest plates we imaged were ~ 6 quintuple layers.

As shown in **Figure 1 (a)**, the point-by-point MFM mode lifts the cantilever at each pixel of the image. The cantilever moves laterally from pixel to pixel along the AFM scan line using force feedback. However, at each pixel position, the cantilever is raised to a set LH and allowed to dither above the surface through several cycles of the oscillator drive. The phase difference between driving frequency and measured frequency is determined at each of these points and plotted as the "phase image." The cantilever is then lowered back to the surface, and the AFM feedback is re-engaged. The cantilever then moves laterally to the next pixel location along the scan line, and the process starts over. Detailed methods for extracting the average phase shift values from the phase image are provided in the supplementary information.

As noted, there are two ways MFM imaging can be carried out. We use point by point scanning in the MFM-AC mode where, for each pixel two measurements are performed. Cantilever dynamics were sensed using slope detection where resonance shifts due to force gradients were observed with the change in vibration amplitude. The first measurement gives topographical information with the tip-sample distance set once the Van der Waals forces become significant. This measurement uses the difference in the amplitude of cantilever oscillation to image the topography. The second measurement is completed by first raising the cantilever to a designated lift height (LH) off the surface and then measuring the phase difference of the cantilever frequency to the driver frequency. At significant LHs, the only forces present should be magnetic. To begin, the LH was set to 0 nm and then the line scan time, scan area, gain, and cantilever driving voltage were adjusted to optimize magnetic contrasts in the images. After settings were established a sequence of images were taken with varying LHs. Polycrystalline, cobalt-coated cantilevers with resonance frequency 60 kHz and force constant $1\mathrm{N}\ \mathrm{m}^{-1}$ were used for the MFM imaging. Silicon cantilevers with resonance frequency of 90 kHz and force constant $1\mathrm{N}\ \mathrm{m}^{-1}$ were used for the AFM imaging. All cantilevers were purchased from NT-MDT Spectrum Instruments and prepared with a 30s re-magnetization with a niobium magnet before insertion into the AFM head. Images shown in this report were taken with lift heights from 0 nm up to 150 nm and scan speeds of 4 µm $s^{-1}$ to 48 µm $s^{-1}$. When the cantilever is driven off its resonance frequency.

**3.1. Tip velocity**

The JEOL SPM 5200 can measure vertical feature sizes ~3 µm with scan speeds up to 4 x $10^{6}$ µm $s^{-1}$. Typically though, high resolution AFM is limited to about 250 µm $s^{-1}$ [21] Ambient imaging where the cantilever is exposed to air, further limits scanning speeds to < 20 µm $s^{-1}$. Again, this is in terms of AC-AFM or "tapping mode." So, we must select scan speeds that result in a lateral tip velocity smaller than that of the cantilever dither's vertical velocity.

In our experiments, due to Faraday effects, it is necessary to understand the role lateral tip velocity plays in observed asymmetries of height (AFM) and $\Delta\phi$ (MFM). These occur in the scan direction (forward and back) of the tip. We examined the lateral tip speed's effect on the image for our scanning geometry and then made comparisons for the scale of the asymmetry we may experience with the induced magnetic response. The topographic responses are shown in **Figure S4**. As the thickness of the plate was known, the topographical images (AFM) obtained from the scan were plotted with respect to the tip scanning speed to compare the variance of the measured height. Although the dithering velocity of the cantilever is greater than that of the lateral motion, varying the lateral speed had a noticeable effect on the topographic image. This is to be expected since the Faraday components change the overall contact force during topographic imaging.

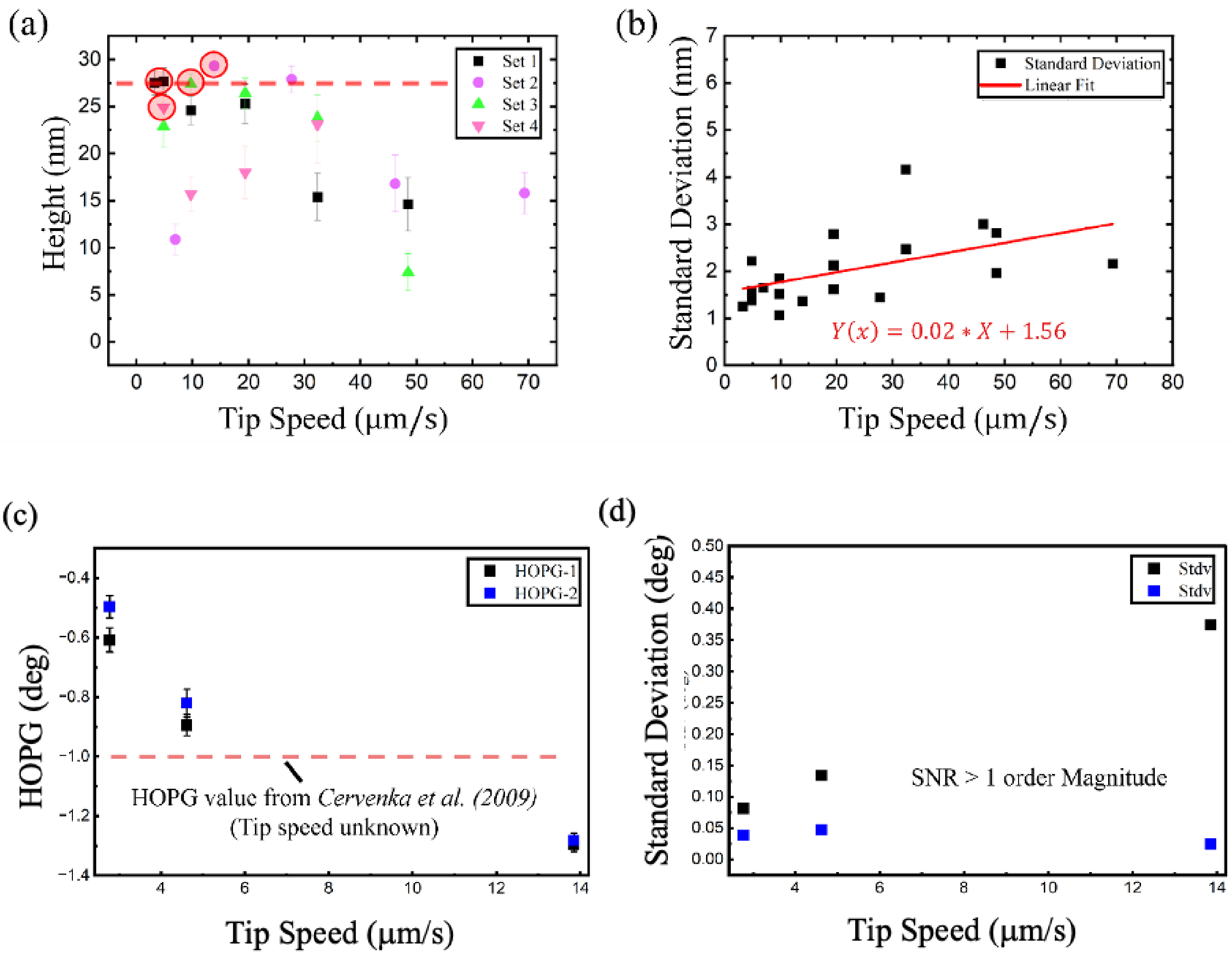

**Figure S4:** (a) Scatterplot of lateral tip speed versus plate mean height in topographical mode and its (b) standard deviation. Each image was taken sequentially on the same plate. Values were calculated by taking six, line plots across the center of the plate rotated by 45 degrees. Mean height variation across individual line plots is represented by one standard deviation from the mean. Points circled in red represent the lowest standard deviation per set. The dotted horizontal red line in (a) is the average of four points circled in red and represents the most probable lift height. (c) and (d) are the phase image as a function of scan velocity.

The microscope produces symmetric imaging with tip speeds < 5-15 μm $s^{-1}$. Accuracy was determined by implementing a statistical analysis of the images and ranking the measurements by their standard deviation. It should be noted that the microscope settings (gain, cantilever driving voltage, etc.), the dimensions and characteristics of the cantilever used, the support substrate, and the surface nature of the plate can also affect the quality of the image, but not the scan axis height asymmetry. In the data shown, all sets were taken sequentially to maintain the same conditions as the lateral tip speed was varied.

### 3.2 Cantilever Dynamics

The bimorph of the JEOL 5200 SPM driving the cantilever coupled with the cantilevers used in these studies, oscillation amplitudes (the cantilever swing) can be varied up to ~ 50 nm. [22] As this amplitude of oscillation is varied, the overall tip velocity through the equilibrium position also varies v ~ ± ωA. Thus, for a given height, we can vary the overall dΦ/dt term in any Faraday induction caused by the tip dither. This will of course also mean that dF/dt → ΔF/Δt so the Δϕ measurement is a little less sensitive to the field gradient. For small changes in A this should not be significant. To use this in understanding our system, we fix the lift height at 100, 150 and 200 nm and varying the cantilever amplitude. This is shown in **Figure S5** where the probe tip is placed over the edge of a $Bi_2Te_3$ crystal.

The cantilever amplitude can be expressed as a function of applied cantilever voltage [22].

$$A(V) = \frac{A_{Bimorph} * V * Q}{\sqrt{Q^2\left(1-\left(\frac{\omega}{\omega_0}\right)^2\right)^2+\left(\frac{\omega}{\omega_0}\right)^2}} \quad (1)$$

Here, $A_{Bimorph}$ is the bimorph amplitude, V is the applied voltage, Q is the quality factor ω is the frequency of oscillation, and $\omega_0$ is the natural resonance frequency of the cantilever. The ratio of $\frac{\omega}{\omega_0} = 0.965$ was used for the data collection in **Figure S5**.

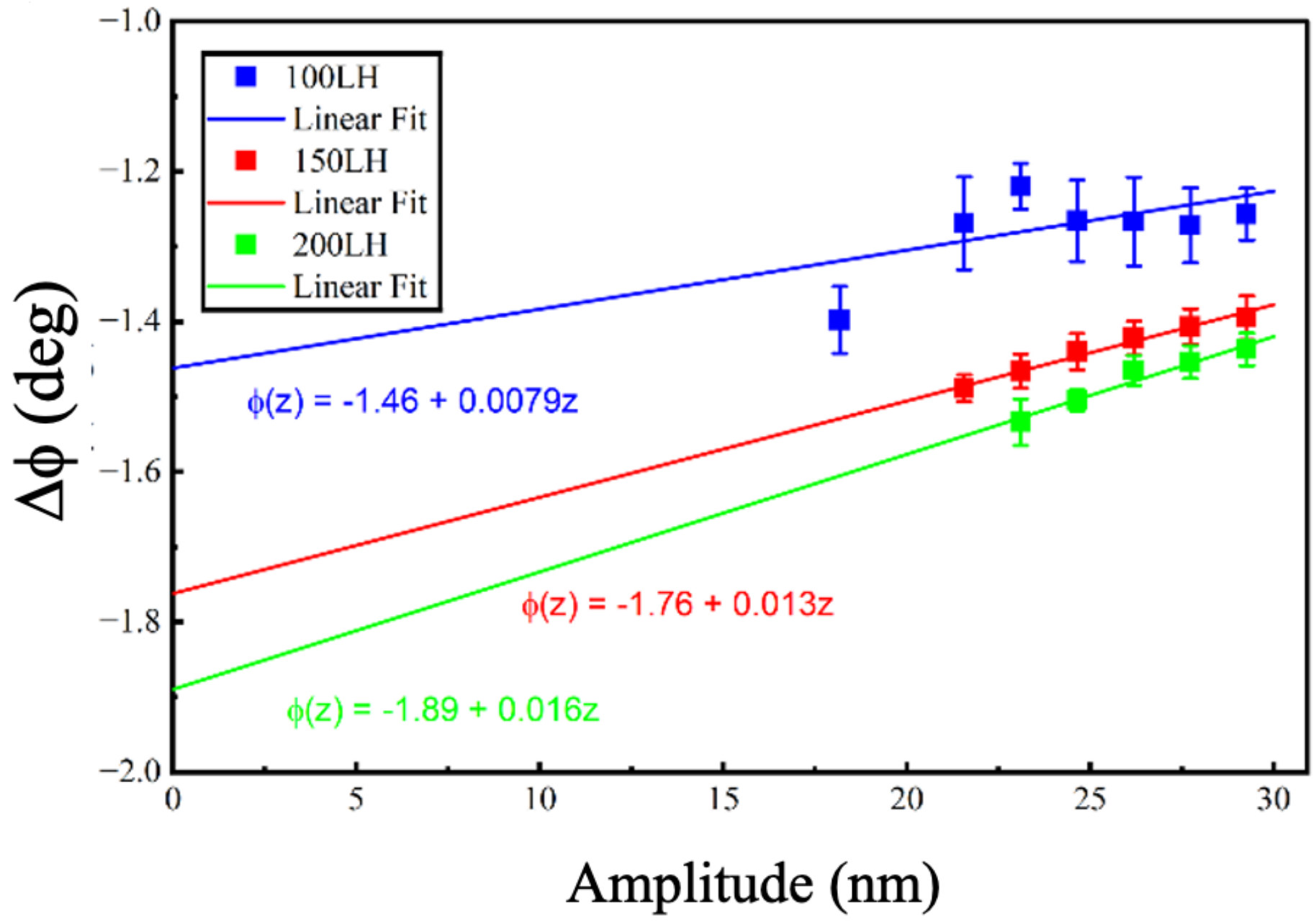

**Figure S5:** Cantilever amplitude vs. edge phase with lift heights 100, 150, and 200 nm. A straight line has been fit to each if the lift heights. For 150 nm and 200 nm LH, maintain a phase varies linearly with applied voltage to bimorph (A). For 100 nm this is less clear.

As noted, at the lower lift height of 100 nm, this linear relationship seems to be harder to establish. However, we expect the field from the tip to be relatively small and at large lift heights, there should be little induction from the tip into the loop formed by the crystal's edge. As the tip gets closer, the stronger time dependent field creates larger induced currents in the edge states of the $Bi_2Te_3$. This modifies the linear relationship between Δϕ and amplitude.

Δϕ reflects these changes at 100 nm LH not only because the tip field is stronger overall at this distance, but also because the changing components of the tip velocity now matter more. These tip velocity components are seen in **Figure S6**. The $V_d$ contribution was calculated by multiplying the resonance frequency by double the cantilever amplitude. The $V_z$ contribution was calculated by using the time per line scan, divided by the number of pixels, and then divided by the number (3) of steps per pixel. The $V_y$ and $V_x$ contributions were taken from the metadata saved on the image file. It should be noted that the $V_z$ contribution will change by $\pm 21$ μm $s^{-1}$ with a LH change of $\pm 10$ nm.

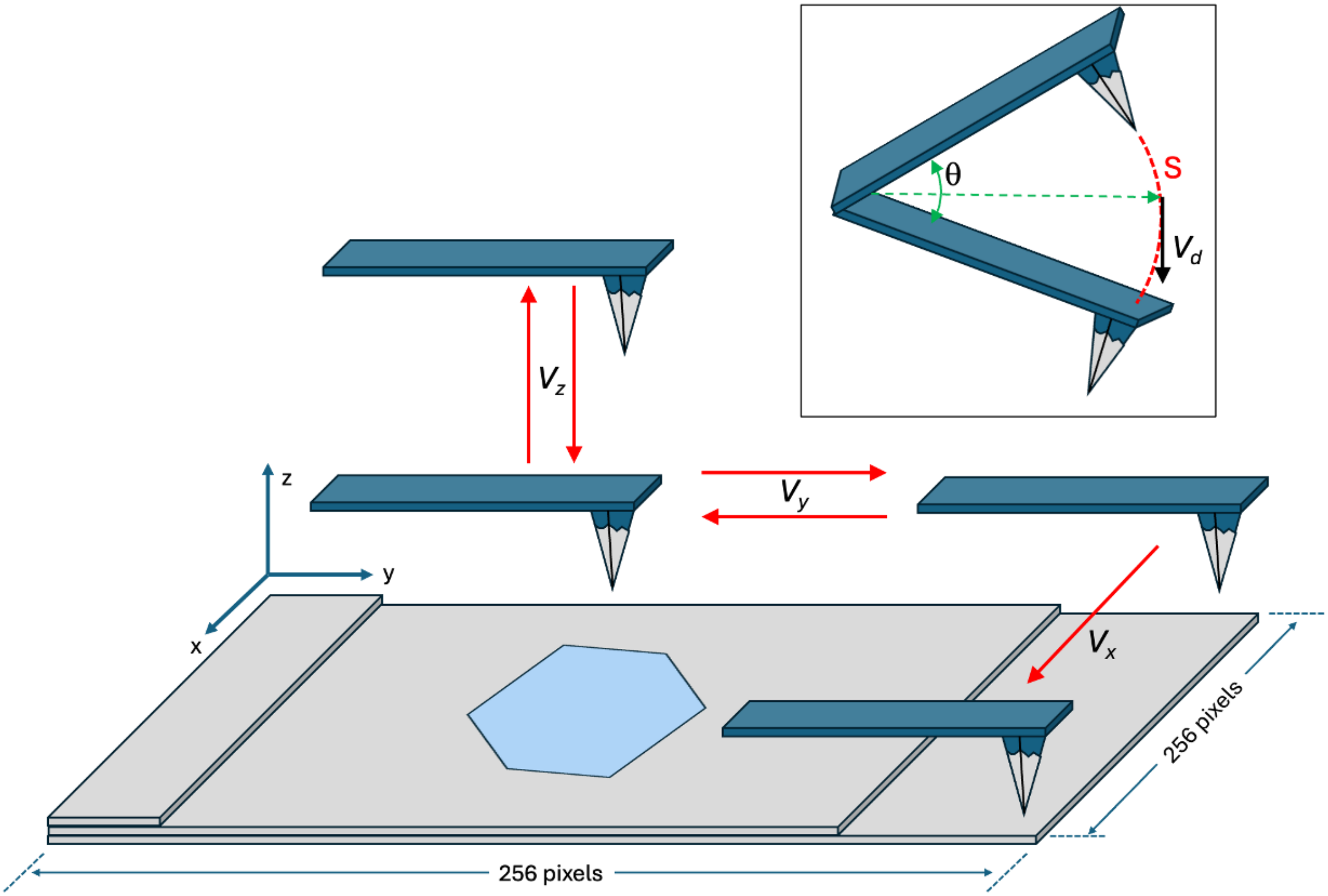

**Figure S6:** For a given lift height (100 nm for example) and drive frequency of 77.7 kHz tip velocity is calculated by setting the scan area to 3.5 μm x 3.5 μm with a resolution of 256 x 256 pixels and varying the scan line time. For 0.333 s $V_x = 0.189$ *μm/s*, $V_y = 9.697$ *μm/s*, $V_z = 210$ *μm/s*, $V_d = 5749$ *μm/s*.

## 4.0 A Simple Model

Contrast (the change in $\Delta\phi$ from support substrate background) on the edges of the $Bi_2Te_3$ plates is due to two current sources of magnetic field: $I_{Faraday}$ and $I_{persistent}$. But there is more. Electrostatic attraction and loss mechanisms must be considered together with these induced currents. Our approach includes the following.

1. The large hemispherical tip forms a capacitor with the substrate. There is an electrostatic attraction due to the small contact potential that comes from all the different materials in the microscope system. When the tip is over the dielectric of the sample, the capacitor has a wider separation distance between contacts by the thickness of the sample, the voltage across the capacitor stays the same, and there is a dielectric response. This yields a different electrostatic attraction creating contrast with the support substrate.
2. Because the tip is dithered, the plates of the tip-substrate capacitor change distance at the drive frequency. The voltage that comes from contact potential stays the same, but the applied field density changes as this distance is varied. This gives a dielectric loss at kHz driving frequencies and over the $Bi_2Te_3$ this is significant. Similarly, the tip scan also experiences a change in dielectric loss at the plate edges. However, for reasonable scan speeds, as used in this study, the dielectric response is faster than the change in tip position so the loss will be vanishingly small.
3. There is a persistent current which comes about due to magnetic scattering in the edge channels. This happens because the localized tip field gets very close to the crystal edge during imaging. This is analogous to the scattering seen when magnetic nanoparticles are placed near a topological insulator. [23, 24]
4. There is an induced current due to the Faraday effect. This effect is associated with the $d\Phi/dt$ of the up and down dither, and the $d\Phi/dt$ of the lateral scanning tip. However, we typically use slow scanning speeds (except where otherwise noted) so this second component is small and the dithering dominates. We propose that the scattering by the tip in #3 creates a resistance within the channel across which the Faraday EMF is dropped. So, we anticipate that this effect is different from that of the Faraday effect in non-dissipative systems such as superconductors. [25]
5. There is a loss due to dissipation in the edge channel. This is caused by the tip-scattering which gives an "effective" magnetoresistance in the edge band. We speculate that this "effective" magnetoresistance results from the multiple scattering events associated with the dithering of the probe tip which creates a time dependent scattering potential. However, the details of the magnetoresistance mechanism are the subject of further studies.

Just as in more standard models, our approach can be split into forces and dissipations. The electric forces are straight forward results of the metal coating on the tip and cantilever which forms a capacitor with the substrate, and the magnetic forces are correlated to the specifics of imaging with the MFM probe. The dissipative part of the imaging will change the overall Q of the cantilever oscillation which leads to contrast in phase.

The schematic layup of our imaging geometry is shown in **Figure S7**. Magnetic tips were Si NT-MDT that are Co/Cr coated and magnetized before use by using a large permanent magnet. The schematic shows the capacitor and the crystal edge vs. background substrate layout.

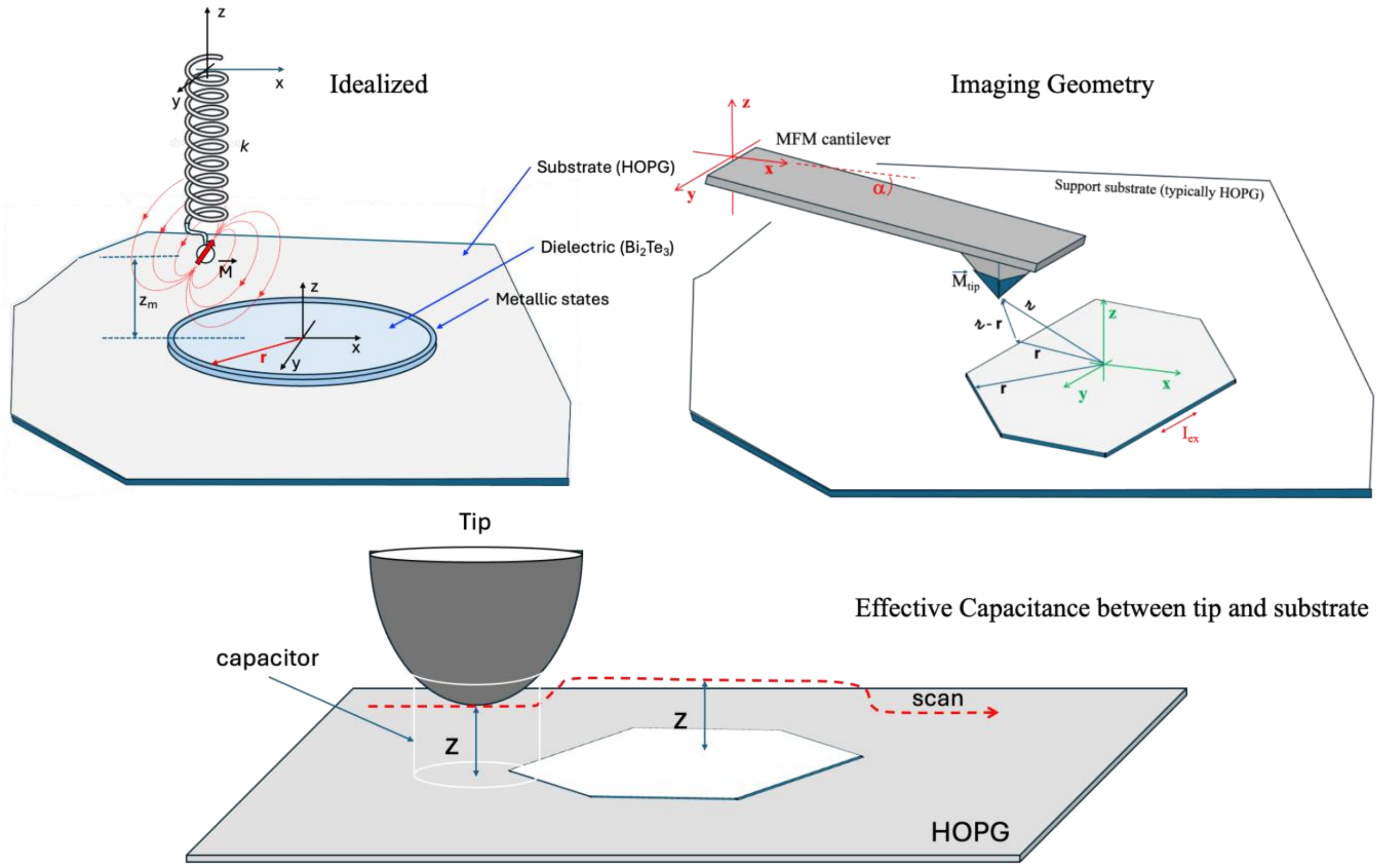

**Figure S7: (top left)** Idealization of tip-substrate geometry (right) as a spring with a magnetic mass suspended over the crystal plate. The current loop source is spatially extensive (~ 1micron diameter) with a magnetic field that interacts with the whole of the conical tip. We imagine this interaction as being localized to a small dipole interacting with the loop, where the tip has an effective dipole moment $m_{eff}$. When r >> LH we can treat the local magnetic field at the edge of the plate as that of a wire segment. Further, a capacitor is formed between the tip and the metallic support substrate (Co/Cr hemisphere and a flat plane of HOPG) [26]. Note z ~ the bottom of tip, $z_m$ is the location of the effective magnetic dipole of the tip.

### 4.1 Electric/Capacitive Forces

To examine the contributions of electric/capacitive forces to the observed MFM contrast, we imaged $Bi_2Te_3$ plates with a non-magnetic tip as a function of lift height. The probes were NT-MDT NSG03 Silicon with 10nm radius, a resonance frequency $\omega_o \sim 90$ kHz and force constant $k \sim 1$ N m$^{-1}$. They have no magnetic coating over the tip, and so magnetic effects are eliminated. Capacitive effects are not eliminated however because the Si tip is conductive due to the Sb doping and the Au reflective back coating, though the capacitor is different from that of the coated tip. Never-the-less, this represents a reasonable approximation of the order of magnitude and z-height behavior of the capacitive force. AFM images of the plates are contrasted with the phase images of the tip dithered above the plate in **Figure S8**. The phase images show some small contrast across the plate. However, interpretation of the change in phase with lift height is a little trickier since the background of the system (ie. the $\Delta\phi$ over HOPG) changes. The actual $\Delta\phi$ over the $Bi_2Te_3$ is ~ constant for this lift range. The dark (negative) $\Delta\phi$ observed suggests this is due to attractive surface forces between these layered materials and the cantilever as this would yield a negative gradient at distance from the source of the field.

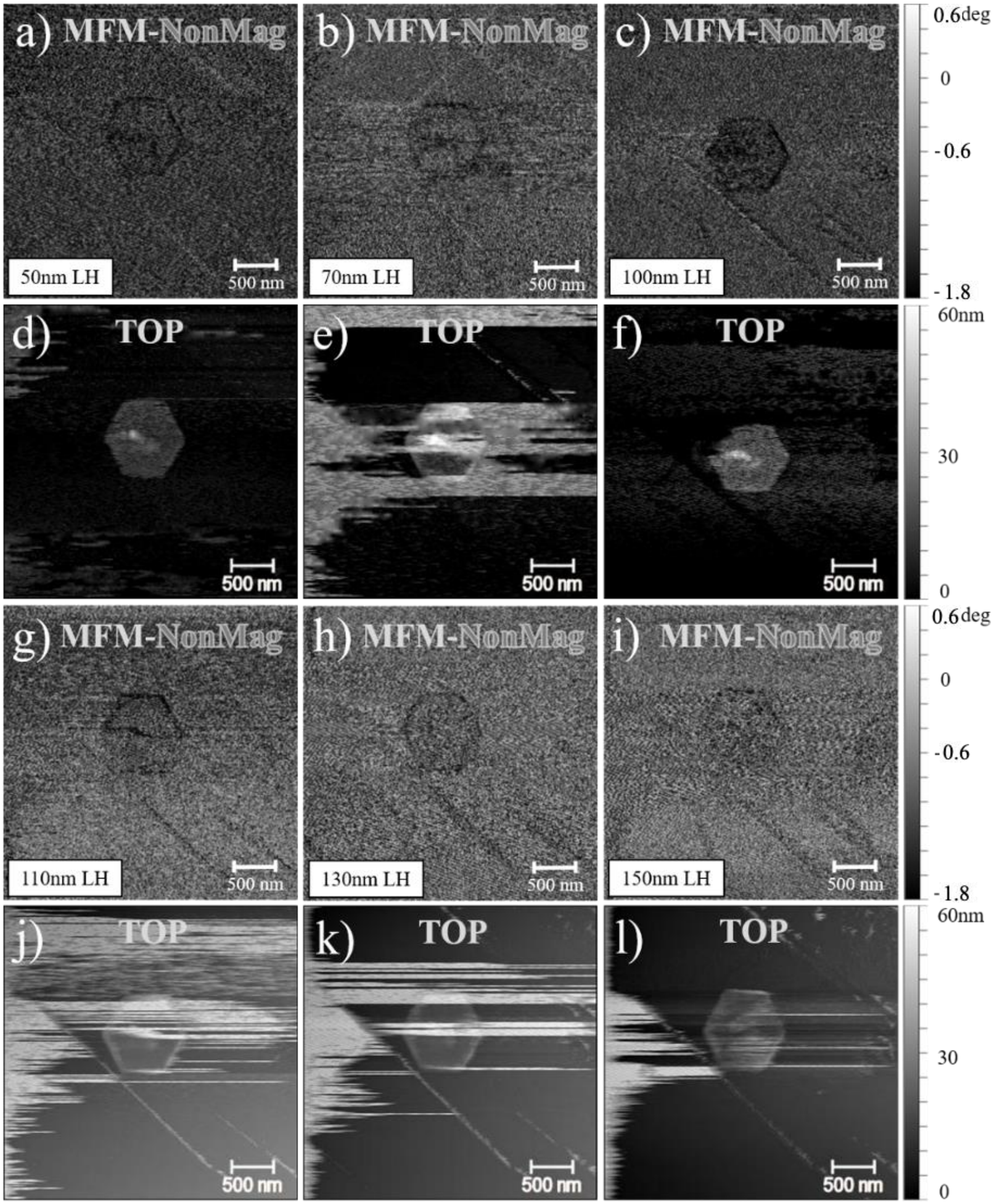

**Figure S8: "MFM" on $Bi_2Te_3$ with nonmagnetic AFM tip.** Phase (a-c, g-i) and topographical (d-f, j-l) images taken of a $Bi_2Te_3$ plate with a non-magnetic tip in MFM mode with lift heights 50-150 nm. Scanning parameters such as gain, feedback, cantilever tip velocity, and lift amp voltage were kept constant throughout. Cleaved graphite was used as the support substrate.

***Van der Waals forces***

We note here that when the tip is very close to the substrate, there are other electronically based forces to consider. There are forces associated with moisture on the condensed on the surface as well as dipole-dipole forces. These are all expressed in terms of a van der Waals term. Thus, the electrostatic forces in the MFM are $F_{electro} \sim F_{VdW} + F_{cap}$ where we use the geometry associated with **Figure S7:**

$$F_{VdW} = -\frac{A_H R}{6z^2} \qquad (2)$$

$A_H$ is the Hamaker constant, R is the radius of a spherical tip, and z is the height from a planar surface [26]. This term can be avoided in the phase image for lift heights above ~ 20 nm or so.

***Capacitive forces***

For the capacitive derived force:

$$F_{cap} = -\frac{\pi\, \varepsilon_o U^2 R}{z} \qquad (3)$$

when the probe is over the graphite surface. It is:

$$F_{cap} = -\frac{\pi\, \varepsilon_{eff} U^2 R}{z} \qquad (4)$$

when over the $Bi_2Te_3$ plate. The R in both equations is the radius of the tip over a flat surface. The dielectric constant changes as the probe scans from the graphite to the $Bi_2Te_3$ sample giving contrast. Using and *effective dielectric* approximation, the dielectric constant $\varepsilon_{eff}$ is written as the ratio: $\varepsilon_{eff} = \{[\varepsilon_{BiTe}\ t] + [\varepsilon_o\ z]\}/(z+t)$, where z is the scan height at which the tip is dithered above the substrate/sample, t is the thickness of the plate. z+t is the distance from the tip to the underlying substrate. t = 0 over the HOPG. The dielectric constant of $Bi_2Te_3$ ($\varepsilon_{BiTe}$) is taken as the bulk value. U is taken to be the contact potential plus any applied voltage between tip and sample. $U = U_{contact} + U_{applied}$ [22].

$$F_{electro} = -\frac{A_H R}{6z^2} - \frac{\pi\, \{\varepsilon_{BiTe} t + \varepsilon_o z\}(U_c + U_a)^2 R}{z\,(z+t)} \qquad (5)$$

In this expression when t goes to zero, the tip is over the HOPG and $A_H$ is that of HOPG. When over the $Bi_2Te_3$, the ε and $A_H$ is that of $Bi_2Te_3$.

$$\frac{dF_{electro}}{dz} = \frac{A_H R}{3z^3} - \frac{\pi\, R\,(U_c + U_a)^2[-\varepsilon_o z^2 - 2\varepsilon_{BiTe} tz - \varepsilon_{BiTe} t^2]}{z^2(z+t)^2} \qquad (6a)$$

$$= \frac{A_H R}{3z^3} + \frac{\pi\, \varepsilon_{BiTe} R\,(U_c + U_a)^2[\frac{1}{\varepsilon_r} z^2 + 2tz + t^2]}{z^2(z+t)^2} \qquad (6b)$$

Looking at the relative contributions of the components, we note that: $A_{H\,(HOPG)} \sim$ 23.8 x $10^{-20}$ J, $\varepsilon_{r(BiTe)} \sim$ 37, and $\varepsilon_{BiTe} \sim \varepsilon_o \varepsilon_r \sim$ (8.854 x $10^{-12}$ F/m) x (37). So, we can approximate:

$$\frac{d\mathrm{F}_{electro}}{dz} \sim \frac{\pi\, \varepsilon_{BiTe} R\, (U_c + U_a)^2 [\frac{1}{\varepsilon_r} z^2 + 2tz + t^2]}{z^2 (z+t)^2} \qquad (7)$$

Over HOPG ($t \rightarrow 0$ and $\varepsilon_{BiTe} \rightarrow \varepsilon_o$) so, generally, this looks like:

$$\frac{d\mathrm{F}_{elec}}{dz} \sim \frac{\pi\, \varepsilon_o R\, (U_c + U_a)^2}{z^2} \qquad (8)$$

Over $Bi_2Te_3$ (for low lift heights $z << t$):

$$\frac{d\mathrm{F}_{elec}}{dz} \sim \frac{\pi\, \varepsilon_{BiTe} R\, (U_c + U_a)^2}{t^2} \qquad (9)$$

***Dielectric loss***

However, we must also consider losses in such a system. These tend to reduce the effective Q of the oscillator system and thus they play an important role in contrast mechanisms generally. We note:

$$\Delta\phi = -\frac{Q}{k_s}\left[\left.\frac{dF(x,y,z)}{dz}\right|_{mag} + \left.\frac{dF(x,y,z)}{dz}\right|_{elec}\right]$$

Q is defined as the ratio of the resonance frequency to the resonance bandwidth, or $2\pi$ times the stored energy divided by the energy lost per cycle. So, one can achieve large contrast by going from a dissipative sample to a non-dissipative substrate. Similarly, we can expect modifications of $k_s$ with field strength due to the addition of torsional cantilever modes as fields get strong. Therefore, neither of these are truly constants, they both depend on z.

For electrostatic/capacitive interactions we ignore changes in $k_s$ since they do not typically lead to torsional forces. However, the cantilever-substrate system forms a capacitor with a $Bi_2Te_3$ dielectric partially filling it. And there is always some contact potential $U_c$ present. Therefore, dithering the tip means that we are applying a (kHz) time varying field across the dielectric. For a dielectric filling a high frequency capacitor $P_d \approx \omega \epsilon_0 \epsilon'_r \tan(\delta)\, E^2 \sim \omega \epsilon_0 \epsilon'_r \tan(\delta)\, (U_c/t)^2$, where $t >> z$ is the effective thickness of the capacitor, $\epsilon'_r$ is the real part of the dielectric and $\tan(\delta)$ is the loss tangent. Notice however, if we raise the distance to the upper electrode (the probe in this case), the lossy part of this system, which is the dielectric, fills the capacitor by a proportionally smaller amount. So, $P_d \sim \omega \epsilon_0 \epsilon'_r \tan(\delta)\, (U_c/t+z)^2$ where z is the distance the cantilever is raised above the sample. So, the dielectric fills less of the capacitor at greater lift heights and the field across the dielectric gets smaller.

This is what happens to first order. Contact potentials between substrate and sample imply some field dependent band bending, $\epsilon'_r$ and $\tan(\delta)$ will also have a field dependence. So if we use applied fields we would have to account for effects due to the large field strengths on the dielectric and on the contact.

The energy stored in the mechanical cantilever system is $E_s \sim ½ kA^2$. For defect free crystals $\tan(\delta)$ vs. E is flat. But for nanoscale materials, polymers, disordered, and defective materials, a weak linear dependence occurs. [27]. For $U_c$ small and no $U_a$, $\tan(\delta) \sim$ constant across a wide range of z. For low lift heights we substitute into Equation 9 giving the simple *approximation*:

$$\Delta\phi_{elec} \sim \frac{\pi^2 A^2 R}{\omega \tan(\delta)} \qquad (10)$$

This means that close to the substrate; the capacitive derived phase doesn't depend strongly on the imaging height. But it depends on the thickness as 1/t. This curious since we would expect that the more of the lossy dielectric we have, the greater the phase shift. However, in this case the increasing t also means a decreasing field across the dielectric. Notice in **Figure S7**, this seems to be approximately correct for the LH range > 30 nm – 40 nm. Δϕ over the HOPG doesn't include the dielectric loss mechanism and so it varies only as ~ $1/r^2$ and this gives contrast between the plate and the substrate. However, our approximation is for small lift heights. For lift heights (z > t) we use the full expression of Equation 7 and include the dependence of $Q_{BiTe}$ on z:

$$\Delta\phi_{elec} \sim \frac{Q_{BiTe}}{k_s}\frac{\pi\, \varepsilon_{BiTe} R\,(U_c + U_a)^2[\frac{1}{\varepsilon_r}z^2 + 2tz + t^2]}{z^2(z+t)^2}$$

$$= 2\pi\,\frac{½\, k_s A^2\,(z+t)^2}{k_s \omega \varepsilon_{BiTe}\tan(\delta)(U_c+U_a)^2}\,\frac{\pi\, \varepsilon_{BiTe} R\,(U_c + U_a)^2[\frac{1}{\varepsilon_r}z^2 + 2tz + t^2]}{z^2(z+t)^2} \qquad (11)$$

$$= \frac{\pi^2 A^2 R}{\omega \tan(\delta)}\,\frac{[\frac{1}{\varepsilon_r}z^2 + 2tz + t^2]}{z^2}$$

$$= \frac{\pi^2 A^2 R}{\omega \tan(\delta)}\left[\frac{1}{\varepsilon_r} + \frac{2t}{z} + \frac{t^2}{z^2}\right]$$

For charge sensitive probes like Kelvin probe force microscopy (KPFM), capacitive forces goes roughly as $(U_c + U_a)^2$ as seen in Equation 8. [22] However, when we image with large dielectric losses, we see that the voltage dependences fall out. Remember that for moderate-to-large driving fields before saturation, $\tan(\delta)$ can exhibit a nearly linear dependence on field strength because hysteretic loop area and domain-wall frictional losses scale quadratically with field while stored energy scales linearly: $\tan(\delta) \sim U/z$. This happens for $U_a$ >100keV/cm or 0.1 V/nm. At LH 100 nm this occurs at 10V.

$$contrast = \frac{Q_{HOPG}\pi\, \varepsilon_o R\,(U_c + U_a)^2}{k_s z^2} - \frac{\pi^2 A^2 R}{\omega \tan(\delta)}\left[\frac{1}{\varepsilon_r} + \frac{2t}{z} + \frac{t^2}{z^2}\right]$$

For conductive, non-magnetic probes $Q_{HOPG}$ approaches $Q_{free}$. Let $\alpha = Q_{HOPG}\pi\, \varepsilon_o R/k_s$, $\beta = \frac{\pi^2 R}{\omega \tan(\delta)}$ and take $U_c^{Bi2Te3} \sim U_c^{HOPG}$ because of the 2D nature. So:

$$contrast = \alpha \frac{(U_c + U_a)^2}{z^2} - \beta A^2 \left[\frac{1}{\varepsilon_r} + \frac{2t}{z} + \frac{t^2}{z^2}\right] \tag{12}$$

For an order of magnitude estimate: [$\varepsilon_o \sim 10^{-11}$ F/m, $Q \sim 10^2$, $k_s \sim 1$N/m] and $\alpha \sim 10^{-9}$ R, [$\tan(\delta) \sim 1$, $A^2 \sim 10^{-14}$, $\omega \sim 10^3$] and $\beta A^2 \sim 10^{-16}$R. For $U_a = 0$ and $U_c \sim 10^{-4}$V then $\alpha U_c^2$ approaches $\beta A^2$. Otherwise, the first term dominates. Since $z^2 \sim 10^{-18}$, then converting to nm for z:

$$\Delta\phi_{contrast} \sim (U_c + U_a)^2 / z^2 \text{ (nm)} - 10^{-7} \left(\frac{1}{\varepsilon_r} + \frac{2t}{z} + \frac{t^2}{z^2}\right) \text{ (nm)}.$$

***Estimates and data for electric forces***

This estimate of phase due to electrostatic/capacitive effects is presented in **Figure S9.** Remember that this is only an estimate, but it does show a phase shift of the order that we observe. It also emphasizes the dominance of the leading term.

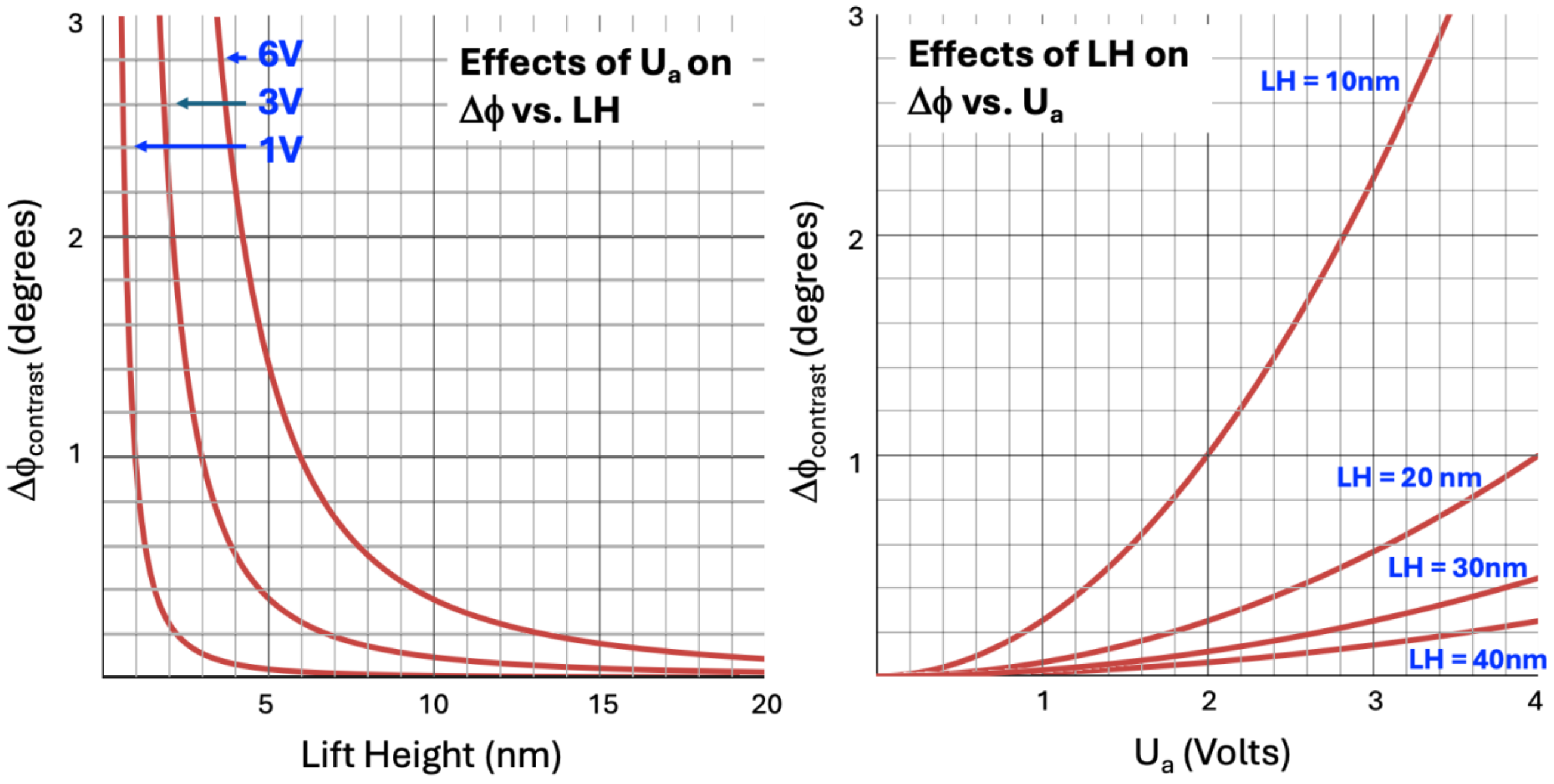

**Figure S9:** A simple estimate of phase shifts from the equation above. Note that these are order of magnitude estimates. LH can be off by a constant factor. However relative changes are well represented.

In **Figures S10 and S11** we see if this is true for our system. In this case a magnetized NT-MDT MFM01tip with a Co/Cr coating with a k ~ 1N/m and resonant frequency ~ 30 kHz as well as an Al reflective coating was used. We assume here that the substrate and plate are relatively weakly conductive and so magnetic responses from them should be equally as weak. **Figure S10(a)** shows the contrast between HOPG and the center plate of the $Bi_2Te_3$ for small lift heights as voltage is applied between the probe and substrate. The experiment applies voltage over a range of 0 – 6 volts. The lift heights we examine are: 0nm LH (as close as possible without touching), 5nm LH, and 10nm LH. The data is plotted as $U_a^2$ resulting in a linear relationship as predicted by Equation 11. For these voltage values and lift heights, the first term in the expression clearly dominates. We note also that the phase shifts observed are of the order that we expect as well.

The plate being imaged in **Figure S10** is ~ 6 nm thick, so at 0nm LH the capacitor is essentially all dielectric, at 5nm LH the capacitor is roughly 11 nm in spacing, and the dielectric portion is roughly ½ of the space between plates. At 10nm LH, the spacing of the capacitor is ~ 16 nm, and ~1/3 of the space is filled with dielectric. For each of these LH values, the change in capacitance relative to the HOPG due to the average dielectric response within the gap, leads to the marginal changes in slope. However, the $U^2$ dependence is maintained.

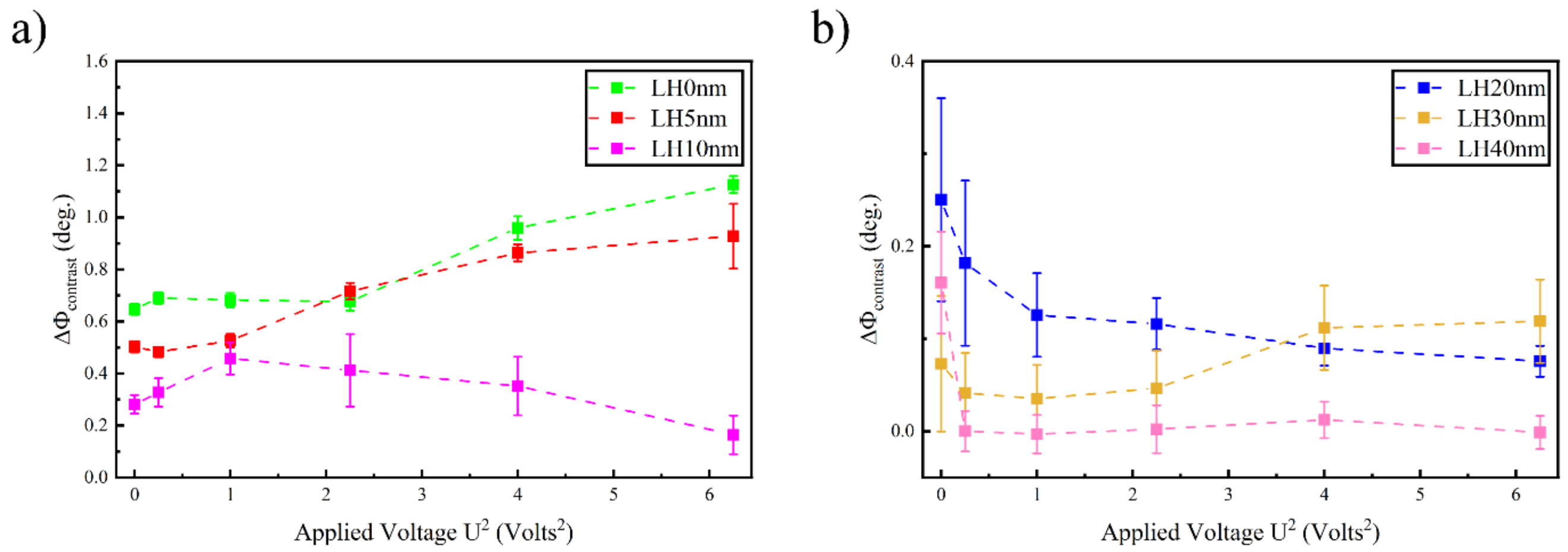

**Figure S10: (a)** The phase contrast: $\Delta\varnothing_{HOPG}$ (over the HOPG) – $\Delta\varnothing_{BiTe}$ (over the plate) = $\Delta\varnothing_{Tot}$, as a function of applied voltage for three low lift height imaging conditions. The $U^2$ dependence is clearly seen as predicted in Equation S11 above. **(b)** For larger lift heights, the phase contrast with $U^2$ is linear for $U^2$ > ~0.25V.

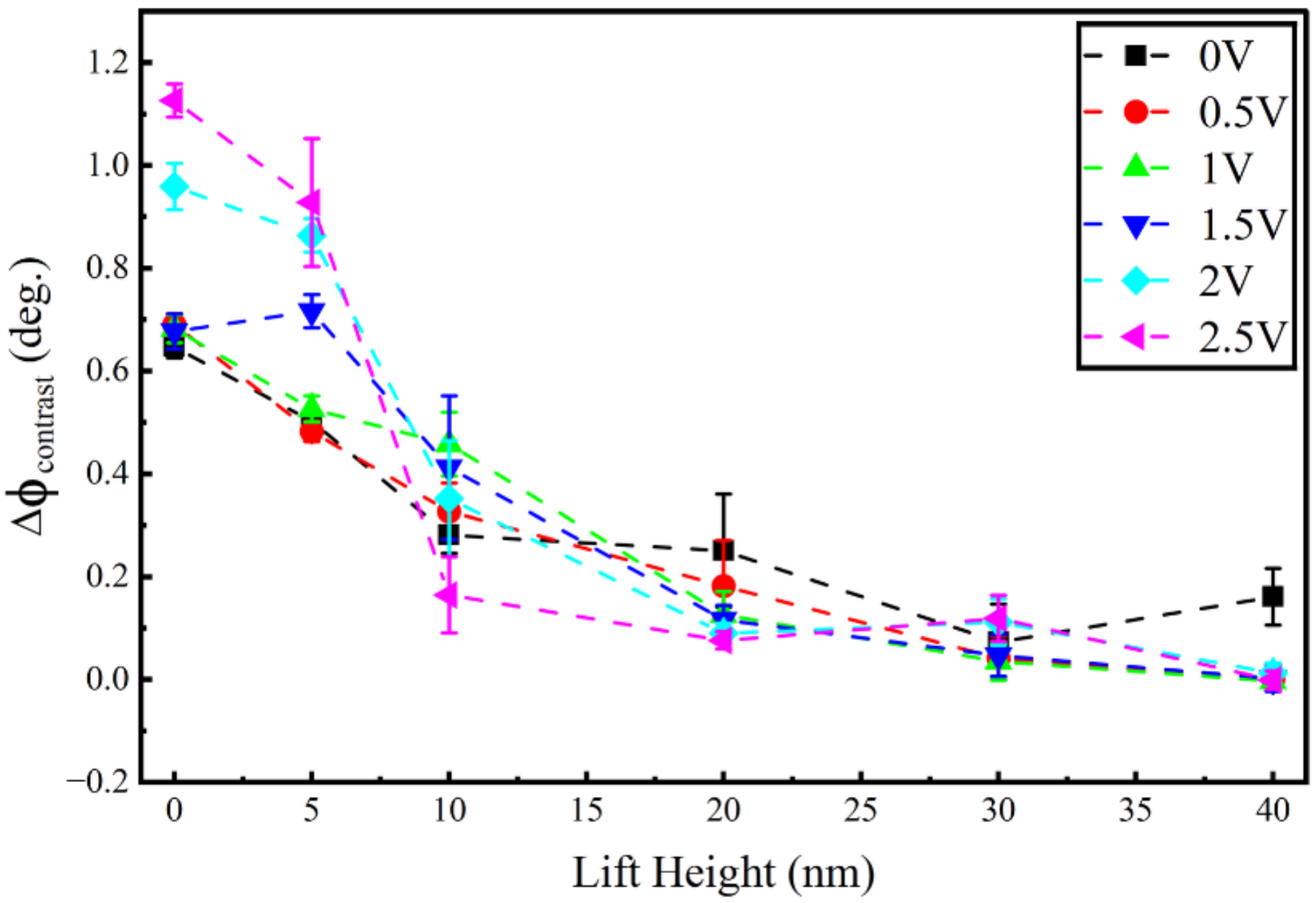

**Figure S11:** Phase contrast over the crystal plane as a function of LH when a voltage ($U_a$) is applied. This follows the predicted dependence expressed in equation 12.

This increasing linear trend is not what is observed for larger LHs. As seen in **Figure S10 (b)**, the mean phase contrast is plotted vs. $U_{app}^2$ for LHs of 20nm, 30nm, and 40nm. For LH ~ 30 nm and 40 nm, $\Delta\phi_{contrast}$ with $U_a^2$ is linear, but relatively flat except for right at 0V. Between: $0 < U_a^2 < 0.1V$ the applied field has a step that could be associated with the cancelling of the contact potential $U_c$. Otherwise, $\Delta\phi_{contrast} / \Delta U_a^2 \rightarrow 0$. 20nm LH seems to be more of a transition height, and in this case a linear decrease with $U_a^2$ is observed. This behavior is easy to understand in terms of **Figure 10**. Notice that the $\Delta\phi_{contrast}$ vs. z curves, for any of the applied voltages, flatten out beyond about 20 nm in LH. This is simply because the forces over the HOPG and over the HOPG with a very thin $Bi_2Te_3$ layer begin to look the same from far away. The $Bi_2Te_3$ is only a small perturbation on the probe-HOPG capacitor at LHs above 30 nm.

We further examine this interpretation in **Figure S11** with $\Delta\phi_{contrast}$ as a function of z with increasing applied voltages. We note that there appears to be two distinctly different regimes of operation: the first high voltages 2-2.5V. In this case the phase has an abrupt downturn at a lift height between 10 nm and 20 nm. However, for smaller $U_a$ ~ 0-1.5V this downturn is shift to LHs between 20 nm and 30 nm.

***Magnetic effects across the edge***

Unlike contrast of the plane or bulk in the center of the crystal from HOPG, the contrast between the crystal edge and HOPG behaves differently. This behavior, measured again as described above, is seen in **Figure S12** and **Figure S13.** There are two completely different height regimes of behavior and to understand them we need to estimate how magnetic effects due to currents in the highly conductive edges, will generate phase shifts. Notice also we have included variations as a function of applied potential so we can determine characteristic length scales of both electrostatic and magnetic forces.

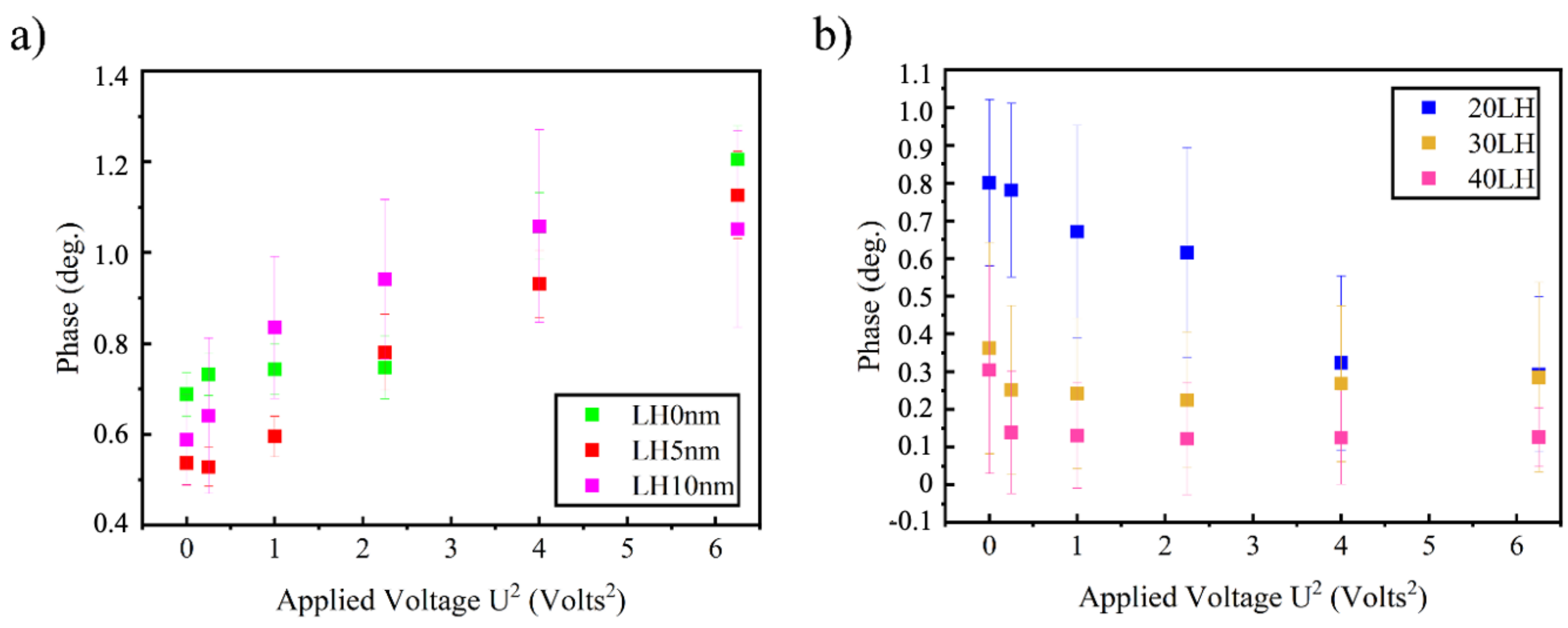

**Figure S12:** Phase contrast over the crystal edge as a function of $U_a^2$.

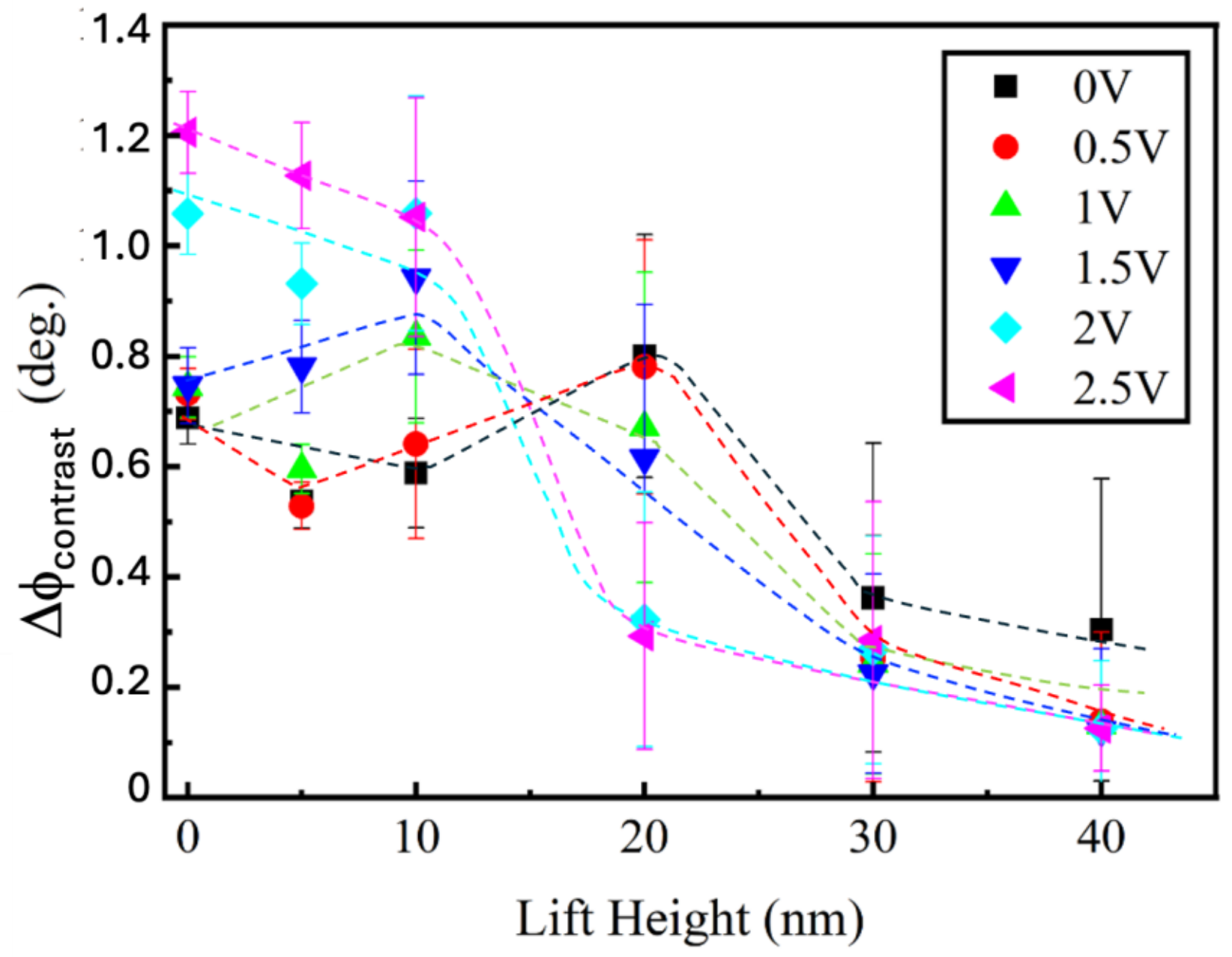

**Figure S13:** Phase contrast over the crystal edge as a function of LH when a voltage ($U_a$) is applied.

Note that in both **Figure S13** and **S14**, the Δϕ being plotted is a contrast, that is a difference between the phase measured at the LH over HOPG and then over the sample.

## 4.2 Magnetic Forces

The images used for the above analysis were, in the later cases, taken using a Co/Cr coated tip magnetized as prescribed by the manufacturer NT-MDT. We should anticipate that HOPG as well as the center of the 2D topological plates will have little magnetic response. There will be some damping due to induced eddy currents since both are slightly conductive. However, the highly conductive edges should certainly express a strong magnetic character. In fact, this is the very point of our work.

To estimate the magnitude of magnetic interactions, we will need to make a few assumptions:

1. We use a "weak tip" approximation. In this case, this means LH values must be greater than or equal to the tip dimensions. [22] We use MFM01 magnetic tips from NT-MDT with an aspect ratio of ~ 430 and tip diameters ~ 35nm. From the electrostatic estimate above, at LH's above this the electrostatic effects will fall off.

2. The plates are ~ 1 μm diameter. We argue that below 200 nm we can approximate the line segments of the outer edge of the plates as being relatively isolated, straight, current carrying wires. The field encountered is thus approximately: $\vec{B}_{edge}(s) \approx \left(\frac{\mu_0}{2\pi} I_{edge}\right) s^{-1}\hat{\varphi}$, as seen in **Figure 3**.

3. For "tapping mode" imaging on most commercial AFMs, the system geometry is setup to be sensitive to forces and gradients along the z-axis of the cantilever. This is specifically arranged so that it corresponds with some x or y-axis of the detector as in **Figure S14.**

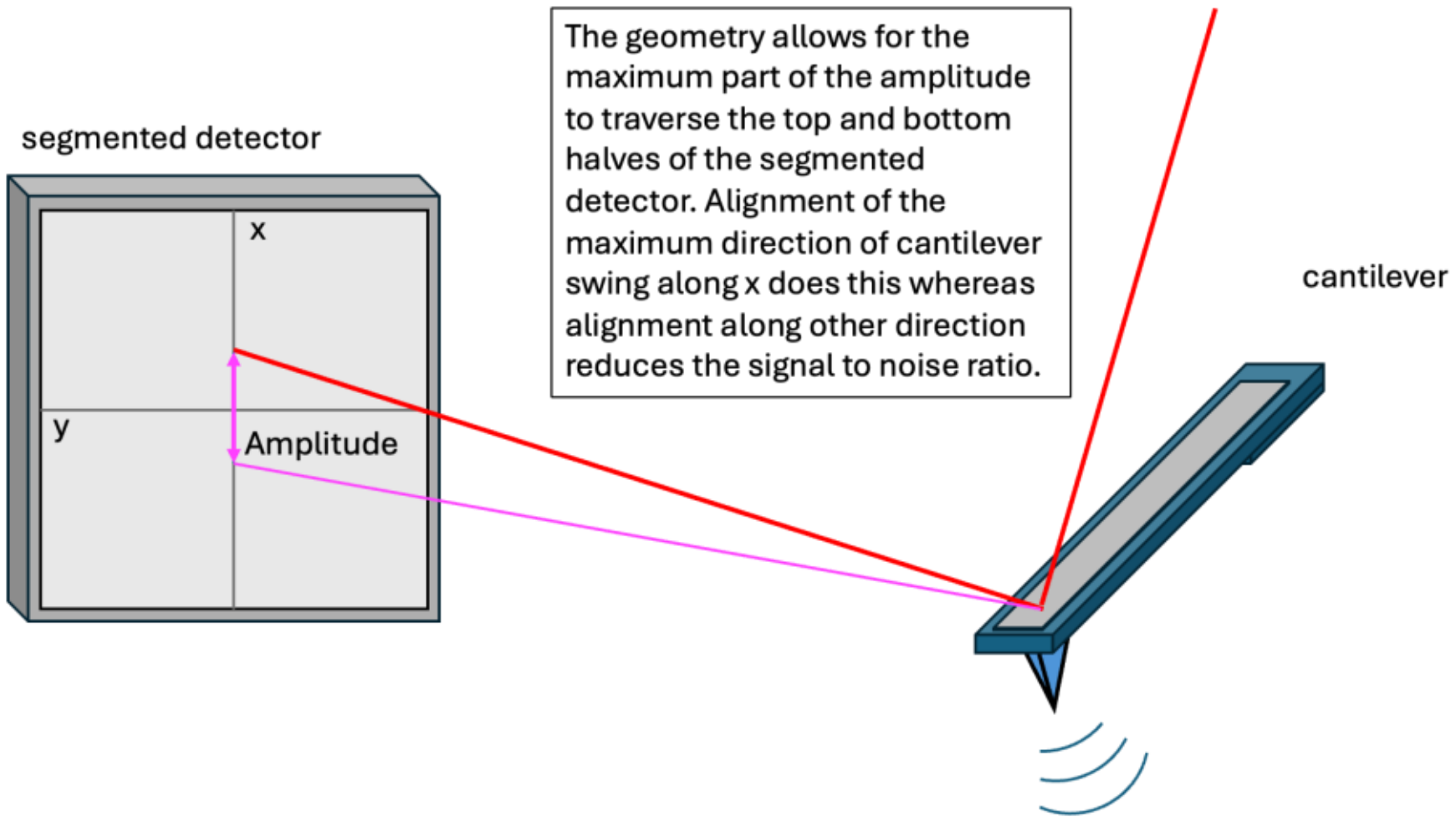

**Figure S14:** the geometry of alignment. Note that for greatest signal to noise the reflected beam needs to move across the segmented detector from the top half to the bottom half. The larger this part of the motion is, the better the signal. But this means that the system is then most sensitive to motion in the Z-direction of the cantilever. Lateral torsion modes and torques are far more difficult to measure in AC-AFM.

So, for small forces and small driving amplitudes we consider only the cantilever's primary bending mode in z. This will ignore torques placed on the tip by field components not aligned with the magnetic moment of the tip (ie. parallel to the surface of the sample). Generally, they act to increase the bending mode amplitude of the cantilever in z. In this case, we claim that the dipole-field interaction is too small to effectively bend the cantilever significantly in z by twisting the tip in around the x/y axis.

4. This simple model estimates induction due to the cantilever dither, and this means we must define the overall geometry of current paths. In our case, the perimeter of a plate is a highly conductive channel. As seen in the microscopy, the plates are faceted and sometimes can be quite irregular. But to estimate flux we assume a circular loop around this perimeter and then add a factor to account for the effects of geometry and nonaxial tip position.

### 4.2.1 General Flux Estimate

For a dipole placed in the center of the loop of radius $r$, and oscillating at an equilibrium height $z_m$ above the loop, the flux through the loop is given by:

$$\Phi = \int_0^r H_z^{dipole} 2\pi r' dr' \tag{6}$$

$$H_z^{dipole} = \frac{m_{eff}}{4\pi} \frac{2z_m^2 - r^2}{(z_m^2 + r^2)^{5/2}} \tag{7}$$

And following the example of reference [29] SI units will be used throughout, but the unit of magnetic moment adopted here is Wb/m, rather than A/m$^2$.

Note, however, for a dipole placed at the edge of the loop, the integral can be approximated as $\Phi_{Faraday} = \frac{1}{4}\gamma\Phi$. This is demonstrated geometrically in **Figure S15**. The factor $\gamma$ accounts for differences in field density in each area segment.

Sample plates vary in shape and size. And we are interested in measuring the phase response when the tip (the magnetic source of the Faraday flux) is dithered near the plate edge, not at the center of the area as the integral above calculates. This makes the overall flux messy to compute for each sample, and it doesn't provide a general statement of flux behavior. Instead, a reasonable estimate of the generalized flux can be made by starting with a symmetrically placed source and then sectioning the surrounding area into inscribed polyhedral shapes that approximate the real samples. Such an estimate allows us to express the time dependent flux in two terms, one of which depends only on the height and symmetry of the source (imaging parameters), the other is only dependent on the geometry of the plate. This means comparisons between samples of different sizes and shapes can be made more easily by varying imaging parameters.

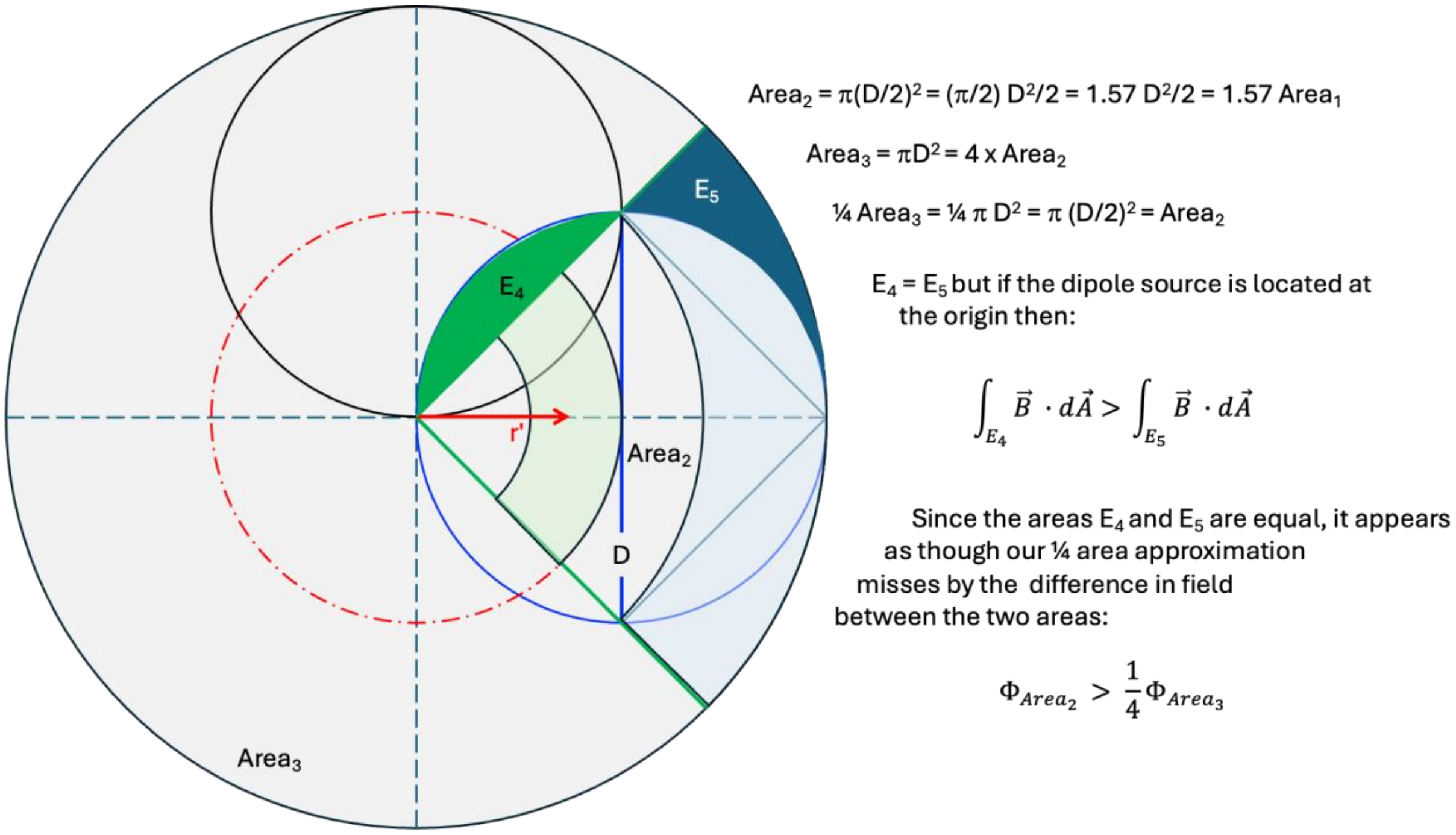

**Figure S15:** Schematic showing geometrical relationships. Areas inscribed within larger areas can be used to provide a good estimate of the flux of a dipole source at the edge of a circular shaped plate.

Using this approach, note that in **Figure S15** the small circle ($Area_2$) is inscribed within $Area_3$ (the large circle) and its diameter ($2r = D$) is the radius of $Area_3$. In this case $Area_2$ is $1/4Area_3$ and $Area_4 = Area_5$. But while these areas are equal, the fluxes through $Area_4$ and $Area_5$ are different because the shapes of the areas are different, and they occur at different distances from the dipole source.

It is easier to account for this by comparing the ¼ pie-slice defined by green lines with the inscribed circle of $r = D/2$. While the two areas are the same, the pie-slice has more of its area located at the edge of the large $Area_3$ whereas the circle has more of its area near the center of $Area_3$. The field from the dipole is much stronger near the center of $Area_3$ than at the edge of $Area_3$. This means that even though the two shapes have the same area, the inscribed circle has a larger flux than the pie-slice.

Thus, if we allow the limits of integration above $r \rightarrow D$ then we get the flux associated with the area $\pi D^2$. We relate the fluxes between the symmetric larger area and the off-axis placement on a smaller circle as:

$$\Phi_{E_5} \quad = \quad \int_{E_5} H_5^{||}\, dA \tag{10a}$$

and

$$\Phi_{E_4} \quad = \quad \int_{E_4} H_4^{||}\, dA \tag{10b}$$

But both Φ's compose $Area_3$ and so the integral can be written as:

$$\Phi_{sample} = \frac{1}{4}\Phi_D - \Phi_{E_5} + \Phi_{E_4} \quad (11a)$$

$$= \frac{1}{4}\gamma\Phi_D \quad (11a)$$

as shown in **Figure S15**. Put this into the formula for flux and we get:

$$\Phi_{sample} = \gamma \int_0^r H_z 2\pi r' dr' = \frac{\gamma m_{eff} r^2}{(z_{\mathrm{m}}^2 + r^2)^{3/2}}.$$

There is an alternative approach to computing this flux. We want the flux through a loop of radius r from a dipole located on the edge of the loop ($r,0,z_m$). This is messy but following *W.R. Smythe in STATIC AND DYNAMIC ELECTRICITY II ed. by, McGraw-Hill, New York, 1950. P. 271*, we can apply the reciprocity theorem (otherwise known as the mutual inductance theorem) to substitute another problem based on a unit current flowing through the current loop and determining the flux off axis at the position of the dipole. This problem is a bit easier, and the solution is given in Smythe:

$$\Phi = m_{eff} \frac{z_m}{\sqrt{(2r)^2 + z_m^2}} [-K(k) + E(k)]$$

Where:

*K(k)* = complete elliptic integral of the first kind

*E(k)* = complete elliptic integral of the second kind

*k* = elliptic modulus parameter:

$$k = \sqrt{\frac{4r^2}{(2r)^2 + z_m^2}}$$

For k ~ 0 ($z_m$ .. r):

Series expansion:

$$K(k) = \frac{\pi}{2} \sum_{n=0}^{\infty} \left[\frac{(2n-1)!!}{(2n)!!}\right]^2 \frac{k^{2n}}{(2n-1)}$$

Series expansion:

$$E(k) = \frac{\pi}{2} \sum_{n=0}^{\infty} \left[\frac{(2n-1)!!}{(2n)!!}\right]^2 \frac{-k^{2n}}{(2n-1)}$$

and

$$\Phi = m_{eff} \frac{z_m}{((2r)^2 + z_m^2)^{1/2}} \left[ -\frac{\pi}{2} \sum_{n=0}^{\infty} \left[\frac{(2n-1)!!}{(2n)!!}\right]^2 \frac{\left(\frac{4r^2}{(2r)^2 + z_m^2}\right)^n}{(2n-1)} + \frac{\pi}{2} \sum_{n=0}^{\infty} \left[\frac{(2n-1)!!}{(2n)!!}\right]^2 \frac{\left(-\frac{4r^2}{(2r)^2 + z_m^2}\right)^n}{(2n-1)} \right]$$

$$\Phi = m_{eff} \frac{z_m}{((2r)^2 + z_m^2)^{1/2}} \frac{\pi}{2} \begin{bmatrix} -1 \\ -\frac{1}{4}\left(\frac{4r^2}{(2r)^2 + z_m^2}\right)^2 \\ -\frac{9}{64}\left(\frac{4r^2}{(2r)^2 + z_m^2}\right)^4 \\ - \cdots \\ +1 \\ -\frac{1}{4}\left(\frac{4r^2}{(2r)^2 + z_m^2}\right)^2 \\ -\frac{3}{64}\left(\frac{4r^2}{(2r)^2 + z_m^2}\right)^4 \\ - \cdots \end{bmatrix}$$

For k~1:

$$k' = (1 + k^2)^{1/2}$$

$$\Lambda = ln\left(\frac{4}{k'}\right)$$

Series expansion 1st kind

$$K(k) = \Lambda + \frac{1}{4}(\Lambda - 1)k'^2 + \frac{9}{64}\left(\Lambda - \frac{7}{6}\right)k'^4 + \frac{25}{256}\left(\Lambda - \frac{37}{30}\right)k'^6 + \cdots (\sim ln\left(\frac{4}{k'}\right)\ for\ k \sim 1)$$

$$\sim ln\left(\frac{4}{(1+\frac{4r^2}{(2r)^2+z_m^2})^{1/2}}\right) = ln\,4 - \frac{1}{2}ln(1+\frac{4r^2}{(2r)^2+z_m^2})$$

$$\ln(x) \sim \sum_{n=1}^{\infty}(-1)^{n+1}\frac{x^n}{n} \sim x - \frac{x^2}{2} + \frac{x^3}{3} - \frac{x^4}{4} + \cdots$$

Series expansion 2nd kind

$$E(k) = 1 + \frac{1}{2}\left(\Lambda - \frac{1}{2}\right)k'^2 + \frac{3}{16}\left(\Lambda - \frac{13}{12}\right)k'^4 + \frac{15}{128}\left(\Lambda - \frac{6}{5}\right)k'^6 + \cdots \text{ (~ 1 for } k \text{ ~1)}$$

$$\Phi = m_{eff}\,\frac{z_m}{((2r)^2 + z_m^2)^{1/2}}\left[-\left(ln4 + \frac{1}{2}ln\left(1 + \frac{4r^2}{(2r)^2 + z_m^2}\right)\right) + 1\right]$$

$$\Phi = m_{eff}\,\frac{z_m 4r^2}{((2r)^2 + z_m^2)^{3/2}} - \left[\frac{0.2}{((2r)^2 + z_m^2)^{1/2}}\right]$$

$$\Phi = m_{eff}\,\frac{z_m 4r^2}{((2r)^2 + z_m^2)^{3/2}} - \left[\frac{0.2((2r)^2 + z_m^2)}{((2r)^2 + z_m^2)^{3/2}}\right]$$

$$\Phi = m_{eff}\,\frac{z_m^2 - z_m 4r^2 + 0.2(4)r^2}{((2r)^2 + z_m^2)^{3/2}}$$

In our case: $z_m << r$, so:

$$\Phi \sim m_{eff}\,\frac{0.2(4)r^2}{((2r)^2 + z_m^2)^{3/2}}$$

Which is essentially:

$$\Phi_{sample} \sim \gamma\Phi_o$$

As we had estimated above!

### 4.2.2 Discussion of Error

It should be recognized that these approximations are used to make the model calculations simpler, such as the angle of the tip dipole, and the prefactor for the flux-area equation (**Figure S15**). Ultimately, using propagation of errors in the model above, these assumptions can vary the outcome of these results by an estimated factor of ~ ± 2√2. However, such errors do not represent order of magnitude changes in the predicted values. Further, we have limited these analyses to the dipole interaction approximation. This also makes analytical expressions easier to obtain. We can also make an estimate of the possible error associated with this approximation by making use of a multipole expansion for both the tip magnetic field as well as the source field. In doing so we find that 2nd order corrections for LHs > 35 nm are far more substantial than those at LH ~ 100 nm. Indeed, corrections fall off as $1/z^5$ with the prefactor for the correction being of the same order as the effect at 35 nm. These estimates of error have integrated into the data set presentation as the uncertainty bars on the data points. A final source of error in the analysis comes from the use of the feedback system to adjust the equilibrium position of the cantilever as the amplitude is adjusted. Note that we have written the tip velocities evaluated at the equilibrium positions but have plotted everything as a function of amplitude directly. This introduces an error because the source of the inducing field changes its position in space (its height above the ring) as the amplitude changes. A comparison between small and large amplitude changes is made in **Figure S8**. The effect on the outcome for small amplitude changes is slight (< 10% in phase).

### 4.2.3 An Expression for Phase

The force dependence on lift height z on our weak tip in a nonuniform magnetic field is given by the separated magnetic charge model [22] as expressed by $\vec{F} = \nabla(\vec{m}\cdot\vec{\mathrm{B}})$. Taking the $F_z$ component alone and using the coordinate system of **Figure S7** and the dipole field above:

$$F_z = \frac{\partial(\vec{m}\cdot\vec{H})}{\partial z} = \frac{m_{eff}\,\partial B_{total}(r)}{\partial z}$$
$$(this\ assumes\ m_{eff}\ isnt\ perfectly\ aligned\ with\ z)$$

and (12)

$$\Delta\phi_{mag}(r) = \frac{Q_o}{k_s}\frac{dF_z}{dz}$$

Again, for our estimate the section of the current carrying edge “appears” to the dipole as a straight wire. In this case:

$$F_z = \frac{\partial(\vec{m}\cdot\vec{H})}{\partial z} = \frac{I\,m_{eff}}{2\pi}\left[\frac{y^2-2yz-z^2}{(y^2+\,z^2)^2}\right] \qquad (13)$$

This gives a phase shift from free cantilever to directly over the edge of the sample of:

$$\Delta\phi_{mag} = \frac{Q_{far}}{k}\frac{dF_z}{dz} \sim -\frac{I\,m_{eff}}{2\pi}\frac{Q_o}{k_s}\left[\frac{1}{z^3}\right]\ ;\ x=0 \qquad (14)$$

Equation 14 is naïve expression for the change in the cantilever’s phase relative to its driving force as the tip is positioned over the edge of the plate and the lift height, $z_m$, is small compared to the diameter of the plate. $m_{eff}$ is the effective dipole moment of the magnetic tip. $Q_o$ is for the tip positioned far from the surface and assumed to be a constant.

Considering the small electrostatic component at this tip position, we expect $\Delta\phi_{edge}$ vs. $1/z^3$ to be approximately a straight line with a slope from which we can solve for the current I that causes this phase shift. We expect the electrostatic $U_c/z^2$ correction to be small because $U_c$ will be small. This is justified by comparison imaging of non-topological chalcogenide plates which show little to no edge contrast under the same imaging conditions. (see below)

Our expectation of the edge phase shift as it compares with data is shown in **Figure S16**. Here there appear to be two different slopes, for different imaging height ranges. Our $\Delta\phi \sim U_c/z^2 - 1/z^3$ estimate should be continuous and close to a straight line. Calculating the numbers from Equation 5 we get: $I \sim 79$ nA for 20 nm < LH < 40 nm and $I \sim 1.04$ μA for 40 nm < LH < 80 nm: a shift of two orders of magnitude within a modest change in LH. Moreover, the larger current seems to occur with the current inducing tip far from the substrate. Equation 5 is classical induction and well proven for MFM imaging of metals. [22] And, the tip hasn't been moved far enough away to make our assumptions no longer valid. Even at 80 nm, the loop appears large to the tip dipole. At 40 nm the dipole is still isolated and weak.

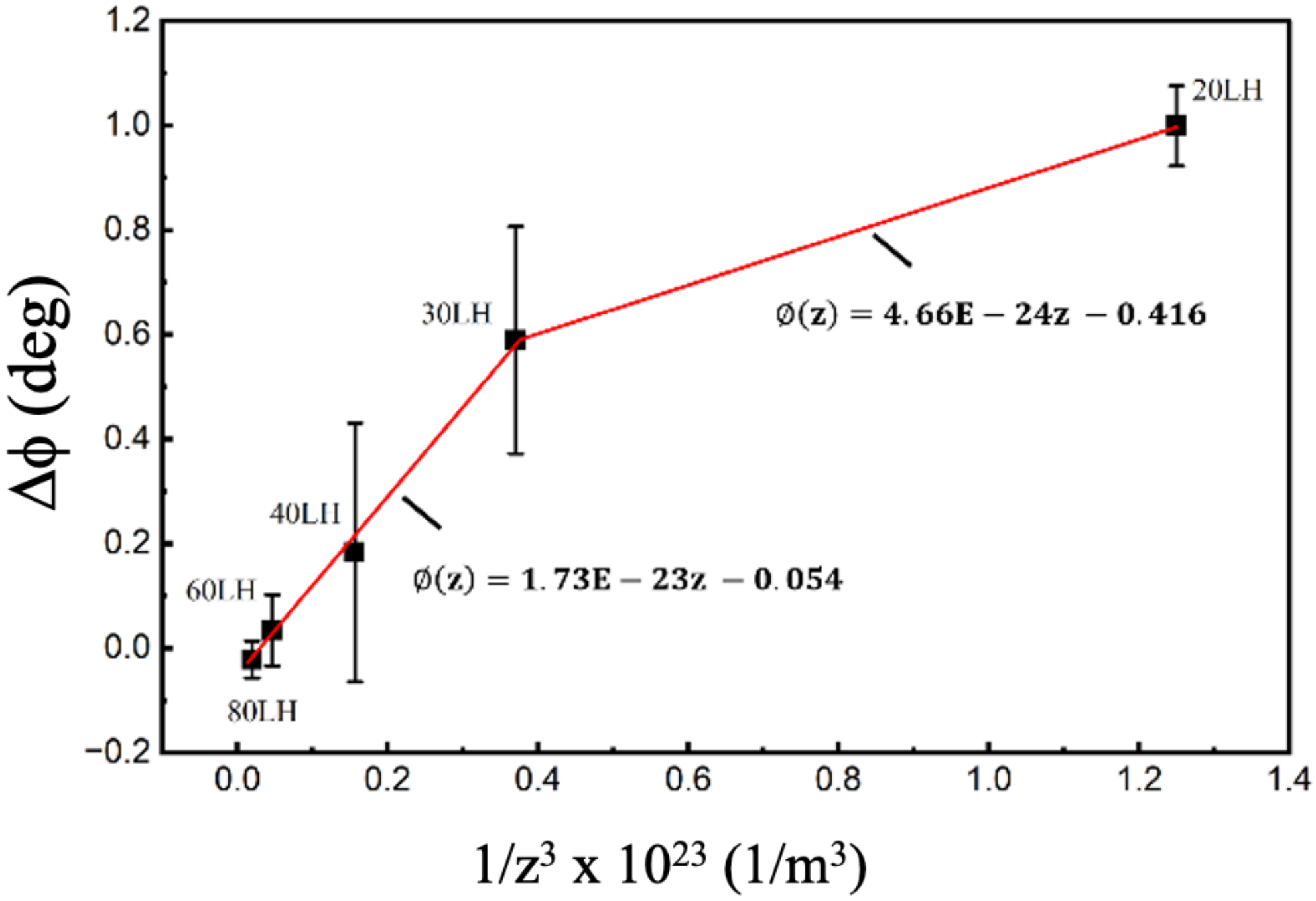

**Figure S16:** $\Delta\phi$ as a function of $1/z^3$ should yield a straight line based on expectations from basic published models. Data however gives a bimodal relationship between slope and current values.

***Including more terms***

**Figure S16** suggests we must modify our model and include in $I$ induced components as well: Faraday induction + magnetic scattering + persistent currents (if they occur). Recall from our estimates in Section 1, we do not anticipate having significant "topological persistent current" due to pure Zeeman symmetry breaking from the static presence of a well removed tip alone. Thus, most non-Faraday components would come from localized scattering along the edge-loop. The current due to the tip motion (dither) is Faraday induced and occurs in any edge channels as well as the conduction band electrons of the plate. However, the mechanisms for creating such non-Faraday currents we suggest come from localized scattering and thus it will introduce back-scattering into the otherwise ballistic edge channels as discussed in literature. [28]

Therefore, we anticipate dissipation losses for tip positions over the plate edge, as well as over the plate itself, and over the HOPG. As discussed in reference [29] the dissipative forces experienced by the tip/cantilever appear as a reduction in the $Q$ factor: $Q_o \rightarrow Q_{edge}$, and $Q_o \rightarrow Q_{mag\,plate}$ etc. Reference [29[29]] calculates these changes due to damping over a simple metal film. They typically depend on a Drude resistivity ($\rho$) imaging height, tip velocity, etc. In our case the resistivity term changes from substrate to sample edge to sample plane. Over the sample's edge loop as we have just discussed, it hides a more complicated physics of magnetic scattering.

Using reference [29] the phase shift of HOPG and the $BiTe_3$ plate (not its edge) with their mobile carriers at room temperature, can be estimated to be:

$$\Delta\phi_{\mathrm{mag\ HOPG/BiTe}} \sim -\frac{1}{2}\frac{z_{\mathrm{c}}^4}{z_{\mathrm{m}}^4}$$

$$z_{\mathrm{c}}^4 \equiv \frac{3}{64\pi}\frac{\omega_o}{k_s}\frac{{m_{eff}}^2}{\rho_{HOPG/BiTe}}.$$

$\omega_o$ is the resonant frequency of the system, $k_s$ is the spring constant. $\rho_{HOPG/BiTe}$ is the resistivity of HOPG or $Bi_2Te_3$ respectively.

***Estimates for a plate with conductive edge***

The imaging configuration we have is different and more challenging. Returning to the magnetic geometry of **Figure S7**, the cantilever as a magnetic dipole mass on a spring coupled to an inductive loop representing the samples' edge.

Any free current, including induced currents, generate a field from the loop. At the edge where the tip is located, $(r,0,z_m)$, this field looks like:

$$H_z = \frac{I}{2r}\frac{1}{\pi}\left[4 + \frac{z_m^2}{r^2}\right]^{-1/2}\left[E(k)\frac{-\left(\frac{z_m}{r}\right)^2}{4 + \frac{z_m^2}{r^2} - 4} + K(k)\right]$$

It too is expressed in terms of elliptic integrals. If we assume: $r^2 \sim (r^2 + z_m^2)^{1/2}$. $k = [4r^3/(4r^2+z_m^2)]^{1/2}$, then our approximations for elliptic integrals above give:

$$H_z = \frac{Ir}{2\pi[4r^2 + z_m^2]^{1/2}}[-E(k) + K(k)]$$

$$k' = (1 + k^2)^{1/2}$$

$$\Lambda = ln\left(\frac{4}{k'}\right)$$

$$\Lambda \sim \left(\frac{2}{(4+\frac{z_m^2}{r^2})^{1/2}}\right)$$

***For a persistent current***

If the current in this loop is a persistently flowing current then we get:

$$H \sim \frac{\mathrm{I}r}{2\pi[4r^2 + \; z_m^2]^{1/2}}\left[\frac{-2r}{(4r^2+z_m^2)^{1/2}} - 1\right]$$

$$F_{persistent} \sim m_{eff}\frac{d}{dz_m}H \sim -\frac{\mathrm{Ir}}{2\pi}\;\frac{d}{dz_m}\;\left[\frac{2r}{(4r^2+z_m^2)} + \frac{1}{[4r^2 + \; z_m^2]^{1/2}}\right]$$

$$F_{persistent} \sim m_{eff}\frac{\mathrm{Ir}}{2\pi}\;\left[\frac{4rz_m}{(4r^2+z_m^2)^2} + \frac{z_m}{[4r^2 + \; z_m^2]^{3/2}}\right]$$

$$\Delta\phi_{persistent} \;=\; \frac{Q_o}{k}\;m_{eff}\frac{\mathrm{Ir}}{2\pi}\;\frac{d}{dz_m}\;\left[\frac{4rz_m}{(4r^2+z_m^2)^2} + \frac{z_m}{[4r^2 + \; z_m^2]^{3/2}}\right]$$

$$\Delta\phi_{persistent} \;=\; \frac{Q_o}{k}\;\frac{m_{eff}\mathrm{Ir}}{2\pi}\left[-\frac{16r{z_m}^2}{(4r^2+z_m^2)^3} - \frac{3{z_m}^2}{[4r^2 + \; z_m^2]^{\frac{5}{2}}} + \frac{1}{[4r^2 + \; z_m^2]^{\frac{3}{2}}} + \frac{4r}{[4r^2 + \; z_m^2]^2}\right]$$

Thus, the field vs. I at the edge of a loop of wire looks very different from our expectations of a sheet of metal (above).

***Including a Faraday current***

If the I isn't persistent but instead is induced by the tip dither, then we might express our expectations as:

$$\Phi_{sample} = \frac{\gamma\pi m_{eff}r^2}{(z_\mathrm{m}^2 + r^2)^{3/2}}$$

Resistance R = 2πrρ

$$I_{Faraday} = \frac{3m_{eff}\gamma z_\mathrm{m}r}{8\pi\rho(z_\mathrm{m}^2 + r^2)^{5/2}}\dot{z}_\mathrm{m}$$

$$H_z \sim \frac{\frac{3m_{eff}\gamma z_\mathrm{m}r}{8\pi\rho(z_\mathrm{m}^2 + r^2)^{5/2}}r}{2\pi[4r^2 + \; z_m^2]^{1/2}}\left[\frac{-2r}{(4r^2+z_m^2)^{1/2}} - 1\right]\dot{z}_\mathrm{m}$$

$$H_z \sim \frac{3m_{eff}\gamma r^2}{16\pi^2\rho}\frac{z_{\mathrm{m}}}{(z_{\mathrm{m}}^2 + r^2)^{5/2}[4r^2 + z_m^2]^{1/2}}\left[\frac{-2r}{(4r^2 + z_m^2)^{1/2}} - 1\right]\dot{z}_{\mathrm{m}}$$

$$F_{Faraday} \sim \frac{3m_{eff}{}^2\gamma r^2}{16\pi^2\rho}\left[\frac{(4r^2 + z_m^2)^{\frac{1}{2}}(z_m^2 r^2 + z_m^4) + 12rz_m^4 + (z_m^2 + 4r^2)^{3/2}(4z_m^2 - r^2) + 34r^3 z_m^2 - 8r^5}{(4r^2 + z_m^2)^2(r^2 + z_m^2)^{7/2}}\right]\dot{z}_{\mathrm{m}}$$

From this we can get an expression for the $\Delta\phi$ of the Faraday effect in the edge-loop in two ways; using the expression for the $\Delta\phi_{persistent}$ and substituting I with the "Faraday current."

$$\Delta\phi_{Faraday} = \frac{Q_o}{k}\frac{3m_{eff}^2}{16\pi^2\rho}\left[-\frac{16r^2 z_m{}^3}{(z_{\mathrm{m}}^2 + r^2)^{5/2}(4r^2 + z_m^2)^3} - \frac{3r^2 z_m{}^3}{(z_{\mathrm{m}}^2 + r^2)^{5/2}[4r^2 + z_m^2]^{\frac{5}{2}}} + \frac{z_{\mathrm{m}}\mathrm{r}^2}{(z_{\mathrm{m}}^2 + r^2)^{5/2}[4r^2 + z_m^2]^{\frac{3}{2}}} + \frac{4z_{\mathrm{m}}\mathrm{r}^3}{(z_{\mathrm{m}}^2 + r^2)^{5/2}[4r^2 + z_m^2]^2}\right]\dot{z}_{\mathrm{m}}$$

We let the geometrical factor $\gamma \sim 1$. With a bit of work these two approaches can be shown to be the same.

$$\Delta\phi_{Faraday} = -\frac{Q_o}{k_s}\frac{3m_{eff}{}^2 r^2}{16\pi^2\rho}\frac{3z_m\left(10z_m^8 + \sqrt{z_m^2+4r^2}\left(28rz_m^6+150r^3z_m^4+156r^5z_m^2-176r^7\right)+101r^2z_m^6+316r^4z_m^4+204r^6z_m^2-336r^8\right)}{(z_m^2+r^2)^{\frac{9}{2}}(z_m^2+4r^2)^{\frac{7}{2}}}\dot{z}_{\mathrm{m}}$$

To deal with $\dot{z}_{\mathrm{m}}$, $z = A\cos\omega t + z_m$ where $z_m$ is held at a constant for each LH, A cos(ωt) represents the tip dither around z and $\dot{z}_m = -A\omega\sin\omega t$.

$$\Delta\phi_{Faraday} = -\frac{Q_o}{k}\frac{3m_{eff}^2}{16\pi^2\rho}\left[-\frac{16r^2 z_m{}^3}{(z_{\mathrm{m}}^2 + r^2)^{5/2}(4r^2 + z_m^2)^3} - \frac{3r^2 z_m{}^3}{(z_{\mathrm{m}}^2 + r^2)^{5/2}[4r^2 + z_m^2]^{\frac{5}{2}}} + \frac{z_{\mathrm{m}}\mathrm{r}^2}{(z_{\mathrm{m}}^2 + r^2)^{5/2}[4r^2 + z_m^2]^{\frac{3}{2}}} + \frac{4z_{\mathrm{m}}\mathrm{r}^3}{(z_{\mathrm{m}}^2 + r^2)^{5/2}[4r^2 + z_m^2]^2}\right]\mathrm{A}\omega$$

$$\sim -\frac{Q_o}{k}\frac{3A\omega m_{eff}^2}{16\pi^2\rho}\left[-\frac{16r^2 z_m{}^3}{r^{11}\left(\frac{z_{\mathrm{m}}^2}{r^2}+1\right)^{5/2}\left(4+\frac{z_m^2}{r^2}\right)^3} - \frac{3r^2 z_m{}^3}{r^{10}\left(\frac{z_{\mathrm{m}}^2}{r^2}+1\right)^{5/2}\left[4+\frac{z_m^2}{r^2}\right]^{\frac{5}{2}}} + \frac{z_{\mathrm{m}}r^2}{r^8\left(\frac{z_{\mathrm{m}}^2}{r^2}+1\right)^{5/2}\left[4+\frac{z_m^2}{r^2}\right]^{\frac{3}{2}}} + \frac{4z_{\mathrm{m}}\mathrm{r}^3}{r^9\left(\frac{z_{\mathrm{m}}^2}{r^2}+1\right)^{5/2}\left[4+\frac{z_m^2}{r^2}\right]^2}\right]$$

So far, we have included only $Q_o$ here, which is not a function of the LH, $z_m$. This expression gives the phase shift due to an undamped dipole cantilever over the sample including a "Faraday current" and a "persistent current" plus currents induced into the substrates:

$$\Delta\phi_{mag} = \Delta\phi_{persistent} + \Delta\phi_{Faraday} + \Delta\phi_{HOPG/BiTe}$$

From this we can express ($\Delta\phi_{mag}$) either as a contrast HOPG substrate to $Bi_2Te_3$ sample, or as absolute shift from free oscillator well above the sample. However, these expressions are lengthy and hard to work with.

***Some Rough Estimates and Approximations***

A few numbers for a magnetic tip: NT-MDT MFM01: m ~ 100 ng ~ $10^{-10}$ kg, $\omega_o \sim 10^4$ $s^{-1}$, r ~ $10^{-6}$ m, $m_{eff}$ ~ ($10^{-17}$ – $10^{-18}$) Wb/m , ρ ~ ?, $Q_o \sim 10^2$, $L_{cantilever} \sim 10^{-4}$ , $\dot{z}_m \sim 10^{-1}$ m/s (max). k ~ 1 N/m:

$$\Delta\phi_{persistent} = \frac{Q_o}{k_s}\frac{m_{eff}\mathrm{I}r}{2\pi}\left[-\frac{16rz_m{}^2}{(4r^2+z_m^2)^3}-\frac{3z_m{}^2}{[4r^2+\ z_m^2]^{\frac{5}{2}}}+\frac{1}{[4r^2+\ z_m^2]^{\frac{3}{2}}}+\frac{4r}{[4r^2+\ z_m^2]^2}\right]$$

$$\Delta\phi_{persistent} = \frac{Q_o}{k_s}\frac{m_{eff}\mathrm{I}r}{2\pi}\left[-\frac{16z_m{}^2}{r^5\left(4+\frac{z^2}{r^2}\right)^3}-\frac{3z_m{}^2}{r^5\left[4+\frac{z^2}{r^2}\right]^{\frac{5}{2}}}+\frac{1}{r^3\left[4+\frac{z^2}{r^2}\right]^{\frac{3}{2}}}+\frac{4r}{r^2\left[4+\frac{z^2}{r^2}\right]^2}\right]$$

$$\Delta\phi_{persistent} = \frac{Q_o}{k_s}\frac{m_{eff}\mathrm{I}}{2\pi}\left[\frac{1}{4}+\frac{1}{8r^2}-\frac{18z^2}{32r^4}\right]$$

Orders of magnitude we get $\Delta\phi_{persistent} \sim 10^5$ I – 10 I $z^2$ (in nm) for z<<r.

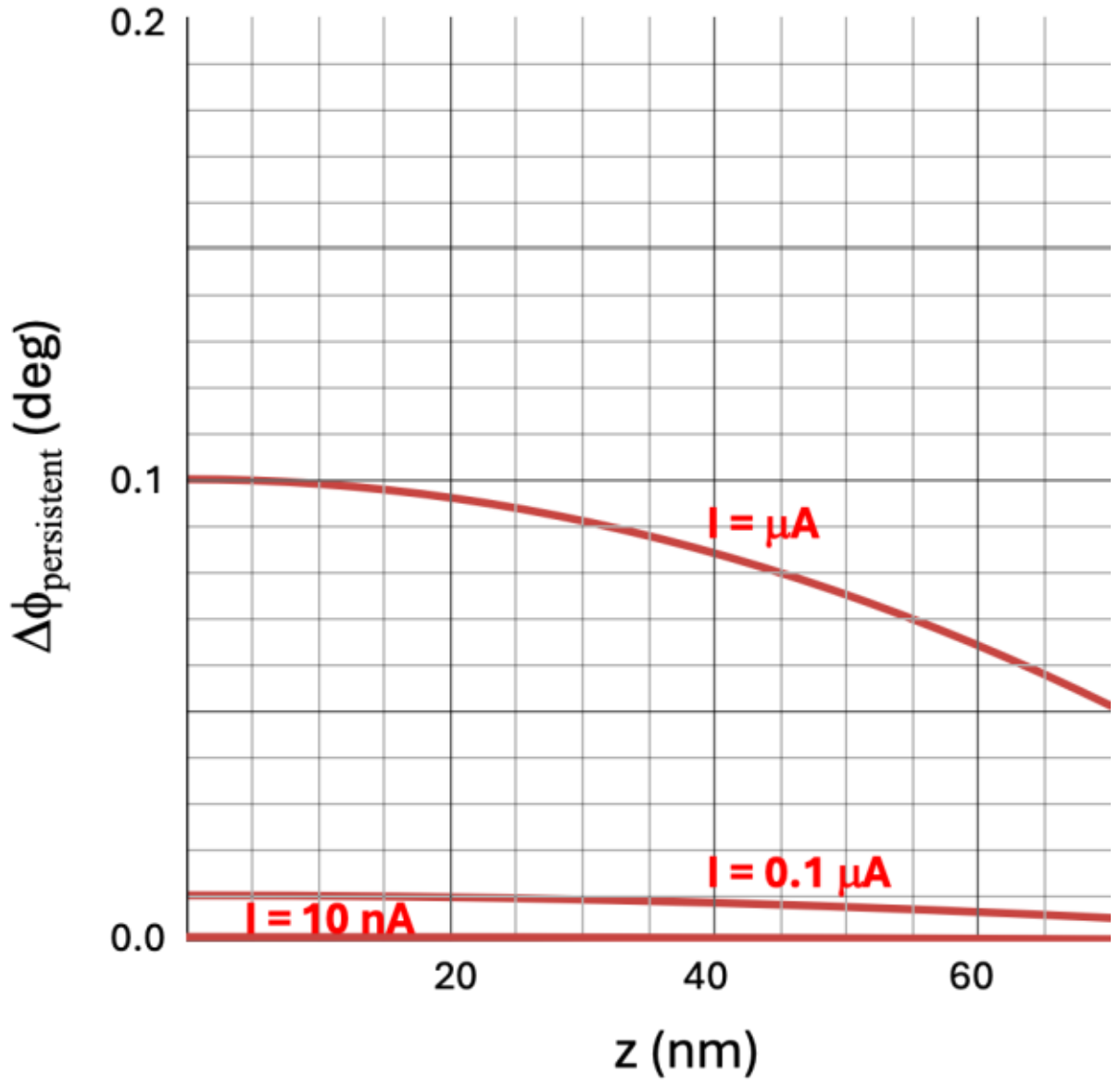

**Figure S17:** a plot of our estimate for the phase shift due to a current in a loop, at the position of the loop edge and in the limit of $z_m/r$ <<1. The shift is small even for currents as high as 1 µA. Notice that we have chosen a current direction giving a positive phase shift.

As $z_m$ approaches 100 nm this shift goes away and our approximation begins to fail. As we move further from the surface, a better approximation for the $z_m$ dependence is given by:

$$\Delta\phi_{persistent} = \frac{Q_o}{k_s}\frac{m_{eff}\mathrm{I}}{2\pi}\left[\frac{4r^2}{[4r^2+\ z^2]^2}+\frac{r}{[4r^2+\ z^2]^{\frac{3}{2}}}-\frac{16r^2z^2}{(4r^2+z^2)^3}-\frac{3rz^2}{[4r^2+\ z^2]^{\frac{5}{2}}}\right]$$

Doing the same exercise for the Faraday component of the phase shift:

$$\Delta\phi_{Faraday} \sim -\frac{Q_o}{k}\frac{3A\omega m_{eff}^2}{16\pi^2\rho}\left[-\frac{16\frac{z_m}{r^3}^3}{r^6\left(\frac{z_m^2}{r^2}+1\right)^{5/2}\left(4+\frac{z_m^2}{r^2}\right)^3}-\frac{3\frac{z_m}{r^3}^3}{r^5\left(\frac{z_m^2}{r^2}+1\right)^{5/2}\left[4+\frac{z_m^2}{r^2}\right]^{\frac{5}{2}}}+\frac{\frac{z_m}{r}}{r^5\left(\frac{z_m^2}{r^2}+1\right)^{5/2}\left[4+\frac{z_m^2}{r^2}\right]^{\frac{3}{2}}}+\frac{4\frac{z_m}{r}}{r^5\left(\frac{z_m^2}{r^2}+1\right)^{5/2}\left[4+\frac{z_m^2}{r^2}\right]^2}\right]$$

$$\sim -\frac{Q_o}{k}\frac{3A\omega m_{eff}^2}{16\pi^2\rho}\left[\left(\frac{\frac{z_m}{r}}{r^5\left(\frac{z_m^2}{r^2}+1\right)^{5/2}\left[4+\frac{z_m^2}{r^2}\right]^{\frac{3}{2}}}\right)+\left(\frac{4\frac{z_m}{r}}{r^5\left(\frac{z_m^2}{r^2}+1\right)^{5/2}\left[4+\frac{z_m^2}{r^2}\right]^2}\right)-\left(\frac{16\frac{z_m}{r^3}^3}{r^6\left(\frac{z_m^2}{r^2}+1\right)^{5/2}\left(4+\frac{z_m^2}{r^2}\right)^3}\right)-\left(\frac{3\frac{z_m}{r^3}^3}{r^5\left(\frac{z_m^2}{r^2}+1\right)^{5/2}\left[4+\frac{z_m^2}{r^2}\right]^{\frac{5}{2}}}\right)\right]$$

$$\sim -\frac{Q_o}{k}\frac{3A\omega m_{eff}^2\, r^{-5}}{16\pi^2\rho}\left[\left(\frac{x}{(x^2+1)^{5/2}[4+x^2]^{\frac{3}{2}}}\right)+\left(\frac{4x}{(x^2+1)^{5/2}[4+x^2]^2}\right)-\left(\frac{16x^3r^{-1}}{(x^2+1)^{5/2}(4+x^2)^3}\right)-\left(\frac{3x^3}{(x^2+1)^{5/2}[4+x^2]^{\frac{5}{2}}}\right)\right]$$

$z_m \sim 10^{-9}$m – $10^{-7}$m , r ~ $10^{-6}$ m and x ~ $z_m/r \sim 10^{-1}$ at its largest, so $(z_m/r)^2 \sim 10^{-2}$.

$$\sim -\ 10^{-4}\, s\left(\frac{1}{(x^2+1)^{\frac{5}{2}}}\right)\left(\left(\frac{x}{(4+x^2)^{\frac{3}{2}}}\right)+\left(\frac{4x}{(4+x^2)^2}\right)-\left(\frac{16x^3r^{-1}}{(4+x^2)^3}\right)-\left(\frac{3x^3}{(4+x^2)^{\frac{5}{2}}}\right)\right)$$

s ~ 1/ρ. $m_{eff} \sim 10^{-15}$ A $m^2$

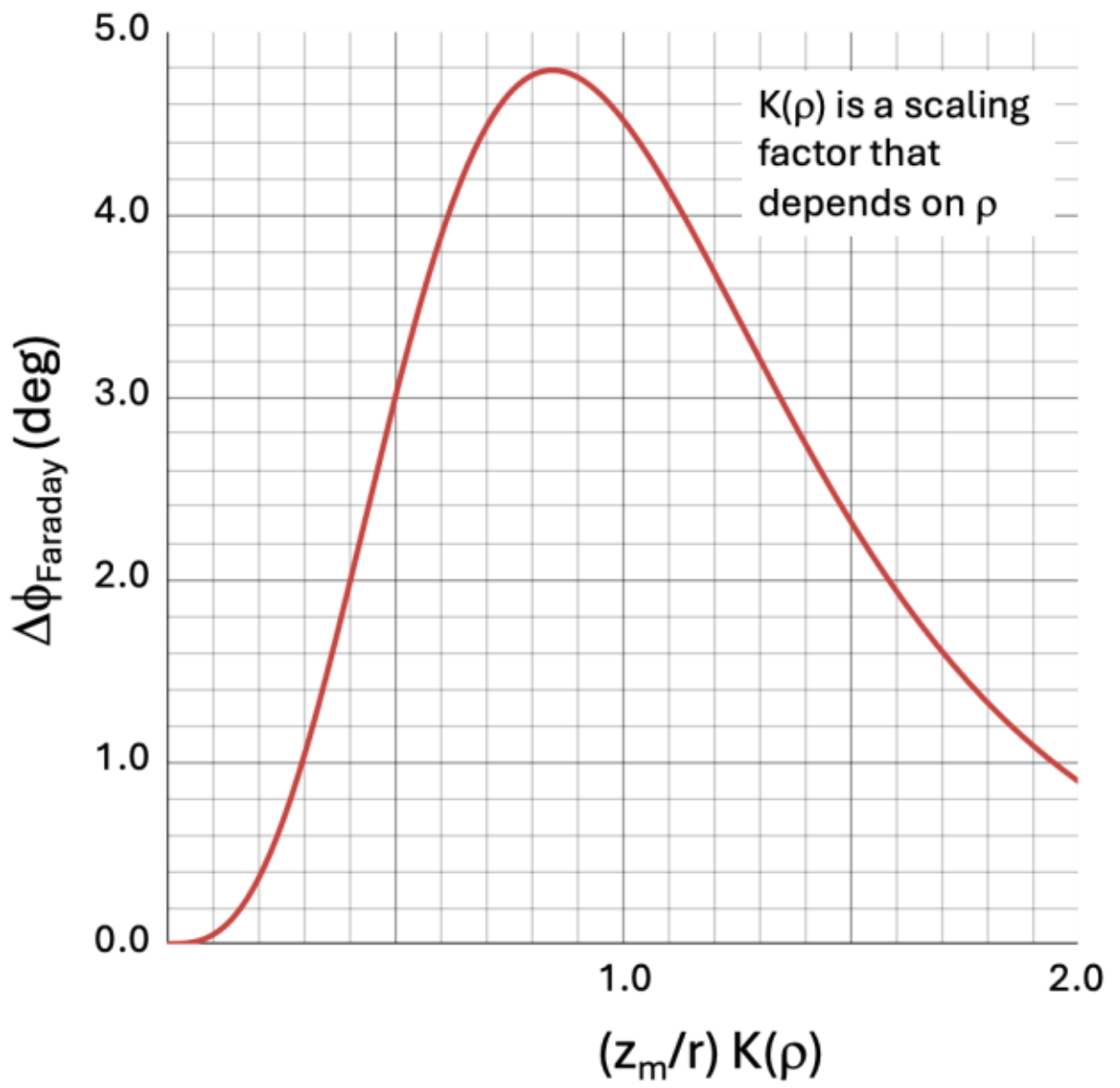

**Figure S18:** the phase shift from the Faraday component is scaled by a geometrical factor that depends on the resistivity of the system ρ. As with all the simulations and estimates presented, we have taken ρ ~ 1 for display.

Our estimate for approximate phase shift in the Faraday component is:

$$\Delta\phi_{Faraday} \sim -\frac{Q_o}{k}\frac{3A\omega m_{eff}^2\, r^{-5}}{16\pi^2\rho(\mathrm{x}^2+1)^{5/2}}\left[\left(\frac{\mathrm{x}}{[4+\ \mathrm{x}^2]^{\frac{3}{2}}}\right)+\left(\frac{4\mathrm{x}}{[4+\ \mathrm{x}^2]^2}\right)-\left(\frac{16x^3r^{-1}}{(4+\mathrm{x}^2)^3}\right)-\left(\frac{3x^3}{[4+\ \mathrm{x}^2]^{\frac{5}{2}}}\right)\right]$$

Finally:

$$\Delta\phi_{\mathrm{mag\ HOPG/BiTe}} \sim -\frac{3}{128}\frac{z_{\mathrm{c}}^4 {m_{eff}}^2\omega_o}{\rho_{HOPG/BiTe}k_s z_{\mathrm{m}}^4}$$

Our estimates are consistent with measurement, phase shifts are in the low single digits, and this occurs near the sample.

***$Q_o \rightarrow Q_l$***

To get the modified Q's such as $Q_{edge}$, we use the approach as in reference [22]. Starting with the equation of motion:

$$m\ddot{z}_{\mathrm{m}} = -k[(z_{\mathrm{m}}-z_{\mathrm{m0}})-(u-u_0)]-\gamma_s\dot{z}_{\mathrm{m}}-\frac{3{m_{eff}}^2\gamma r^2}{16\pi^2\rho}\left[\frac{(4r^2+z_m^2)^{\frac{1}{2}}(z_m^2r^2+z_m^4)+12rz_m^4+(z_m^2+4r^2)^{3/2}(4z_m^2-r^2)+34r^3z_m^2-8r^5}{(4r^2+z_m^2)^2(r^2+z_m^2)^{7/2}}\right]\dot{z}_{\mathrm{m}}$$

If the crystal is nicely faceted and large relative to the lift height, we can estimate the geometric factor γ to be ~1. $z_{mo}$ is equilibrium, $\gamma_s$ are viscous losses in the cantilever in air, $u = u_0 + a\cos\omega t$ is the driving force, and the solutions look like: $z_{\mathrm{m}} = z_{\mathrm{m0}} + A\cos(\omega t+\phi)$. We are only interested in the modification of Q from the "in air" Q far from the surface. Thus, from reference [22] we write $Q_{l}$ as:

$$\frac{m\omega_o}{Q_l} \sim \gamma_s + \frac{3{m_{eff}}^2r^2}{16\pi^2\rho}\left(\frac{(4r^2+z_m^2)^{\frac{1}{2}}(z_m^2r^2+z_m^4)+12rz_m^4+(z_m^2+4r^2)^{\frac{3}{2}}(4z_m^2-r^2)+34r^3z_m^2-8r^5}{(4r^2+z_m^2)^2(r^2+z_m^2)^{\frac{7}{2}}}\right)$$

$$\frac{1}{Q_l} \sim \gamma_s + \left(\frac{3{m_e}^2r^2((4r^2+z^2)^{\frac{1}{2}}(z^2r^2+z^4)+12rz^4+(z^2+4r^2)^{\frac{3}{2}}(4z^2-r^2)+34r^3z^2-8r^5)}{16\pi^2\rho m\omega_o(4r^2+z^2)^2(r^2+z^2)^{\frac{7}{2}}}\right)$$

$$\frac{3{m_e}^2r^2((4r^2+z^2)^{\frac{1}{2}}(z^2r^2+z^4)+12rz^4+(z^2+4r^2)^{\frac{3}{2}}(4z^2-r^2)+r^3(34z^2-8r^2)}{16\pi^2\rho m\omega_o(4r^2+z^2)^2(r^2+z^2)^{\frac{7}{2}}}$$

$$\sim \frac{3{m_e}^2r^2(r\left(4+\frac{z^2}{r^2}\right)^{\frac{1}{2}}\left(1+\frac{z^2}{r^2}\right)r^2z^2+12rz^4+r^5\left(\frac{z^2}{r^2}+4\right)^{\frac{3}{2}}\left(4\frac{z^2}{r^2}-\right)+r^5(34\frac{z^2}{r^2}-8)}{16\pi^2\rho m\omega_o r^{11}\left(4+\frac{z^2}{r^2}\right)^2\left(1+\frac{z^2}{r^2}\right)^{\frac{7}{2}}}$$

$$\frac{1}{Q_l} \sim \gamma_s + \frac{3{m_e}^2((4+x)^{\frac{1}{2}}(1+x)x+12x^2+(x+4)^{\frac{3}{2}}(4x-1)+(34x-8))}{16\pi^2\rho m\omega_o r^4(4+x)^2(1+x)^{\frac{7}{2}}}$$

For an order of magnitude estimate we fill in literature numbers for the constants:

$$\frac{1}{Q_l} \sim \frac{\gamma_s\, \rho(4+x)^2(1+x)^{\frac{7}{2}}}{\rho(4+x)^2(1+x)^{\frac{7}{2}}} + \frac{(4+x)^{\frac{1}{2}}(1+x)x + 12x^2 + (x+4)^{\frac{3}{2}}(4x-1) + (34x-8)}{\rho(4+x)^2(1+x)^{\frac{7}{2}}}$$

$$\frac{1}{Q_l} \sim \frac{\gamma_s\, \rho(4+x)^2(1+x)^{\frac{7}{2}} + (4+x)^{\frac{1}{2}}(1+x)x + 12x^2 + (x+4)^{\frac{3}{2}}(4x-1) + (34x-8)}{\rho(4+x)^2(1+x)^{\frac{7}{2}}}$$

$$Q_l \sim \frac{\rho(4+x)^2(1+x)^{\frac{7}{2}}}{\gamma_s\, \rho(4+x)^2(1+x)^{\frac{7}{2}} + (4+x)^{\frac{1}{2}}(1+x)x + 12x^2 + (x+4)^{\frac{3}{2}}(4x-1) + (34x-8)}$$

In the approximation of r >> z. This looks like:

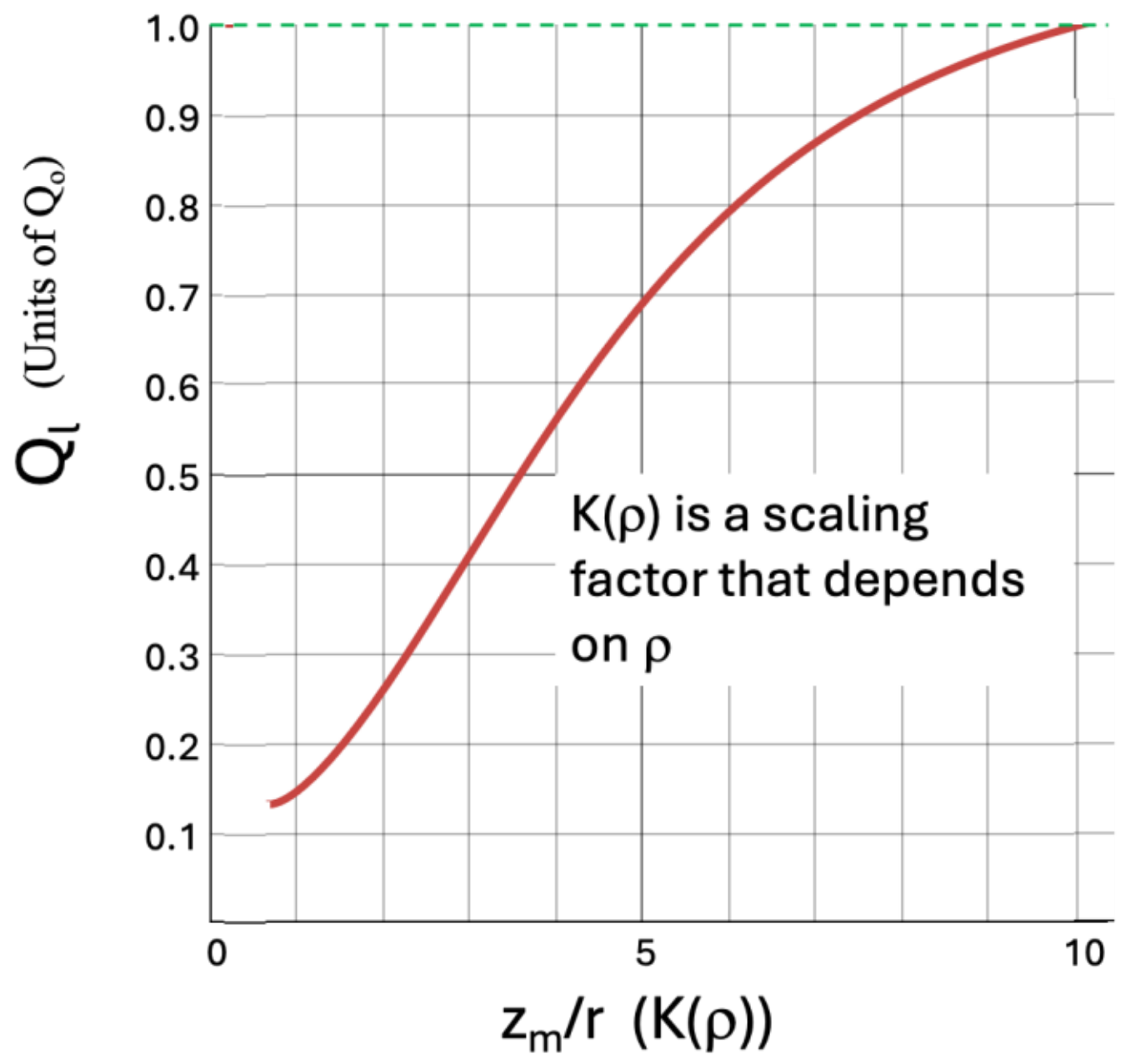

**Figure S19:** This is an order of magnitude estimate of the behavior of Q with damping and Faraday induction along the edge of the plate. Depending on ρ, $m_{eff}$, $\omega_o$, the overall shift in Q can be significant and generally it will dominate contrast mechanisms as we have seen from these estimates. The graph shown here uses reasonable numbers found in literature and is normalized to a ρ of 1. Note that z (or $z_m$) ≠ $z_t$ (the bottom of the tip) so this graph is related to but not equal to the lift height of imaging.

Very close to the sample we may use the simplification:

$$Q_l \sim \frac{4\pi^2 \rho m \omega_o r^6}{{m_e}^2(4z^2 - r^2)}$$

As noted, these expressions are only approximations of phase and cantilever dynamics for $z_m/r << 1$. For higher LHs, capacitive forces are not well imagined by our simplified imaging geometry in **Figure S9**. However, within the window of applicable approximations, i.e. for the LH's of these experiments, they provide guidance to understanding the contrast mechanisms and allow for a more direct interpretation of the magnetic components of ambient imaging.

### 3.5. Cantilever oscillation amplitude

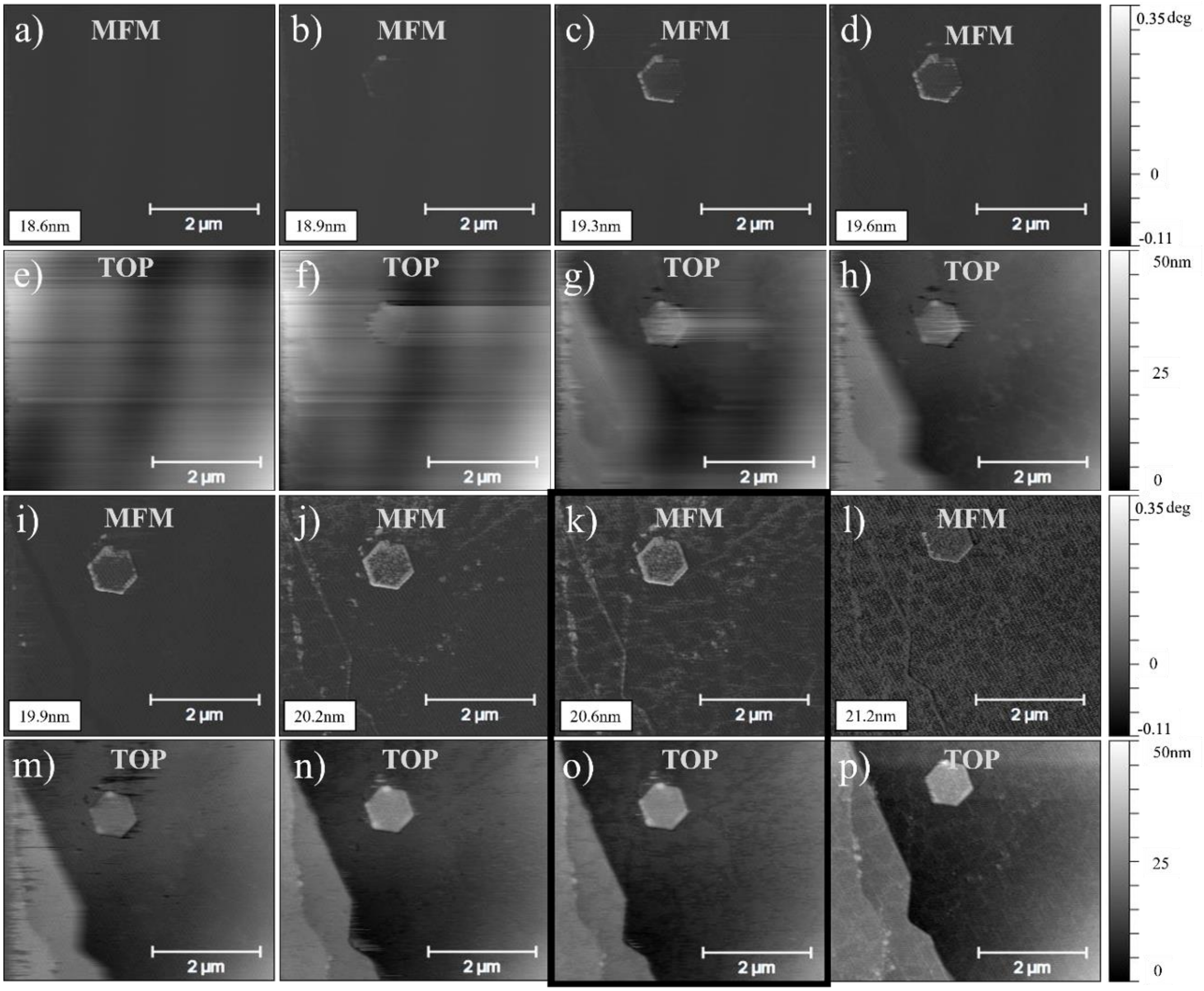

**Figure S20:** MFM on Bi2Te3 with varying cantilever amplitude. Magnetic (a-d, i-l) and topographical (e-h, m-p) images taken of a $Bi_2Te_3$ plate with a magnetic tip in MFM mode with a constant lift height of 100 nm. The cantilever voltage was varied from 570 mV to 660 mV which resulted in cantilever amplitudes from 17.9 nm to 29.3 nm. The system will auto set the cantilever's driving voltage when it determines the resonance frequency of the tip. In this case the system set the voltage to 650 mV and the corresponding

images are highlighted in the box. The quality factor of the MFM tip was measured to be 181.809. Cleaved graphite was used as the support substrate.

### 3.6 MFM "tomography" data

In **Figure S21**, a 6 nm thick, 1 μm diameter plate is imaged in an LH region for which a strong $U^2$ dependence is expected. In the left panel sequence (**Figure S21 (a)**), 5LH, so the electrostatic interaction ($U_c$) influences the Δϕ of the whole plate, dominating other contrast mechanisms. Moreover, the contrast between the edge and the center is not observed even at a -2 V bias, suggesting that a van der Waals interaction is dominant at this LH. In the right panel (**Figure S21 (b)**), at 10LH, we notice that Δϕ on the edge, as compared to the Δϕ in the plate bulk, is now much more pronounced. At 0 V and 1 V, the effect of the capacitive forces 'washes out' the edge contrast. However, the contrast difference between the center of the plate and the edge increases with increasing $U_a$, suggesting that the contact potential difference is minimal and the dominant force is magnetic. Imaging at a greater LH for another set of plates is shown in **Figure S22**. At these heights, we anticipate the dominant contrast mechanisms to involve magnetic interactions. Curiously, the thickness of the plates has little effect on the edge contrast. Moreover, electrostatic interactions are negligible at this LH, suggesting that our identification of the edge contrast as the signature of a *persistent current* in a topologically protected state is reasonable.

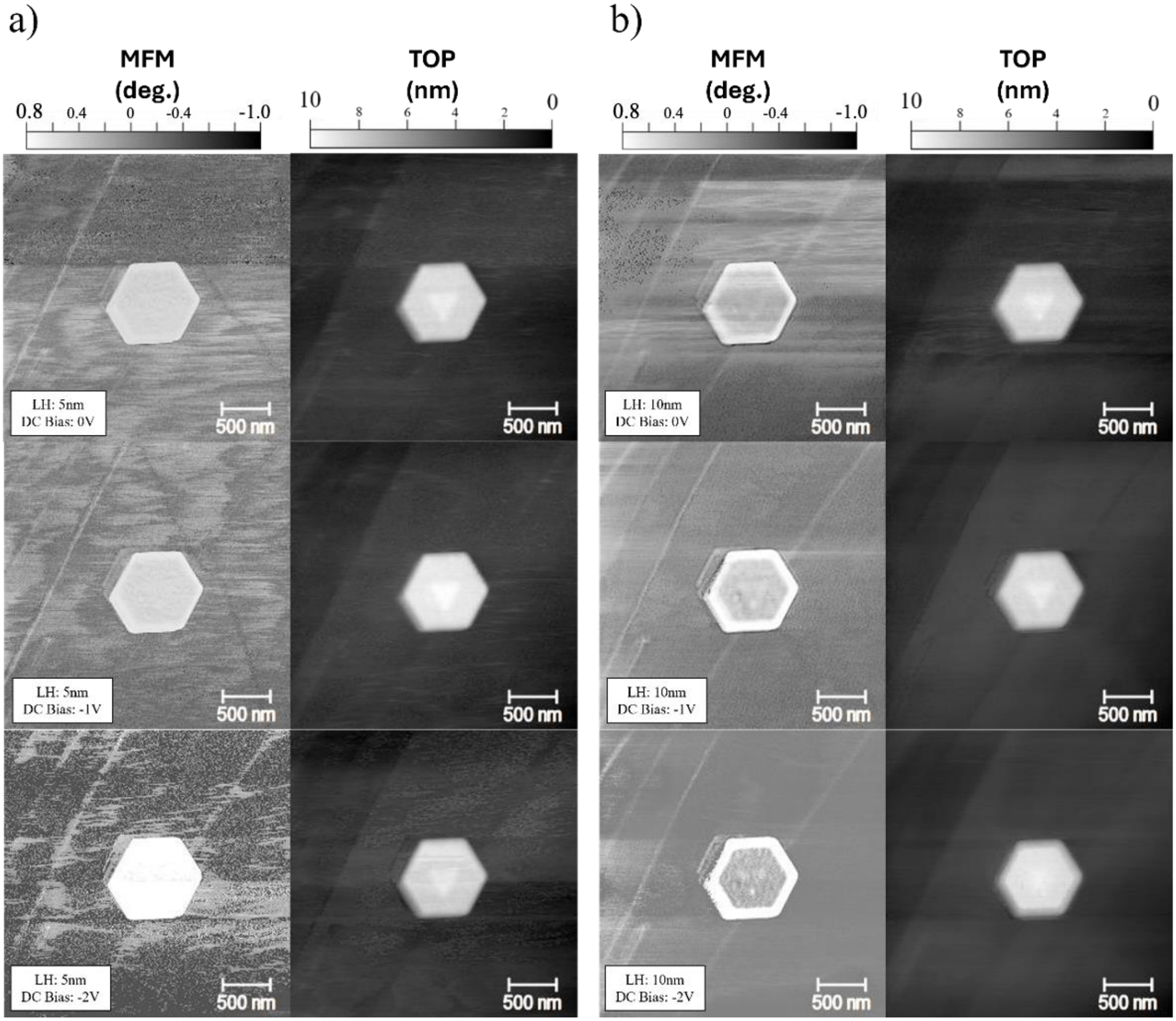

**Figure S21:** MFM images of the same $Bi_2Te_3$ plates on HOPG as in **Figure 2 (b)** but with varying $U_a$ and LH. (a) MFM and topographic (TOP) images of a $Bi_2Te_3$ plate with increasing $U_a$ (0 to -2 V) with 5 nm cantilever LH. (b) Same measurement as in (a) but with 10 nm cantilever LH on the same plate. Clearly, the edge state can be observed now. In particular, the background noise has also reduced significantly. Both observations indicate that the magnetic force gradient is dominating.

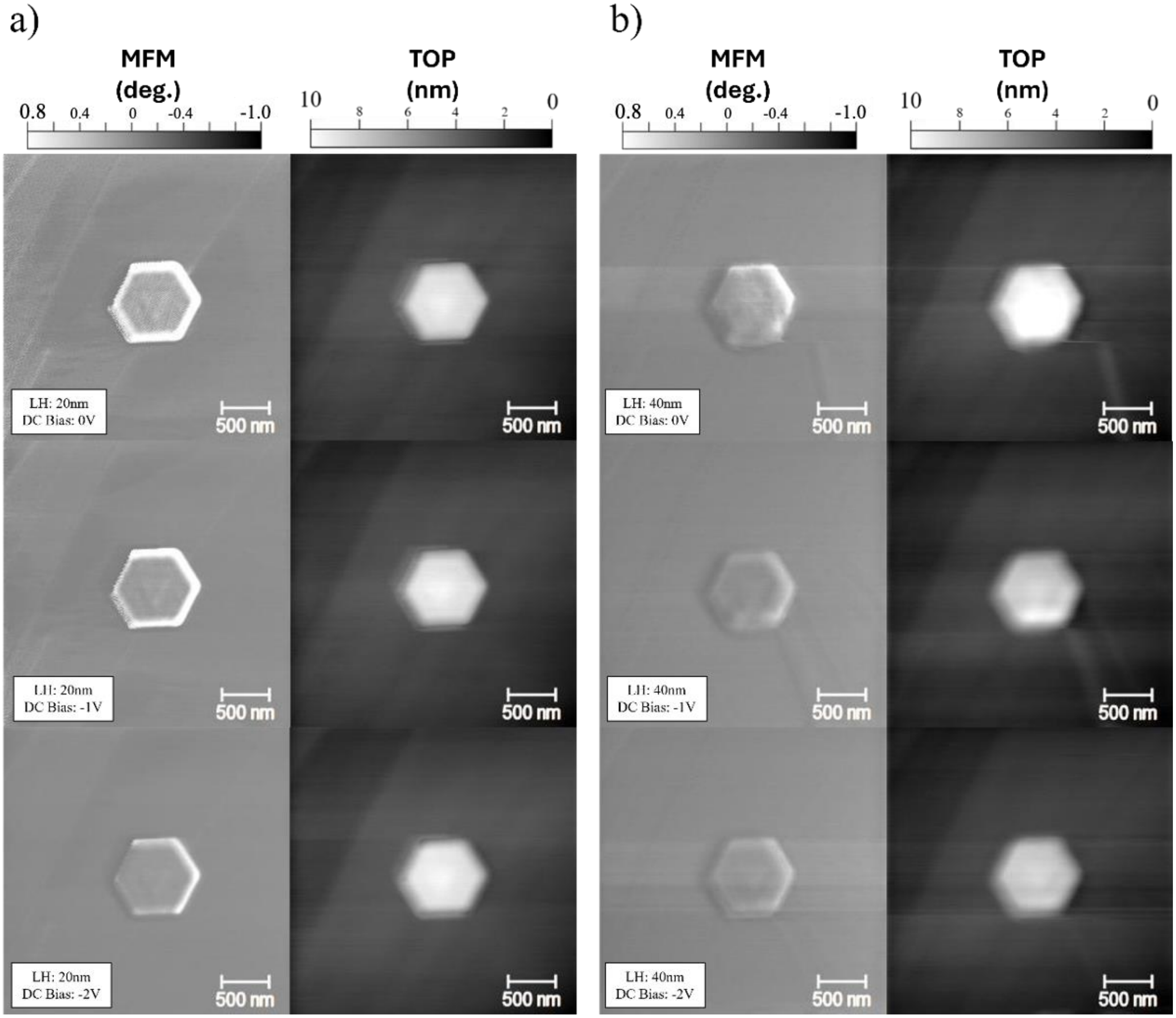

**Figure S22:** MFM images of the same $Bi_2Te_3$ plates on HOPG as in **Figure S12 (b)** but with varying $U_a$ and LH. (a) MFM and topographic (TOP) images of a $Bi_2Te_3$ plate with increasing $U_a$ (0 to -2 V) with lift height 20 nm. (b) Same measurement as in (a) but with lift height 40 nm.

## 4.0. Comparison Imaging with MFM tips

An important way to check for instrumental anomalies is to image other materials with similar sizes and shapes but that are topologically trivial. We pursued this with several materials including metals, small band gap dielectrics, and then other topological systems. We also compared the effects of support substrate choice.

### 4.1 Metals

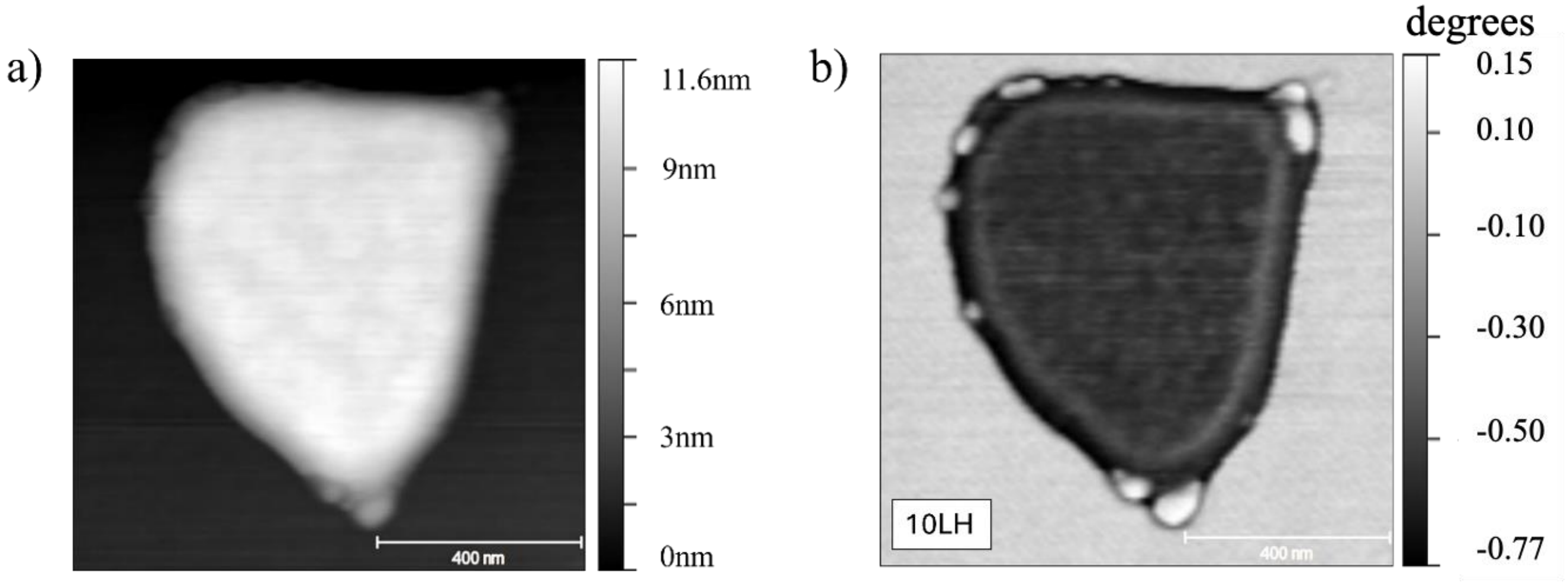

**Figure S23**: (a) Topographical image of gold plate imaged under similar settings as the $Bi_2Te_3$ system. (b) Magnetic image of gold plate at a lift height of 10 nm. Imaged with JSPM 5200

While we have discussed references of Faraday induction in thin metal films, less has been done to understand Faraday induction in non-metallic nano-scale metals. [30] Such Faraday induction in metallic nanoparticles is known generally, and we show an example of this in **Figure S23**. In this MFM image, the contrast is the reverse of the edge states we see in the TI systems. From Lenz's law the Faraday-induced currents in the Au nanoparticle will oppose the introduction of a field. Thus, we anticipate a small repulsive force in the MFM due to the opposing magnetic field that is induced as well as a strong dissipation effect, when imaging a metallic particle as seen in literature. [31] These combined to give the negative phase shift observed as is consistent with reference [29]. But this is NOT what we see in the edge states of the $Bi_2Te_3$ plate. The edges of TIs seem to yield a strongly positive shift in the phase indicating a repulsive force. The center of the TI crystals give a positive phase shift from the background which is also consistent with the capacitive forces discussed above.

### 4.2 Nano-Dielectrics

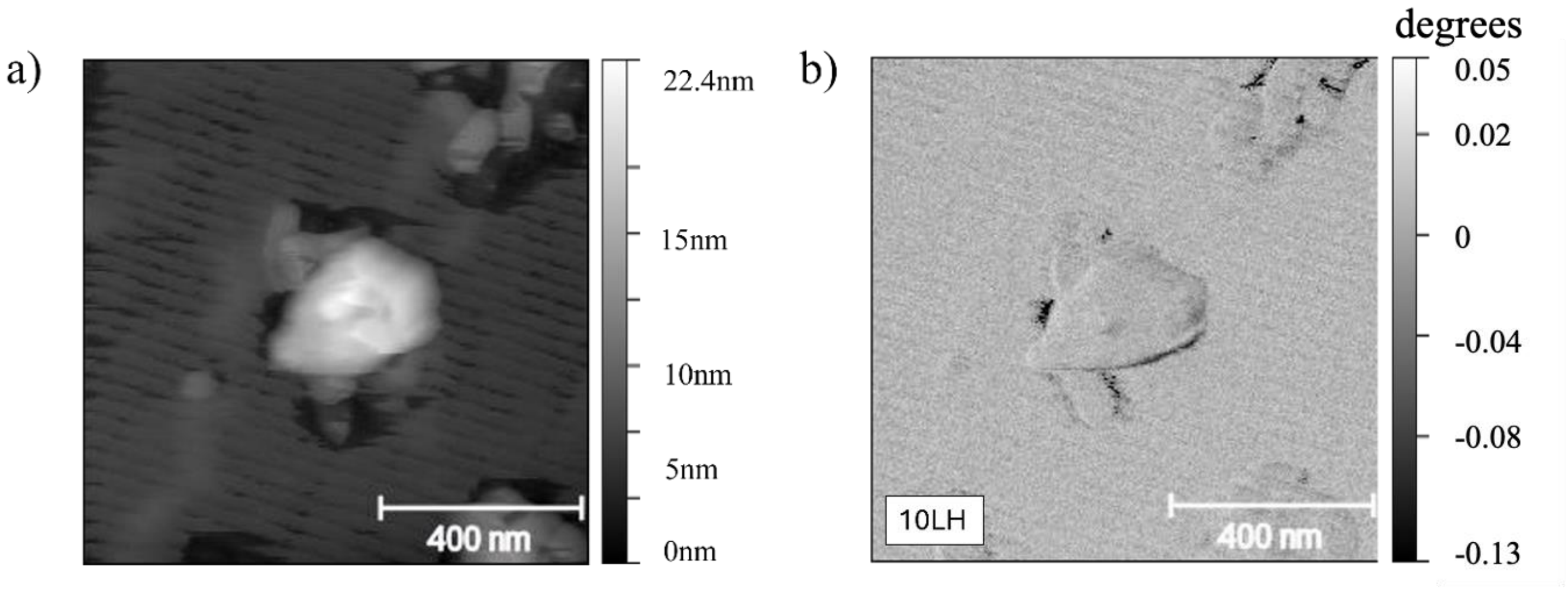

**Figure S24:** (a) Topographical image of $MoS_2$ nanoflake. (b) Magnetic image of $MoS_2$ nanoflake at lift height of 10 nm. Imaged on the JEOL JSPM 5200

Comparisons of MFM contrasts in the imaging of nanoscale dielectric materials are also inconsistent with observations in TIs. For example, the chalcogenide $MoS_2$ has similar chemical reactivity, but is topologically trivial. Shown above in **Figure S24** is a ~0.4 μm crystal flake ~ 20 nm in thickness; on the right is a typical AFM image and on the left the phase image taken at a LH of 10 nm. We note that we do see some small phase change across the flake as expected since it has a different dielectric constant from the support HOPG. However no significant edge enhancement is observed in the phase imaging. This $MoS_2$ flake has dimensions similar to those of the TIs under study here. The role of the dielectric in creating an enhanced capacitive force is evident. But the lack of significant edge contrast enhancement is important because this $MoS_2$ flake is exposed to air and certainly has oxygen contamination around the edges. While this would naturally lead to artefacts in STM/STS [32] it doesn't show up in MFM.

## 4.3 Other topological plates and Support substrates

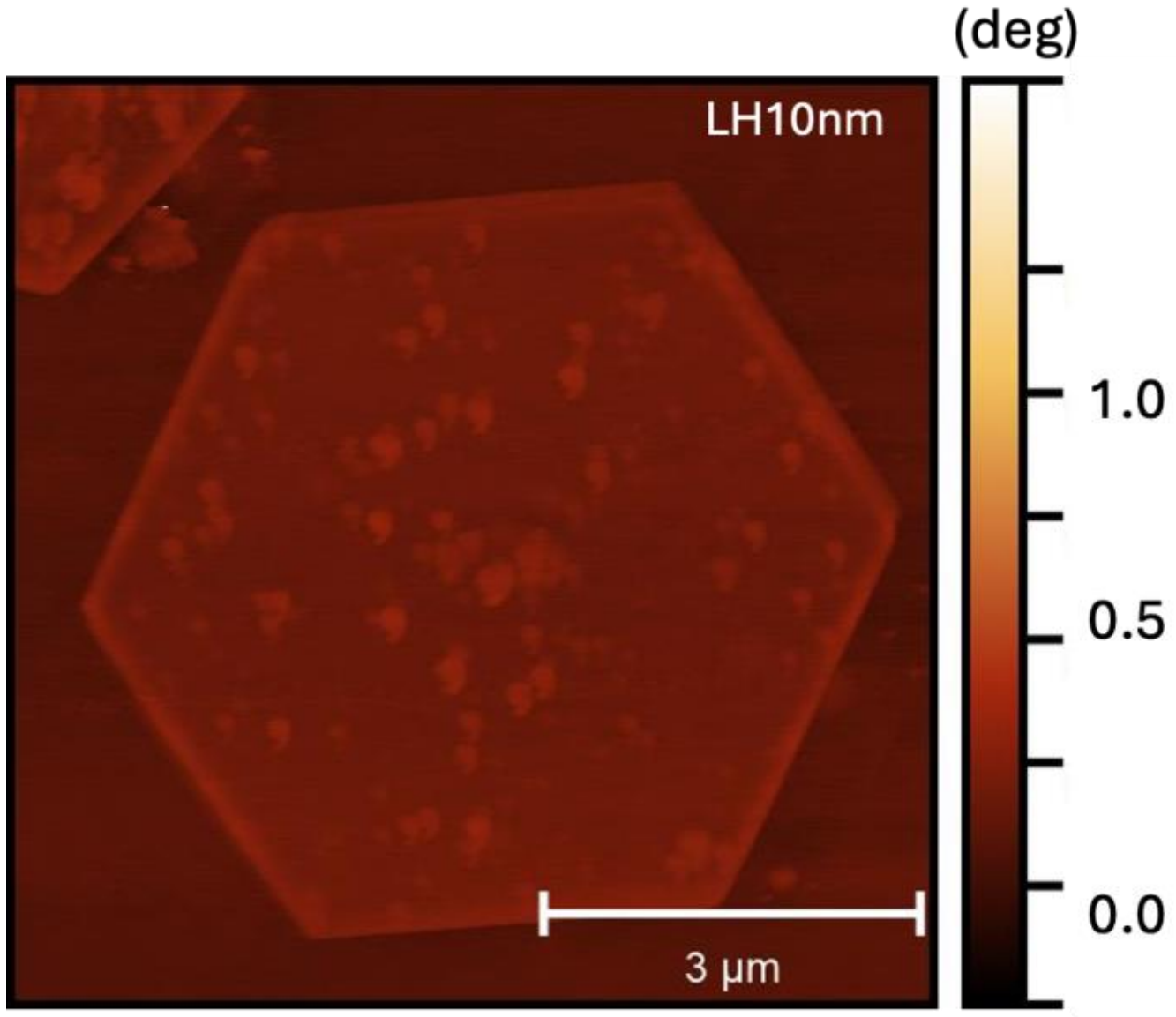

**Figure S25:** Shown is an $Sb_2Te_3$ hexagonal crystal approximately 6 nm thick and ~ 3 μm in diameter. We notice here that this TI also shows edge contrast. Importantly this was imaged on thermally grown $SiO_2$/Si. These images also show the lift height and oscillation amplitude dependence's we observe for $Bi_2Te_3$ on HOPG. The contact potential to the $SiO_2$ is different from that of the HOPG and so the electronic/capacitive contrast to the middle of the plate has changed slightly. This sample was imaged on a Bruker Dimension 3100.

We also compared other TI plates synthesized using the methods outlined in references. [33] In this case $Sb_2Te_3$ was deposited on thermally grown $SiO_2$/Si and imaged using the same techniques as described for the $Bi_2Te_3$ samples. Contrast between the plane of the plate and the background support substrate was slightly different from that observed for HOPG due to differences in the contact potentials. However, the edge contrast is still strong and exhibits the same overall height dependence as seen in $Bi_2Te_3$.

## 5.0 Important experimental note about our data:

AFM topography as well as MFM phase imaging was performed by a JEOL 5200 SPM. These instruments are well known to have an unresolved software issue with its units of degrees. Typically, the microscope gives phase shifts in both ambient and vacuum imaging conditions that exceed 100 degrees! Most researchers simply publish these results as is. It is the comparison of phase shifts from point to point on a substrate that is usually the important variable. [34] But in our case, we would like to know the real values. The issue with the JSPM 5200 arises from two sources, the zero- setpoint (phase slider) that is provided on the internal lockin amplifier, and the conversion between units in the software. (1 rad. = 57.3 deg.) In the JSPM 5200 this software conversion seems to be set backwards. If the zero is set for the initial oscillation well above the substrate, then dividing the output by 57.3 gives the correct units in degrees. This phase shift agrees with the Bruker/DI systems using the same samples and substrates and similar cantilever characteristics. It also agrees with our estimated shifts and those in literature. [35] In this supplement the data sets are all calibrated like this.

[10] It is important to recognize that there is still some debate about plasmons within states of topological origin. Recall that such topological states require spin-momentum locking, so unless there is some mechanism which allows for TRS breaking such as strong magnetic fields, or other interactions with counterpropagating currents, then conditions for plasmon formation cannot be met in these 1D channels. In the transmission electron microscope the samples do encounter both high magnetic fields as well as large current densities as the beam interacts with and populates the conduction band of the materials. See for example:

[13] Zhao, M., Bosman, M., Danesh, M., Zeng, M., Song, P., Yudi Darma, Andrivo Rusydi, Lin, H., Qiu, C.-W., & Loh, K. P. Visible Surface Plasmon Modes in Single Bi2Te3 Plate. Nano Lett. 15, 8331–8335 (2015), DOI: https://doi.org/10.1021/acs.nanolett.5b03966, Schlomer, H., Jiang, Z. & Haas, S. Plasmons in two-dimensional topological insulators. Phys. Rev. B 103, (2021), DOI: https://doi.org/10.1103/PhysRevB.103.115116

[14] Zhao, M., Zhang, J., Gao, N., Song, P., Bosman, M., Peng, B., Sun, B., Qiu, C. W., Xu, Q. H., Bao, Q., & Loh, K. P. Actively Tunable Visible Surface Plasmons in Bi2Te3 and their Energy-Harvesting Applications. Ad. Mater. 28, 3138–3144 (2016), DOI: https://doi.org/10.1002/adma.201506367

[15] Williams, D. B. & Carter, C. B. Transmission Electron Microscopy: Basics I. (Springer Science+Business Media, 233 Springer Street, New York, 1996).

[16] Zhao, M., Bosman, M., Danesh, M., Zeng, M., Song, P., Yudi Darma, Andrivo Rusydi, Lin, H., Qiu, C.-W., & Loh, K. P. Visible Surface Plasmon Modes in Single Bi2Te3 Plate. Nano Lett. 15, 8331–8335 (2015), DOI: https://doi.org/10.1021/acs.nanolett.5b03966, La, H., Brokkelkamp, A., van der Lippe, S., ter Hoeve, J., Rojo, J., & Conesa-Boj, S. Edge-Induced excitations in Bi2Te3 from spatially-resolved electron energy-gain spectroscopy. Ultramicroscopy 254, (2023), DOI: https://doi.org/10.48550/arXiv.2305.03752

[17] Talebi, N., Ozsoy-Keskinbora, C., Benia, H.M., Kern, K., Koch, C.T., van Aken, P.A. Wedge Dyakonov Waves and Dyakonov Plasmons in Topological Insulator Bi2Se3 Probed by Electron Beams. ACS Nano 10, 6417–7224 (2016), DOI: https://doi.org/10.1021/acsnano.6b02968

[18] Hsu, J.-C., Lee, H.-L., Chang, W.-J. Flexural vibration frequency of atomic force microscope cantilevers using the Timoshenko beam model, Nanotechnology 18 285503, (2007), DOI: https://doi.org/10.1088/0957-4484/18/28/285503;%20COUNTRY%20OF%20INPUT:%20INTERNATIONAL%20ATOMIC%20ENERGY%20AGENCY%20(IAEA)

[19] Martínez-Martín, D., Jaafar, M., Pérez, R., Gómez-Herrero, J., & Asenjo, A. Upper Bound for the Magnetic Force Gradient in Graphite Phys. Rev. Lett. 105, 257203 (2010), DOI: https://doi.org/10.1103/PhysRevLett.105.257203

[20] Zahid, F., Lake, R. Thermoelectric properties of $Bi_2Te_3$ atomic quintuple thin Films Appl. Phys. Lett. 97, 212102 (2010) https://doi.org/10.1063/1.3518078

[21] Pedrak, R., Ivanov, Tzv., Ivanova, K., Gotszalk, T., Abedinov, N., Rangelow, I. W., Edinger, K., Tomerov, E., Schenkel, T., & Hudek, P. Micromachined atomic force microscopy sensor with integrated piezoresistive sensor and thermal bimorph actuator for high-speed tapping-mode atomic force microscopy phase-imaging in higher eigenmodes. J. Vac. Sci. Technol B 21, (2003), DOI: https://doi.org/10.1116/1.1614252, Lohau, J., Kirsch, S., Carl, A., Dumpich, G. & Wassermann, E. F. Quantitative determination of effective dipole and monopole moments of magnetic force microscopy tips. Journal of Applied Physics 86, 3410–3417 (1999), DOI: https://doi.org/10.1063/1.371222

[22] Sarid, D. Scanning Force Microscopy: With Applications to Electric, Magnetic, and Atomic Forces. (Oxford University Press, New York, 1994).

[23] Dotdaev, A. S., Rodionov, Y. I., Rozhkov, A. V., Grigoriev, P. D., Semiclassical Scattering by Edge Imperfections in Topological Insulators in a Magnetic Field. Jetp Lett. 120, 675–686 (2024). https://doi.org/10.1134/S0021364024603154

[24] Delplace, P., Li, J., Büttiker, M. Magnetic-Field-Induced Localization in 2D Topological Insulators, Phys. Rev. Lett. 109, 246803 (2012), DOI: https://doi.org/10.1103/PhysRevLett.109.246803

[25] Mironov, S. V., Mel'nikov, A. S., Tokman, I. D., Vadimov, V., Lounis, B., Buzdin, A. I. Inverse Faraday Effect for Superconducting Condensates, Phys. Rev. Lett. 126, 137002 (2021), DOI: https://doi.org/10.1103/PhysRevLett.126.137002

[26] Israelachvili, J. N. Intermolecular and Surface Forces 3rd edition, Elsevier Academic Press, Amsterdam (2011) ISBN: 9780123919274

Jaafar, M., Iglesias-Freire, O., Serrano-Ramón, L., Ibarra, M. R., de Teresa, J. M., & Asenjo, A. Distinguishing magnetic and electrostatic interactions by a Kelvin probe force microscopy-magnetic force microscopy combination. J. Nanotechnol. 2, 552–560 (2011), DOI: https://doi.org/10.3762/bjnano.2.59

Griffiths, D. J. Introduction to Electrodynamics, Cambridge University Press Cambridge U.K. (2017) ISBN: 978-1108420419

[27] This is a surprising relevant reference since the structure we are studying can be considered a composite conductor: Polizos, G., Tuncer, E., Tomer, V., Sauers, I., Randall, C. A., Manias, E., Dielectric Spectroscopy of Polymer-Based Nanocomposite Dielectrics with Tailored Interfaces and Structured Spatial Distribution of Fillers in Nanoscale Spectroscopy with Applications Ed. Sarhan Musa CRC PRESS (2013) ISBM 9781466568532

[28] Braginsky, L. S., Entin, M. V., Edge state magnetoresistance of a two-dimensional topological insulator, EPL, 134 (2021) 37001 doi: 10.1209/0295-5075/134/37001

[29] Wakaya, F. Detection of nonmagnetic metal thin film using magnetic force microscopy, Appl. Phys. Lett. 113, 261601 (2018) https://doi.org/10.1063/1.5079763

[30] Szczytko, J., Vaupotič, N., Madrak, K., Sznajder, P., Górecka, E. Magnetic moment of a single metal nanoparticle determined from the Faraday effect, Phys. Rev. E 87, 033201 (2013), DOI: https://doi.org/10.1103/PhysRevE.87.033201, Neves, C. S., Quaresma, P., Baptista, P. V., Carvalho, P. A., Araújo, J. P., Pereira, E., & Eaton, P. New insights into the use of magnetic force microscopy to discriminate between magnetic and nonmagnetic nanoparticles, Nanotechnology 21 (2010) 305706 (9pp) doi:10.1088/0957-4484/21/30/305706

[31] Denisov, S. I., Lyutyy, T. V., Pedchenko, B. O. & Babych, H. V. Eddy current effects in the magnetization dynamics of ferromagnetic metal nanoparticles. J. Appl. Phys. 116, (2014), Shanks, W. E. Persistent Currents in Normal Metal Rings. (PhD. Thesis, Yale University, 2011).

[32] Trainer, D. J., Putilov, A. V., Giorgio, C. Di, Saari, T., Wang, B., Wolak, M., Chandrasena, R. U., Lane, C., Chang, T.-R., Jeng, H.-T., Lin, H., Florian Kronast, Gray, A. X., Xi, X., Nieminen, J., Bansil, A., & Iavarone, M. Inter-Layer Coupling Induced Valence Band Edge Shift in Mono- to Few-Layer MoS2. Scientific Reports, 7, 40559, (2017), DOI: 10.1038/srep40559.

[33] Kimberly, T. Q., Ciesielski, K. M., Qi, X., Toberer, E. S., & Kauzlarich, S. M. High Thermoelectric Performance in 2D Sb2Te3 and Bi2Te3 Nanoplate Composites Enabled by Energy Carrier Filtering and Low Thermal Conductivity. ACS Applied Electronic Materials, 6(5) (2023), DOI: https://doi.org/10.1021/acsaelm.3c00385

[34] Rivera, M., Rios-Reyes, C. H., Mendoza-Huizar, L. H. Magnetic transition phase diagram of cobalt clusters electrodeposited on HOPG: Experimental and micromagnetic modelling study, Journal of Magnetism and Magnetic Materials 323 997–1000 (2011), DOI: https://doi.org/10.1016/j.jmmm.2010.11.088

[35] Makarova, T. L., Han, K.-H., Graphite under the magnetic force microscope, phys. stat. sol. (b) 244, No. 11, 4138–4142 (2007), DOI: 10.1002/pssb.200776201